\pdfoutput=1
\documentclass[twocolumn]{aastex631}

\newcommand{\cha}{\textit{Chandra}}
\newcommand{\xmm}{XMM-\textit{Newton}}
\newcommand{\nustar}{\textit{NuSTAR}}
\newcommand{\bat}{\textit{Swift}-BAT}

\usepackage{amsmath}
\shorttitle{AASTeX v6.3.1 Sample article}
\shortauthors{Cox et al.}
\graphicspath{{./}{figures/}}

\begin{document}

\title{Chandra Follow-up Observations of Swift-BAT-selected AGNs III}

\author[0000-0003-2287-0325]{Isaiah S. Cox}
\affiliation{Department of Physics and Astronomy, Clemson University, 
Clemson, SC, 29634, USA}

\author[0000-0003-3638-8943]{Núria Torres-Alb\`a}
\affiliation{Department of Physics and Astronomy, Clemson University, 
Clemson, SC, 29634, USA}

\author[0000-0001-5544-0749]{Stefano Marchesi}
\affiliation{Department of Physics and Astronomy, Clemson University, 
Clemson, SC, 29634, USA}
\affiliation{Dipartimento di Fisica e Astronomia, Università degli Studi di Bologna, via Gobetti 93/2, 40129 Bologna, Italy}
\affiliation{INAF-Osservatorio Astronomico di Bologna, Via Piero Gobetti, 93/3, I-40129, Bologna, Italy}

\author[0000-0001-9379-4716]{Peter Boorman}
\affiliation{Cahill Center for Astronomy and Astrophysics, California Institute of Technology, Pasadena, CA 91125, USA}

\author[0000-0002-7791-3671]{Xiurui Zhao}
\affiliation{Department of Astronomy, University of Illinois at Urbana-Champaign, Urbana, IL 61801, USA}

\author[0000-0001-6564-0517]{Ross Silver}
\affiliation{NASA-Goddard Space Flight Center, Code 660, Greenbelt, MD, 20771, USA}

\author[0000-0002-6584-1703]{Marco Ajello}
\affiliation{Department of Physics and Astronomy, Clemson University, 
Clemson, SC, 29634, USA}

\author[0000-0002-7825-1526]{Indrani Pal}
\affiliation{Department of Physics and Astronomy, Clemson University, 
Clemson, SC, 29634, USA}




\begin{abstract}

The cosmic X-ray background (CXB) is dominated by the obscured and unobscured coronal light of active galactic nuclei (AGN). At energies below 10\,keV, the CXB can be well explained by models taking into account the known AGN and the observed distribution of their obscuring, line-of-sight column densities, $N_{\rm H,l.o.s}$. However, at energies around the Compton reflection hump ($\sim30$\,keV), the models fall short of the data. This suggests the existence of a population of as yet undetected Compton-thick AGN ($N_{\rm H,l.o.s}>1.5\times10^{24}$\,cm$^{-2}$) whose X-ray spectra are dominated by the light that has been reprocessed by the obscuring material. In this work, we continue the effort to find and catalog all local ($z<0.05$) Compton-thick (CT) AGN. To this end, we obtained soft X-ray data with \cha\ for six local BAT detected sources lacking ROSAT (0.1-2.4\,keV) counterparts, indicating potential obscuration. We fit their spectra with Bayesian and least squares methods using two different models, \texttt{borus02} and \texttt{UXCLUMPY}. We compare the results of the different models and methods and find that the $N_{\rm H,l.o.s}$ is consistently measured in each case. Three of the sources also were observed with \xmm\ allowing the opportunity to search for variability in soft X-ray flux or $N_{\rm H,l.o.s}$. From this sample, we find one strong CT candidate (NGC\,5759) and one weaker CT candidate (CGCG\,1822.3+2053). Furthermore, we find tentative evidence of $N_{\rm H,l.o.s}$ variability in 2MASX\,J17253053--4510279, which has $N_{\rm H,l.o.s}<10^{22}$\,cm$^{-2}$.

\end{abstract}



\section{Introduction} \label{sec:intro}

With the advent of high-sensitivity, imaging soft X-ray telescopes like \cha\ and \xmm, it is now well-established that active galactic nuclei (AGN) are the main contributors to the cosmic X-ray background \citep[CXB, e.g.,][]{Alexander03,Gandhi03,Gilli07,Treister09,ueda14,Cappelluti17,Ananna19}. The accretion of matter onto the central supermassive black hole in these objects results in UV and optical photons being upscattered to X-ray energies by the high energetic plasma in the corona, producing a power law with a photon index $\sim1.6-2.0$ \citep[e.g.,][]{Haardt93,Nandra94,Reeves2000}. In most sources this power law is attenuated by intervening matter through photoelectric absorption, which causes a change in the spectral shape, as lower energy X-rays are preferentially absorbed. The strength of this effect depends on the amount of intervening matter which is parameterized by the hydrogen column density in the line of sight, $N_{\rm H,l.o.s}$. Therefore, the shape of the CXB depends on the distribution of $N_{\rm H,l.o.s}$ in the AGN population \citep[e.g.,][]{Matt&Fabian94,Ghisellini94,Comastri95,Gilli99}.

Models accounting for the distribution of $N_{\rm H,l.o.s}$ in the AGN population are able to reproduce the CXB emission below 10\,keV \citep[e.g.,][]{Worsley05,Hickox06}. However, the excess in the CXB data at higher energies around the Compton hump $(\sim30$\,keV) indicates the presence of a population of Compton-thick (CT) AGN with very strong absorption \citep[$N_{\rm H,l.o.s}>\sigma_{T}^{-1}\sim10^{24}$\,cm$^{-2}$;][]{Ajello08a}. At these energies, $<40$\,\% of the CXB has been resolved by \nustar\ \citep[e.g.,][]{Harrison16}. Models indicate that the CT fraction in the local universe could be up to 30-50\,\% \citep[e.g.,][]{Gilli07,ueda14,Ananna19}. Hard X-ray, flux limited observations so far have shown that this fraction is only $<20$\,\% \citep[e.g.,][]{DellaCeca08,Burlon11,Vasudevan13,Ricci15,Lanzuisi18,NTA21}. There are currently only 66 AGN\footnote{All 66 sources have $z<0.1$. At $z\gtrsim0.05$, only quasars with $L_{2-10\text{\,keV}}>10^{44}$\,erg\,s$^{-1}$ have been found to be CT.} that have \nustar\ observations confirming their CT nature \citep{Boorman24}. It is important to find and catalog all of the CT-AGN within a volume-limited ($z<0.05$), complete sample to estimate the true CT fraction and confirm their contribution to the CXB. 

Furthermore, the radiation from these objects is typically dominated by the reflection component coming from reprocessing of the primary power law by the surrounding material. This region is commonly referred to as the dusty torus. It is often used to explain infrared observations as well as the observed differences between many classes of AGN \citep[e.g.,][]{antonucci_unified_1993,urry_unified_1995}. The X-ray signature of the torus is usually drowned out by the primary power law originating from the corona. However, in CT-AGN, the absorption at energies $>10$\,keV is strong enough to allow the reflection component to outshine the primary power law and be detected by \nustar\ \citep[e.g.,][]{arevalo14,Bauer15}. 

Physically-motivated torus models such as \texttt{MYTorus} \citep{murphy_x-ray_2009}, \texttt{borus02} \citep{balokovic_new_2018}, and \texttt{UXCLUMPY} \citep{buchner_x-ray_2019} can be used to model the X-ray spectra of CT-AGN and constrain properties of the torus geometry \citep[e.g.][]{Marchesi19,Zhao21}. However, the small number of known CT-AGN limits the conclusions we can make about the torus and different models can sometimes yield conflicting results on the torus parameters such as the covering factor \citep[e.g.][]{Brightman15,Saha22,Kallova24,Boorman24}. By increasing the number of known CT-AGN, we can further understand the nature of the obscuring material in AGN as well as the AGN contribution to the CXB. 

Combining data from both hard X-ray telescopes like \bat\ \citep{Barthelmy05} along with soft X-ray data from \cha\ and/or \xmm, for example, has been shown to be a reliable method for selecting candidate CT-AGN. \cite{marchesi17} and \cite{Silver22a} used a program to identify hard X-ray sources from the 150-month \bat\ catalog and observe them with quick $(\sim10$\,ks) \cha\ exposures resulting in $\sim95$\,\% of sources being obscured and $\sim20$\,\% being CT candidates \citep{Silver22a}. The simultaneous fitting of the \cha\ and \bat\ data provides an estimate of $N_{\rm H,l.o.s}$ which can be confirmed with a follow-up \nustar\ observation \citep[e.g.][]{marchesi17a,Marchesi18,Zhao19a,Zhao19b,Traina21,NTA21,Silver22b,Sengupta23}. By this method, the population of known CT-AGN can be slowly increased and their tori properties can be constrained. 

In this work, we continue the effort of the Clemson-INAF Compton thick AGN project (CI-CTAGN)\footnote{\url{https://science.clemson.edu/ctagn/}} which has so far confirmed $\sim30$ CT-AGN in the local universe. We present the first \cha\ results of 6 AGN selected according to the method explained in \cite{Silver22a} and outlined below. We fit these \cha\ data simultaneously with the \bat\ and \xmm\ data (when available) to provide $1-150$\,keV coverage to constrain the line-of-sight $N_{\rm H,l.o.s}$ and torus geometrical properties. For those sources with \xmm\ and \cha\ data available, we also look for evidence of variability in $N_{\rm H,l.o.s}$ as has been done in recent works \citep[e.g.][]{pizzetti_multi-epoch_2022,NTA23,Pizzetti:2024}.

In Section \ref{sec:data}, we briefly recount the selection criteria from \cite{Silver22a} and present the data and reduction methods. In Section \ref{sec:analysis}, we present the fitting methods and models used to obtain the $N_{\rm H,l.o.s}$ and torus parameters. We present the results of the modeling and provide CT-AGN candidates in Section \ref{sec:results}. In Section \ref{sec:discussion}, we compare the results of the models and fitting methods and discuss potential variability in obscuration. We summarize our results and discuss future work in Section \ref{sec:conclusions}.

We adopt a cosmology with $H_0=70$\,km\,s$^{-1}$\,Mpc$^{-1}$, $\Omega_m=0.27$, and $\Omega_{\Lambda}=0.73$. All reported errors are at a 90\,\% confidence level unless otherwise stated.


\section{Sample Selection and Data} \label{sec:data}

\begin{table*}[t]
\centering
  \caption{Summary of the sample and the related X-ray observations. All \cha\ observations are 10\,ks long.}\label{tab:sample}
\begin{tabular}{ c c c c c c c c }
\hline\hline
 Source Name & RA & Dec & Redshift & Counts\,(Rate)\footnote{These are the total counts (count rate) used in the fitted spectrum.} & Counts\,(Rate)\footnote{These are the sum of the total counts (count rates) used in the fitted spectra from both MOS cameras and the PN camera.} & Obsid & Obsid \\
   & (J2000) & (J2000) &  & \cha & \xmm & \cha & \xmm  \\
   \hline
   2MFGC9836 & 187.206082 & -29.627365 & 0.0276 & 788\,(7.91E-2) & $-$ & 25258 & $-$ \\
   NGC5759 & 221.811739 & 13.456590 & 0.0277 & 36\,(3.36E-3) & 288\,(1.43E-2) & 25257 & 0881840201 \\
   IC1141 & 237.445689 & 12.399175 & 0.0147 & 80\,(7.92E-2) & $-$ & 25256 & 0881840101\footnote{These data are not used due to strong background flaring throughout the exposure.} \\
   2MASXJ17253053--4510279 & 261.377141 & -45.174592 & 0.0196 & 683\,(6.85E-2) & 1846\,(7.84E-1) & 25260 & 0881840501 \\
   CGCG1822.3+2053 & 276.110536 & 20.902621 & 0.0167 & 67\,(6.61E-3) & 547\,(6.95E-2) & 25259 & 0881840401 \\
   MCG+02-57-2 & 335.937290 & 11.835737 & 0.0294 & 1514\,(1.52E-1) & $-$ & 25261 & $-$ \\
   \hline
\end{tabular}
\end{table*}

The sample in this work was selected from hard X-ray sources detected by \bat\ (in $15-150$\,keV) that lack a ROSAT \citep{Voges99} counterpart (in $0.1-2.4$\,keV). Five of the sources are from the 150-month \bat\ catalog (Imam et al. in preparation\footnote{The 150 month catalog can be found here \url{https://science.clemson.edu/ctagn/bat-150-month-catalog/}}) while one source (2MASX\,J17253053--4510279) is from the 157-month \bat\ catalog (Lien et al. in preparation\footnote{The 157 month catalog can be found here \url{https://swift.gsfc.nasa.gov/results/bs157mon/}}) while also being detected in the 105 month catalog \citep{Oh18}. The five sources from the 150-month \bat\ catalog were processed by the \texttt{BAT\_IMAGER} code \citep{segreto2010palermo} and their spectra have been background subtracted and averaged over the entire 150 month exposure time.

The fact that these sources were detected in hard X-rays but not soft X-rays indicates possible X-ray obscuration with column densities $N_{\rm H,l.o.s}>10^{23}$\,cm$^{-2}$ \citep[see e.g.,][and \citealt{koss_new_2016}]{ajello2008swift}. However, \cite{Silver22a} showed that this threshold may be lower ($N_{\rm H,l.o.s}\gtrsim5\times10^{22}$\,cm$^{-2}$) as the 150-month \bat\ catalog goes deeper in flux and detects intrinsically fainter sources. Furthermore, we avoided sources classified as Seyfert 1, since the presence of strong broad lines in the optical spectrum is generally correlated with unobscured AGN \citep[e.g.,][]{Koss17}. Among our sources, only MCG\,+02-57-2 is classified as a Seyfert 1.9 indicating only a weak H$\alpha$ broad line present \citep{smith2020bat} and some Seyfert 1.9 galaxies have large $N_{\rm H,l.o.s}$ values \citep[e.g.,][]{Shimizu18}. We note that while Seyfert 2 galaxies are more obscured than Seyfert 1s on average \citep[see e.g.,][]{marchesi16}, a significant fraction \citep[as high as 66\,\% e.g.,][]{garcet2007xmm} of Seyfert 2 galaxies are found to have unobscured X-ray emission in some samples. It should also be noted that, although we exclude Seyfert 1 galaxies from our selection, they can also be obscured in the X-rays \citep[e.g.][]{Risaliti2001,Risaliti2007,merloni2014incidence,Miniutti2014,Shimizu18,Serafinelli2023}.

\subsection{Chandra}
All six objects in our sample were observed with ACIS-S for $\sim10$\,ks during \cha\ guest observing Cycle 23 (ID: 23700077, P.I. Silver\footnote{\url{https://cxc.harvard.edu/cgi-bin/propsearch/prop_details.cgi?pid=6115}}). The observations can be downloaded at~\dataset[DOI: 10.25574/cdc.321]{https://doi.org/10.25574/cdc.321}. A summary of the sample along with the \cha\ observation date and $1-7$\,keV count rate is shown in Table \ref{tab:sample}. The \cha\ data were reduced and spectra were obtained by following standard procedures using the CIAO version 4.13 software \citep{fruscione_ciao}. We used a $5"$ circular region for the source and an annulus with $r_{\text{in}}=6"$ and $r_{\text{out}}=15"$ for the background region. We used CIAO \texttt{specextract} to extract the spectrum in each of these regions. We then binned the source spectrum to 5\,counts/bin for spectral fitting with the W statistic, cstat \citep{Wachter79}. 

\subsection{XMM-Newton}
\begin{figure}
    \centering
    \includegraphics[width=0.45\textwidth]{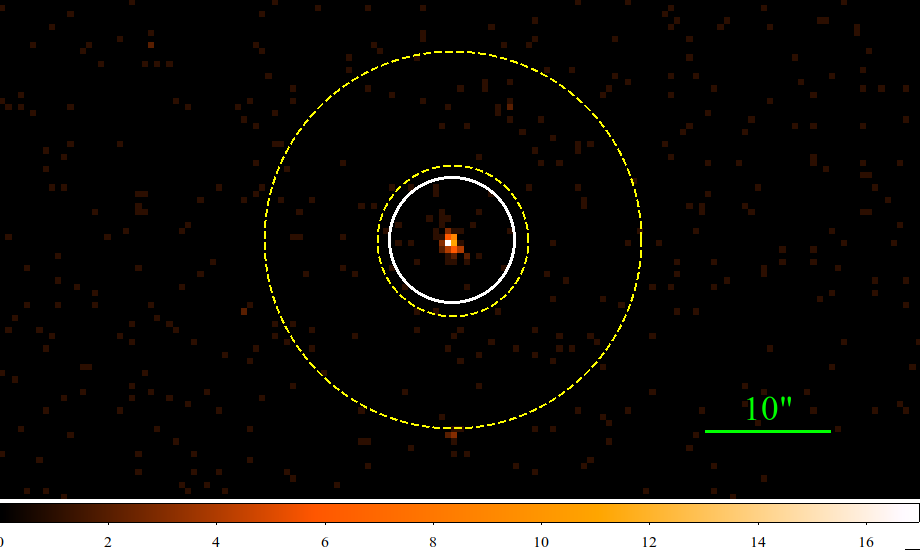}\hfill
    \includegraphics[width=0.45\textwidth]{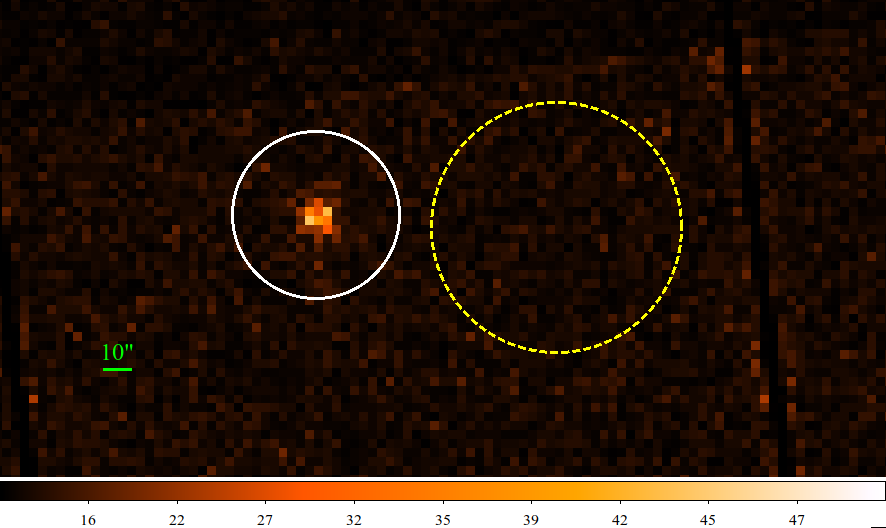}\hfill
    \caption{\textit{Top panel}: Image showing the \cha\ data for CGCG\,1822.3+2053 with the source extraction region indicated by the solid white circle and the background extraction region indicated by the dashed yellow annulus. \textit{Bottom panel}: Source and background extraction regions for the \xmm\ data overlayed on an image from the EPIC-PN camera. In both panels, North is up and East is left.}
    \label{fig:cgc_image}
\end{figure}

Three of the sources in our sample also have available data from \xmm\ (CGCG\,1822.3+2053, 2MASX\,J17253053--4510279, and NGC\,5759) (ID: 088184, P.I. Silver\footnote{\url{http://esdcdoi.esac.esa.int/doi/html/data/astronomy/xmm-newton/088184.html}}). The cleaned exposure times for CGCG\,1822.3+2053 are 8.1\,ks, 8.0\,ks, and 5.3\,ks for the MOS-1, MOS-2, and PN cameras, respectively. For 2MASX\,J17253053--4510279, the cleaned times are 4.1\,ks, 3.9\,ks, and 1.1\,ks, and for NGC\,5759 the cleaned times are 13.7\,ks, 13.5\,ks, and 9.7\,ks. The \xmm\ data were reduced following standard procedures using the SAS version 21.0.0 \citep{jansen_xmm}. To extract the spectra, we used a 30" circular region for the source and a 45" circular region for the background. The background region was chosen as close as possible to the source and at a similar height on the CCD while avoiding contamination from other sources. The spectra were binned to 5\,counts/bin for fitting with cstat. The \cha\ and \xmm\ images of CGCG\,1822.3+2053 with the source and background extraction regions are shown in Figure \ref{fig:cgc_image}, as an example.

\newpage

\section{Spectral Analysis} \label{sec:analysis}

We simultaneosly fit the \cha, \bat\, and \xmm\ (when available) data using two different regression methods and two different torus models. Simultaneously fitting the soft X-ray data from \cha\ and \xmm\ with the hard X-ray data from \bat\ is useful to break the degeneracy between the photon index and the $N_{\rm H,l.o.s}$ in obscured sources. For consistency with the analysis in \cite{Silver22a}, we exclude data below 1\,keV as it contains little information on $N_{\rm H,los}$ for moderately to heavily obscured sources. Excluding these data allow for simpler models while still providing accurate constraints on $N_{\rm H,los}$. Furthermore, the \texttt{borus02} model is tabulated only for energies $>$1\,keV and we decide to keep the fitted energy range consistent across all models.  We describe the regression methods and models in this section.

\subsection{Regression Methods}

We use two different regression algorithms to perform the fits. We describe them in this section.

\subsubsection{Levenberg-Marquardt (LM)}

Most commonly, X-ray spectra are fitted using some variation of the method of least squares. The computational requirements of this method are relatively light, however, the regression can lead to local minima rather than the global minimum. Therefore, the results are potentially sensitive to the initial parameter values. Furthermore, linear approximations are required which can significantly affect results when the model exhibits nonlinearity. Nevertheless, this type of regression is most commonly used in X-ray spectral fitting due to the computational efficiency.

The final results of this type of fitting are an estimate on the value of parameters of interest along with confidence intervals for each parameter. The best parameter estimate is the value of the parameter resulting in the model with the smallest fit statistic that the regression was able to find. The confidence interval for parameters of interest can be calculated with the \texttt{steppar} command which steps through a range of parameter values and computes the fit statistic with the parameter fixed at each step value. We take the 90\,\% confidence interval to be the range of parameter values for which the difference between the fit statistic and the minimum fit statistic is less than 2.7 \citep[e.g.,][]{Bonamente2020}. 

We use the Levenberg–Marquardt (LM) algorithm built in to \texttt{XSPEC} version 12.11.1 for the least squares regression. We fit all data available for each source using the models outlined in Section \ref{sec:models}. We begin each fit with the same set of parameters for each source. Particularly, we start at $\Gamma=1.8$ and $N_{\rm H}/(10^{22}\text{\,cm}^{-2})=10$ with all the torus parameters frozen. We then thaw the torus geometry parameters one at a time and ensure that we reach a stable fit by running \texttt{error} calculations on each free parameter until no new best fits are found. The output of the \texttt{steppar} calculation is used to create confidence contours on $N_{\rm H}$ and $\Gamma$.

\subsubsection{Nested Sampling}

An alternative to the LM algorithm is to use a Bayesian approach which we accomplish using nested sampling \citep{Buchner_2023,buchner2023statisticalaspectsxrayspectral}. This method allows for a global search over complex parameter spaces and therefore avoids getting trapped in local minima. The primary downside of nested sampling is the computation time. For models with more than a few free parameters and multiple observations, this method becomes infeasible without the use of high performance computing clusters.

The final result of this Bayesian approach is the posterior distributions for each parameter of interest. The posterior distribution contains all of the information obtainable about the unknown parameter from the prior knowledge of the parameter and the information in the observed data. We take the best estimate of the parameter to be the most probable value of the posterior distribution. We take the 90\,\% confidence interval (or `credible' interval) to be the interval containing 90\,\% of the posterior, with the mass below and above the interval being equal. That is, we calculate the 5th and 95th percentiles, and exclude the values outside this range. 

To obtain the posterior distributions, we use the Bayesian X-ray analysis tool \texttt{BXA} \citep{buchner14bxa} version 4.0.5 which uses the nested sampling algorithm \texttt{UltraNest} \citep{buchner21ultranest} within the \texttt{PyXSPEC}  \citep{Gordon21} fitting environment. We use uniform priors for the photon index, $N_{\rm H,los}$ and torus geometry parameters. See Table \ref{tab:fit_parameters}. We use log-uniform priors (that is, uniform priors on a log scale) for the normalization, cross-normalization, and scattering fraction parameters.  The fits for each source were performed on Clemson University's high-performance computing cluster, \textit{Palmetto}, using 20 cores and 16GB RAM. With these specifications the computation time required ranged from a few hours (for IC\,1141) up to a few hundred hours (for CGCG\,1822.3+2053).

\subsection{Models} \label{sec:models}

We use two different physically motivated torus models (\texttt{borus02} and \texttt{UXCLUMPY}) to model the reflection component seen in the X-ray spectrum. In most cases, the parameters determining the reflection component are unconstrained so we also use simpler model configurations (\texttt{borus02*} and \texttt{MYTorus}) as a consistency check for the $N_{\rm H,los}$ results. A summary of the initial guesses and priors for parameters of interest is shown in Table \ref{tab:fit_parameters}.

\begin{deluxetable*}{|c|cc|cc|}

\tablecaption{Summary of initial inputs and priors. \label{tab:fit_parameters}}

\tablehead{Model & borus02 & UXCLUMPY & borus02 & UXCLUMPY \\ 
& \multicolumn{2}{c|}{Least Squares Initial Input} & \multicolumn{2}{c|}{Bayesian Prior}
}

\startdata
$\Gamma$ & 1.8 & 1.8 & Uniform(1.4,2.6) &  Uniform(1.4,2.6) \\
$\log(N_{H,\rm tor} /\text{\,cm}^{-2})$ & 24 & $-$ & Uniform(22,25.5) & $-$ \\
$CF_{\rm tor}$ & 0.67 & $-$  & Uniform(0.1,1.0) & $-$ \\
$\theta_{\rm inc}(^{\circ})$ & 45 & 90 & Uniform(19,87) & Uniform(0,90)  \\
$\sigma_{\rm tor}$ & $-$ &  28 &  $-$ & Uniform(6,84) \\
CTKcover & $-$ &  0 &  $-$ & Uniform(0.0,0.6) \\
\hline 
$N_{\rm H} (10^{22}\text{\,cm}^{-2})$ & 10 & 10 & Uniform(0.1,500) & Uniform(0.1,500)  \\
\enddata
\tablecomments{The bounds for torus geometry parameters and photon index span the entire allowable range for each model.}
\end{deluxetable*}

\subsubsection{\texttt{borus02}} \label{sec:borus02}

The first model is \texttt{borus02} \citep{balokovic_new_2018}. This model assumes a uniform density, spherical torus with conical polar cutouts. The opening angle is determined by the covering factor $c_f=\cos(\theta_{\text{torus}})$, which we leave free to vary in the range $0.1-1.0$. The inclination angle with respect to the observer is provided as $\cos(\theta_{\text{inc}})$ and is left free to vary between 0.05 (edge-on) and 0.95 (face-on). The torus column density, $\log(N_{H,\text{torus}}/\text{cm}^{-2})$, is left free to vary between $22.0-25.5$. The model provides the relative abundance of iron as a parameter, however, we chose to leave it fixed at the solar abundance. This torus geometry is illuminated by an X-ray source located at the center, radiating with a photon index, $\Gamma$, and a high-energy cutoff, $E_{\text{cut}}$. We leave $\Gamma$ free to vary between $1.4-2.6$. We leave $E_{\text{cut}}$ fixed at 300\,keV \citep{ueda14,Balakovic20} in all sources.

To account for the line-of-sight component, we add a primary power law (with parameters tied to the X-ray source in the torus compenent) attenuated by photoelectric absorption (\textit{zphabs}) and Compton scattering out of the line of sight (\textit{cabs}). We also add a scattering component which is parameterized as a fraction, $f_s$, of the primary power law to account for elastic scattering of X-rays into the line of sight.

The model is implemented in \texttt{XSPEC} as
\begin{align*}
    \texttt{borus02} = \quad&C\cdot \textit{phabs} \cdot(\text{borus02\_v170323a.fits} \\ &+\textit{zphabs}\cdot\textit{cabs}\cdot\textit{zcutoffpl}\\
    &+f_s\cdot\textit{zcutoffpl}),
\end{align*}
where $C$ is a constant to account for flux variability between observations and \textit{phabs} accounts for Galactic absorption. 

\subsubsection{\texttt{UXCLUMPY}} \label{sec:uxclumpy}

The second model is \texttt{UXCLUMPY} \citep{buchner_x-ray_2019}. This model accounts for the clumpiness of the torus and assumes a distribution of spherical clouds surrounding a central X-ray source producing a cutoff power law similar to \texttt{borus02}. We again leave the photon index free to vary between $1.4-2.6$ and $E_{\text{cut}}$ fixed at 300\,keV. 

The torus cloud geometry is determined by the parameter \textit{TORsigma}, which sets the width of the gaussian distributed cloud distribution. We leave this parameter free to vary between $6^{\circ}-84^{\circ}$. The inclination angle is left free to vary between $0^{\circ}$ (face-on) and $90^{\circ}$ (edge-on). To account for strong reflection in some sources, the parameter \textit{CTKcover} provides an additional Compton-thick reflecting component with a covering factor that we leave free to vary between $0-0.6$. Unlike \texttt{borus02}, the hydrogen column density, $N_{H}$, provided by \texttt{UXCLUMPY} is the $N_{\rm H,l.o.s}$. Both photoelectric absorption and Compton scattering are accounted for in the model. We compare the best-fit value for this parameter to the value found for \textit{zphabs} and \textit{cabs} in \texttt{borus02}. 

The scattering component is also computed by \texttt{UXCLUMPY} as a warm mirror which can scatter the primary power law as well as some reflected emission into the line of sight. These parameters are tied to the main parameters and the strength is set with the scattering fraction $f_s$ similar to \texttt{borus02}. 

The total model is implemented in \texttt{XSPEC} as
\begin{align*}
    \texttt{UXCLUMPY} = \quad&C\cdot \textit{phabs} \cdot(\text{uxclumpy-cutoff.fits} \\
    &+f_s\cdot\text{uxclumpy-cutoff-omni.fits}).
\end{align*}

\subsubsection{\texttt{borus02*}} \label{sec:borus02_fixed}

In general, the data presented in this paper are not of high enough quality to constrain the AGN torus parameters. Therefore, we also fit each source using the \texttt{borus02} model with the parameters $N_{\rm H,tor}$, $c_f$, and $\cos(\theta_{\rm inc})$ frozen at their typical values, as reported in \cite{Zhao21}. In this work, the authors find that for both Compton-thin and CT-AGN, the average torus column density is $\log(N_{\rm H,tor}/\text{cm}^{-2})\sim24$ and the average covering factor is $\sim$0.67 for Compton-thin sources. Since the distribution of torus inclination angles should be uniform between 0$^{\circ}$ and 90$^{\circ}$, we fix the value to 45$^{\circ}$ for lack of a better choice and to minimize the bias of assuming either an edge-on or a face-on scenario.

\subsubsection{\texttt{MYTorus}} \label{sec:mytorus}

We also use the \texttt{MYTorus} \citep{murphy_x-ray_2009} model with $N_{\rm H,los}$ coupled to the $N_{\rm H,tor}$ for the sources with only one soft X-ray observation. The model assumes a true torus geometry with a fixed half-opening angle of  60$^{\circ}$ which corresponds to a covering factor of 0.5. We leave the inclination angle fixed at 90$^{\circ}$ (edge-on).

For the sources with two soft X-ray observations we use the model with $N_{\rm H,los}$ decoupled from the $N_{\rm H,tor}$ to account for variability. We leave the inclination angle fixed at 90$^{\circ}$ for the \texttt{MYtorusZ} component which includes the line-of-sight absorption. The inclination angle for the \texttt{MYtorusS} and \texttt{MYtorusL} components is fixed at 45$^{\circ}$ and the $N_{\rm H,tor}$ is fixed at $1.0\times10^{24}\text{\,cm}^{-2}$ to match the \texttt{borus02*} model.

The total model is implemented in \texttt{XSPEC} as
\begin{align*}
    \text{\texttt{MYTorus}} = \quad&C\cdot \text{\textit{phabs}} \cdot( \\
    &\text{mytorus\_Ezero\_v00.fits}\cdot\text{\textit{zpowerlw}} \\
    &+C_S\cdot \text{mytorus\_scatteredH300\_v00.fits} \\
    &+C_L\cdot \text{mytl\_V000010nEp000H500\_v00.fits} \\
    &+f_s\cdot \text{\textit{zpowerlw}}).
\end{align*} 
where $C_S=C_L=1$. The $N_H$ in the \texttt{MYtorusZ} component is allowed to vary between the \cha\ and \xmm\ observations where applicable, as it is in the other three models.

The \texttt{MYTorus} model as implemented is unable to fit unobscured sources with $N_{\rm H,los}<10^{22}$\,cm$^{-2}$. Our sample includes two sources (2MASX\,J17253053--4510279 and MCG\,+02-57-2) for which that is the case. Since the reflection component is not important for these two sources, we use a simple absorbed powerlaw implemented in \texttt{XSPEC} as 
\begin{align*}
	\text{Simple} = C\cdot\textit{phabs}\cdot\textit{zphabs}\cdot\textit{zcutoffpl}
\end{align*}
in place of \texttt{MYTorus}.


\section{\texttt{Results}} \label{sec:results}

\begin{deluxetable*}{|c|cccc|cccc|}

\tablecaption{Summary of $N_{\rm H,l.o.s}$ $(10^{22}\text{\,cm}^{-2})$ measurements\label{tab:summary_results}}

\tablehead{Model & MYTorus & borus02* & borus02 & UXCLUMPY & MYTorus & borus02* & borus02 & UXCLUMPY \\ 
& \multicolumn{4}{c|}{Least Squares} & \multicolumn{4}{c|}{Bayesian}
}

\startdata
2MFGC\,9836 & 17.2$^{+2.5}_{-2.3}$ & 17.4$^{+1.5}_{-2.7}$ & 18.3$^{+2.5}_{-3.0}$ & 14.4$^{+3.7}_{-1.5}$ 
& 16.9$^{+2.7}_{-2.0}$ & 16.8$^{+2.8}_{-2.0}$ & 17.3$^{+3.9}_{-2.1}$ & 15.9$^{+2.7}_{-2.5}$  \\
NGC\,5759\,(\textit{Ch}) & 50$^{+63}_{-24}$ & 36.1$^{+43.8}_{-19.8}$ & 36$^{+42}_{-16}$ & 50$^{+75}_{-29}$ 
& 81$^{+350}_{-30}$ & 51.2$^{+390}_{-22}$ & 65$^{+390}_{-27}$ & 110$^{+340}_{-48}$  \\
NGC\,5759\,(\textit{XMM}) & $>$52 & $>$49 & $>$53 & $>$49 
& 140$^{+330}_{-50}$ & 117$^{+360}_{-36}$ & 290$^{+190}_{-190}$ & 170$^{+310}_{-57}$  \\
IC\,1141  & 44$^{+15}_{-12}$ & 42.4$^{+19.4}_{-14.6}$ & 43$^{+69}_{-13}$ & 47$^{+23}_{-16}$ 
& 45$^{+14}_{-13}$ & 43.2$^{+21.5}_{-12.5}$ & 49$^{+120}_{-15}$ & 50$^{+92}_{-15}$  \\
2MASX\,J17253053--4510279\,(\textit{Ch}) & 0.38$^{+0.21}_{-0.18}$ & 0.50$^{+0.18}_{-0.20}$ & -- & 0.51$^{+0.17}_{-0.16}$ 
& 0.53$^{+0.17}_{-0.17}$ & 0.54$^{+0.21}_{-0.15}$ & -- & 0.46$^{+0.16}_{-0.14}$  \\
2MASX\,J17253053--4510279\,(\textit{XMM}) & $<$0.15 & $<$0.18 & -- & $<$0.27 
& 0.11$^{+0.09}_{-0.01}$ & 0.12$^{+0.12}_{-0.02}$ & -- & 0.11$^{+0.08}_{-0.01}$ \\
CGCG\,1822.3+2053\,(\textit{Ch}) & 76$^{+58}_{-27}$ & 63.2$^{+38.9}_{-21.2}$ & 87$^{+35}_{-27}$ & 77$^{+48}_{-22}$ 
& 87$^{+220}_{-25}$ & 77$^{+360}_{-22}$ & 87$^{+300}_{-25}$ & 96$^{+100}_{-30}$  \\
CGCG\,1822.3+2053\,(\textit{XMM}) & 46$^{+11}_{-10}$ & 38.7$^{+9.5}_{-6.4}$ & 54$^{+16}_{-11}$ & 50$^{+21}_{-12}$ 
& 48$^{+12}_{-8}$ & 43$^{+9}_{-7}$ & 53$^{+16}_{-8}$ & 55$^{+16}_{-11}$ \\
MCG\,+02-57-2 & $<$0.43 & $<$0.35 & -- & $<$0.29 
& 0.13$^{+0.27}_{-0.02}$ & 0.15$^{+0.28}_{-0.03}$ & -- & 0.13$^{+0.23}_{-0.02}$
\enddata
\tablecomments{The MYTorus column shows the results using the simple absorbed powerlaw model for 2MASX\,J17253053--4510279 and MCG\,+02-57-2.}
\end{deluxetable*}

In this section, we present the fitting results for each individual source and discuss exceptions to the analysis presented in Section \ref{sec:analysis}. Table \ref{tab:summary_results} shows a summary of the measured $N_{\rm H,l.o.s}$ values for each source. Note that the $\chi^2$ fit statistics shown in Tables \ref{tab:cgc_results}$-$\ref{tab:2masx_results} for the Bayesian method are computed using the parameters from the posterior that result in the maximum likelihood. In general, these values are not the same as the most probable value for the parameter, which is shown in the tables. See Section \ref{subsec:posterior_values} for further discussion on this issue. We discuss only the results from the \texttt{borus02} and \texttt{UXCLUMPY} models, since they are consistent with the \texttt{borus02*} and \texttt{MYTorus} results and the latter two models were only used as a consistency check for the former. The results for CGCG\,1822.3+2053 are shown in this section (Figures \ref{fig:cgc_spectra}$-$\ref{fig:cgc_contours}) while the results for the rest of the sources are shown in the Appendix (Figures \ref{fig:2mfgc_results}$-$\ref{fig:mcg_corner}).

\subsection{\texttt{2MFGC\,9836}} \label{sec:2MFGC9836}

2MFGC\,9836 was classified as a Seyfert 2 galaxy by \cite{Chen22} based on optical spectroscopy. The source was observed with \cha\ in December of 2021 for 10\,ks. We ignored channels in the \bat\ data that had a count rate consistent with zero and were much lower than the rest of the data because they affected the fitting of the photon index. The \cha\ spectrum was fitted in the energy range 1-7\,keV.

The X-ray spectrum shows an obscured but decidedly Compton-thin nature. While the $N_{\rm H,l.o.s}$ is measured quite precisely, the torus parameters remain unconstrained in both models. The best-fit parameters and errors for both models and methods are shown in Table \ref{tab:2mfgc_results}. The $N_{\rm H,l.o.s}$ posteriors from the nested sampling and the contours from the LM fit along with the best-fit spectra are shown in Figure \ref{fig:2mfgc_results}. The corner plots for both models are shown in Figure \ref{fig:2mfgc_corner}. There is a 0.0\,\% probability of the data being Compton-thick for both models according to the $N_{\rm H,l.o.s}$ posteriors. We classify this source as an obscured, Compton-thin AGN.

\subsection{\texttt{NGC\,5759}} \label{sec:NGC5759}

NGC\,5759 is part of a galaxy pair with LEDA\,200319. The source was observed with \xmm\ in August of 2021 for 15\,ks and with \cha\ in September of 2022 for 10\,ks. The soft X-ray emission from LEDA\,200319 is primarily below 1\,keV. By extrapolating a powerlaw model from the data, the 15-150\,keV flux is estimated to be around a factor of 1000 less than the flux of NGC\,5759. Therefore, we assume negligible contamination in the \bat\ spectrum from LEDA\,200319. The \cha\ data were fitted in the energy range 1-7\,keV while the \xmm\ data were fitted in the energy range 1-8\,keV. The spectrum shows a heavily obscured nature. The best-fit parameters and errors for both models and methods are shown in Table \ref{tab:ngc_results}. The $N_{\rm H,l.o.s}$ posteriors from the nested sampling and the contours from the LM fit along with the best-fit spectra are shown in Figure \ref{fig:ngc_results}. The corner plots for both models are shown in Figure \ref{fig:ngc_corner}.

The \xmm\ observation yields a lower limit near the Compton-thick threshold in both the contours and the posterior distributions. The posteriors for the \cha\ data are largely consistent with a Compton-thick solution with more than half of the probability mass being in the Compton-thick regime. However the LM contours have a greater confidence in a Compton-thin solution. The posteriors for the \cha\ data show a 59.1\,\% and 60.6\,\% probability of being Compton thick according to the \texttt{borus02} and \texttt{UXCLUMPY} models, respectively. The posteriors for the \xmm\ data show a 84.1\,\% and 86.3\,\% probability. This source is the best Compton-thick candidate in the present sample and future \nustar\ data from cycle 10 (ID: 10209, P.I. Cox)  will help confirm this result. Given the current data, we classify this source as an obscured, Compton-thick AGN candidate.

\subsection{\texttt{IC\,1141}} \label{sec:IC1141}

IC\,1141 is classified as a low-ionization nuclear emission-line region (LINER) based on SDSS spectroscopic data \citep{toba14}. The source was observed in July of 2023 with \cha\ for 10\,ks and in February 2022 with \xmm\ for 13\,ks. The \xmm\ data is very poor quality due to a flare occurring throughout the observation. As a result, we use only the \cha\ and \bat\ data in our analysis. The \cha\ data were fitted in the energy range 1-7\,keV.

The X-ray spectrum shows a moderately obscured nature. The best-fit parameters and errors for both models and methods are shown in Table \ref{tab:ic_results}. The $N_{\rm H,l.o.s}$ posteriors from the nested sampling and the contours from the LM fit along with the best-fit spectra are shown in Figure \ref{fig:IC1141_results}. The corner plots for both models are shown in Figure \ref{fig:ic_corner}. Based on the $N_{\rm H,l.o.s}$ posteriors, there is a 5.5\,\% and 4.4\,\% probability of the source being Compton-thick according to the \texttt{borus02} and \texttt{UXCLUMPY} models, respectively. We classify this source as an obscured, Compton-thin AGN.

\subsection{\texttt{2MASX\,J17253053--4510279}} \label{sec:2MASXJ17253053-4510279}

2MASXJ17253053--4510279 is an AGN that has been previously classified as a Seyfert 2 galaxy \citep{koss22}. \cite{malizia2023update} classified it as a LINER galaxy that is most likely type 1 despite WISE data showing an unusually blue $W1-W2$ color (0.36) that can be indicative of absorption in AGN. They found it to be unabsorbed in the X-rays and noted the presence of a broad HeI line \citep{ricci22} leading to their classification of type 1. 

The source was observed in October 2021 with \xmm\, for 14.6\,ks and in October 2022 with \cha\ for 10\,ks. The \cha\ data were fitted in the energy range 1-7\,keV while the \xmm\ data were fitted in the energy range 1-9\,keV. The best-fit models along with the $N_{\rm H,l.o.s}$ posteriors from the nested sampling and the contours from the LM fit are shown in the appendix. As can be seen, the spectrum is unabsorbed. For this reason, we decided to remove the scattering component from the model and freeze all of the torus parameters as they could not be constrained, and caused the default model to be too computationally intensive for the Bayesian analysis.

The best-fit parameters and errors for both models and methods are shown in Table \ref{tab:2masx_results}. The $N_{\rm H,l.o.s}$ posteriors from the Bayesian analysis and the contours from the LM fit along with the best-fit spectra are shown in Figure \ref{fig:2masx_results}. The corner plots for both models are shown in Figure \ref{fig:2masx_corner}. There is a 0.0\,\% probability of the data being Compton-thick for both models according to the $N_{\rm H,l.o.s}$ posteriors. We classify this source as a Type 1 AGN. Interestingly, there is a strong hint of variability in the obscuration between the \cha\ and \xmm\ observations even though both epochs have $N_{\rm H,l.o.s}<10^{22}\,$cm$^{-2}$. The $N_{\rm H,l.o.s}$ is fairly well constrained in the \cha\ data, whereas the \xmm\ data provide an upper limit around the Galactic absorption, indicated by the dashed horizontal line in the contour plot of Figure \ref{fig:2masx_results}.

Another interesting thing to note about this source is the large difference in the cross-normalization between observations indicating flux variability. The fits show that the \xmm\ cross-normalization is $\sim2$ times higher than \cha, while the \bat\ cross-normalization is $\sim3-4$ times higher than \cha. Since the \bat\ flux should result in the average flux of the source over the 157 months, our soft X-ray observations, which are a full year apart, both occurred during a relatively low flux state. This also implies that the source must have been in a much higher flux state at some point during the 157 months of \bat\ integration time. The \bat\ lightcurve\footnote{The lightcurve data is available at \url{https://swift.gsfc.nasa.gov/results/bs157mon/1504}} indeed shows evidence of variability in the 14-195\,keV flux. The source is quite faint so we used a weighted average to rebin the lightcurve to $\sim$yearly bins which shows the source varies on the order of 1\,mCrab ($\sim 2\times10^{-11}$\,erg/s/cm$^{-2}$). A chi-square test assuming a constant flux results in a p-value of 0.03 indicating flux variability in the hard X-rays. The variable nature of this source could be the reason this source was selected as a potentially obscured AGN. Given the current data, we decide to classify this source as an unobscured, potentially $N_{\rm H,l.o.s}$-variable AGN.

\subsection{\texttt{CGCG\,1822.3+2053}} \label{sec:CGCG1822.3+2053}

\begin{figure}[h!]
    \centering
    \includegraphics[width=0.45\textwidth]{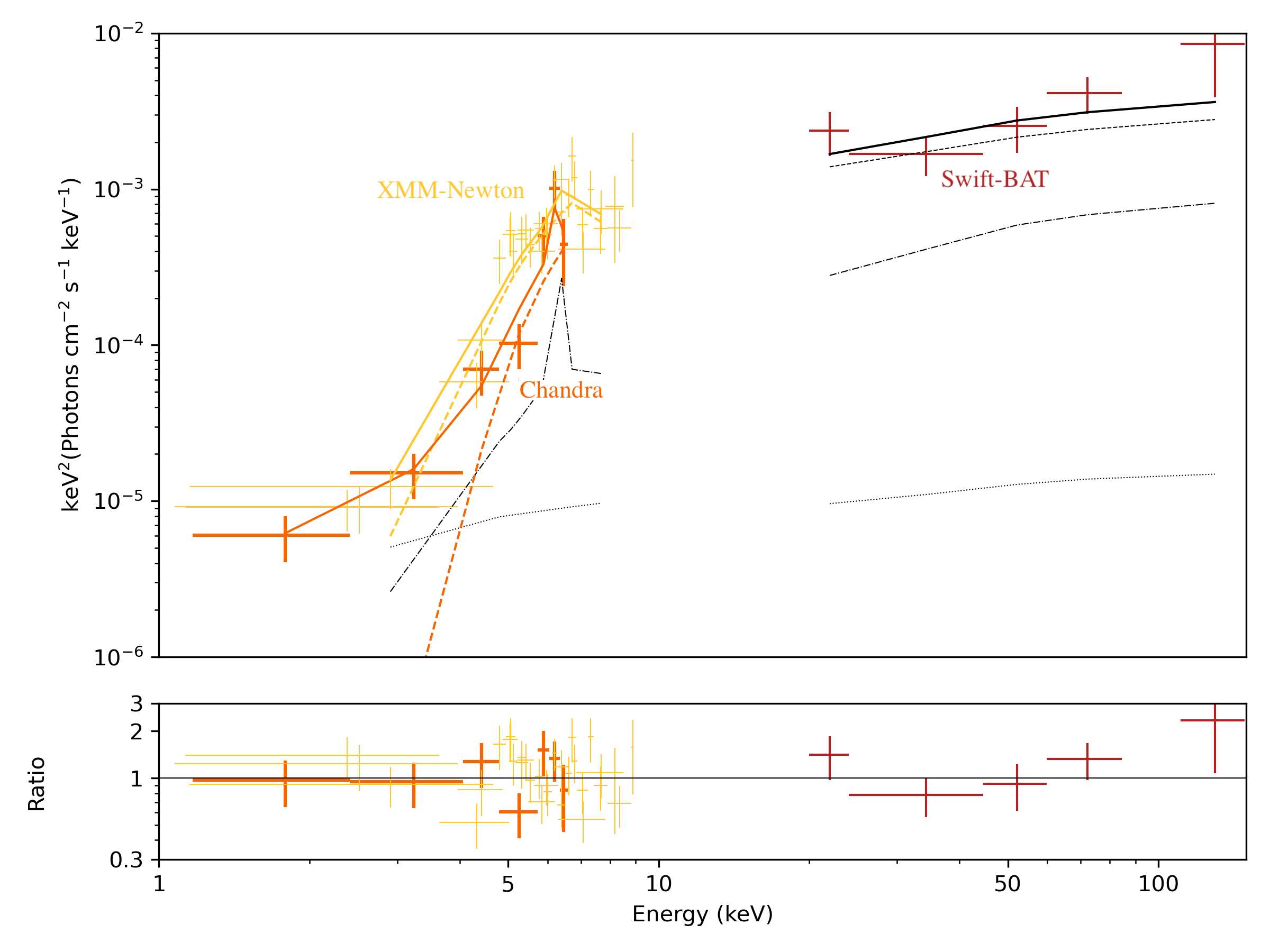}\hfill
    \includegraphics[width=0.45\textwidth]{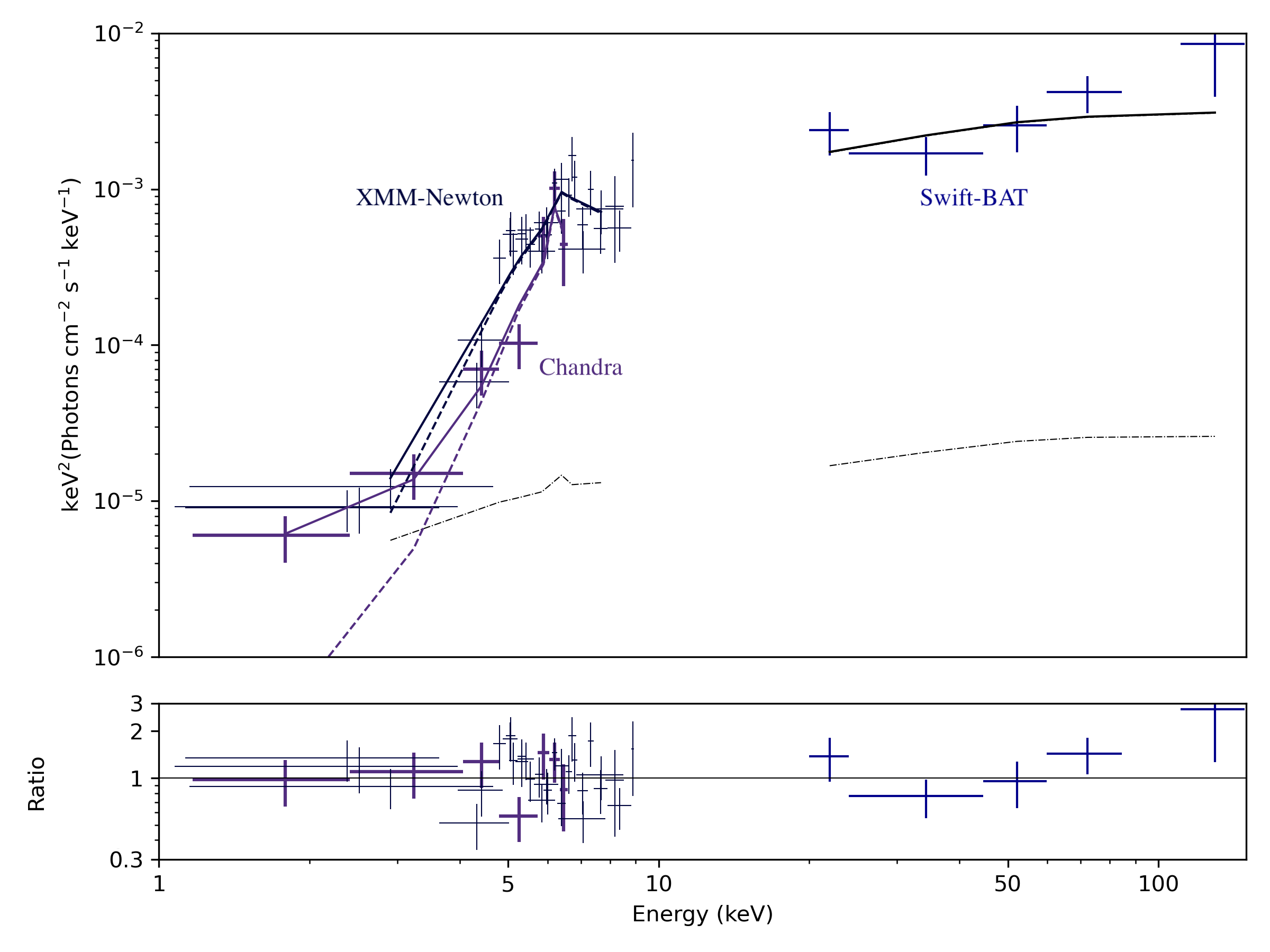}\hfill
    \caption{CGCG\,1822.3+2053. \textit{Top panel}: Best fit with the \texttt{borus02} model (solid lines). The binned \cha\ (thick orange), \xmm\ (thin yellow), and \bat\ (red) data are shown with the intrinsic (dashed lines), reflection (dash-dotted lines), and scattering (dotted lines) components. \textit{Bottom panel}: Best fit with the \texttt{UXCLUMPY} model (solid lines). The dashed lines show the transmitted and reflected components while the dash-dotted lines show the scattered component in \texttt{UXCLUMPY}.}
    \label{fig:cgc_spectra}
\end{figure}

\begin{figure}[h!]
    \centering
    \includegraphics[scale=.55]{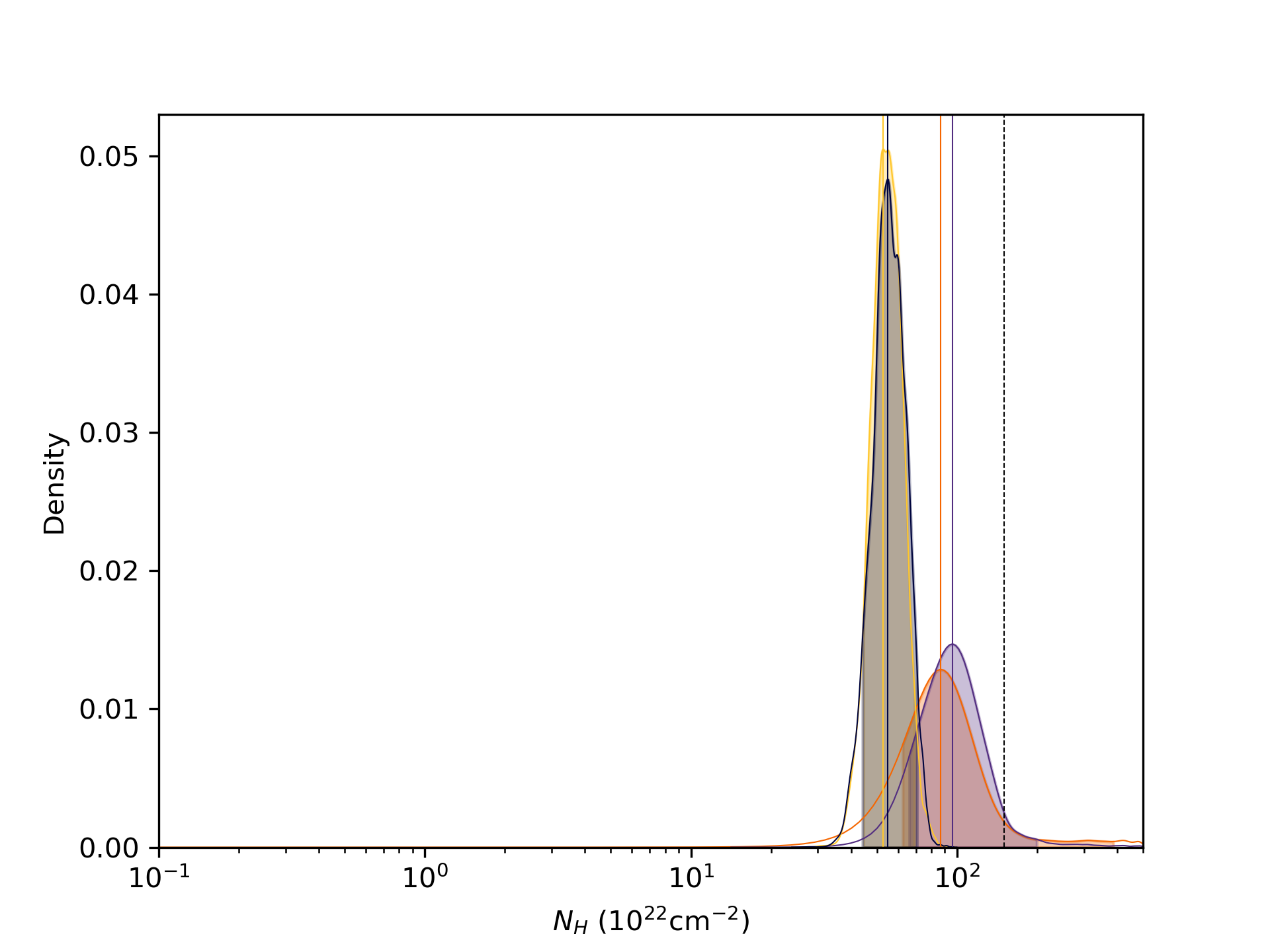}\hfill
    \caption{Posterior probability distribution of the line-of-sight $N_{\rm H,l.o.s}$ of CGCG\,1822.3+2053 for both models and both observations. The solid vertical lines indicate the most probable value for $N_{\rm H,l.o.s}$ and the shaded regions indicate the 90\,\% credible interval for the $N_{\rm H,l.o.s}$ measurement. The colors orange and purple correspond to the \cha\ observations fit with the \texttt{borus02} model and the \texttt{UXCLUMPY} model respectively. The colors yellow and blue correspond to the \xmm\ observations fit with the \texttt{borus02} model and the \texttt{UXCLUMPY} model respectively. The dashed vertical line indicates the Compton-thick threshold.}
    \label{fig:cgc_results}
\end{figure}

\begin{figure*}[h!]
    \centering
    \includegraphics[scale=.55]{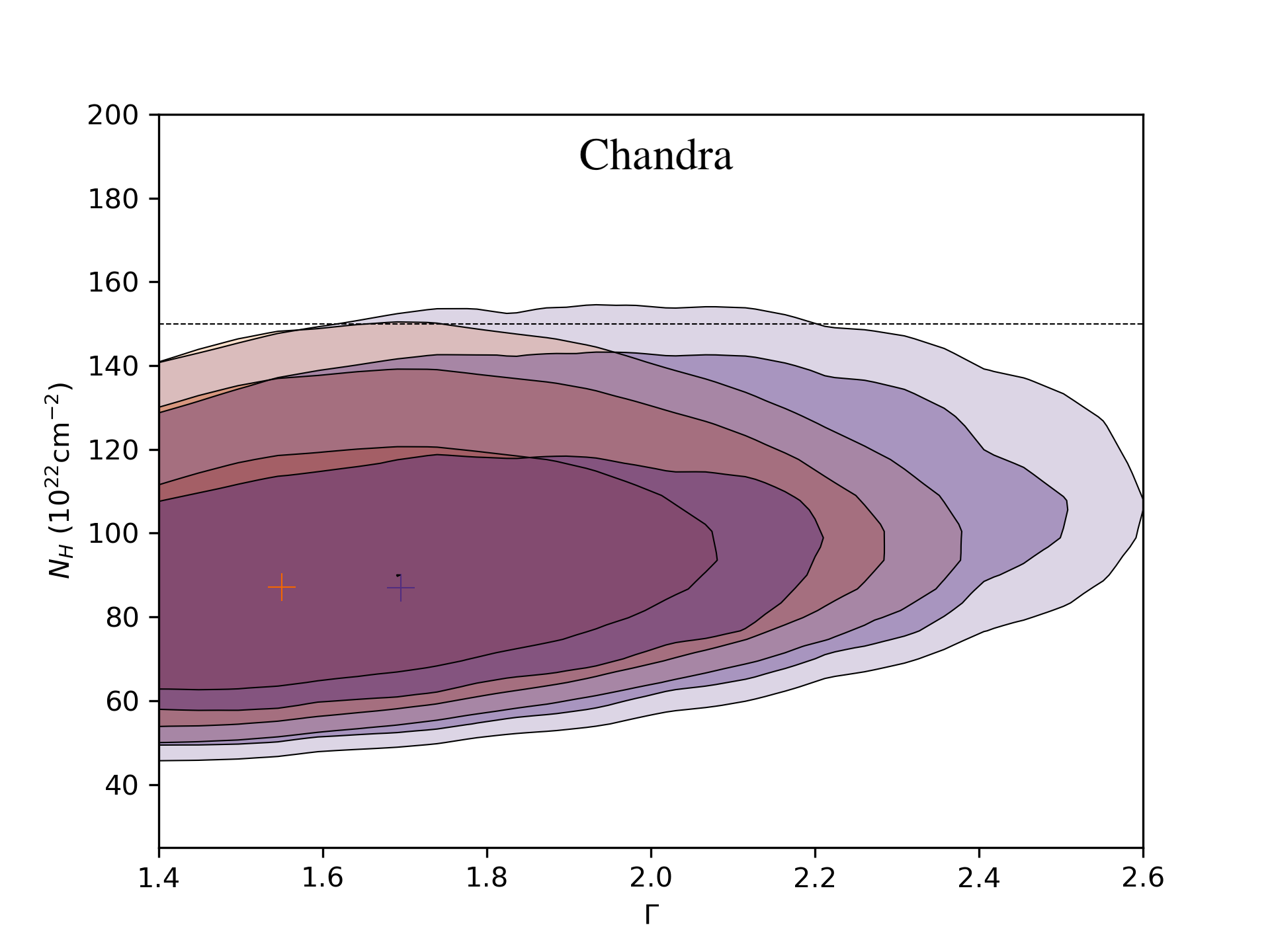}
    \includegraphics[scale=.55]{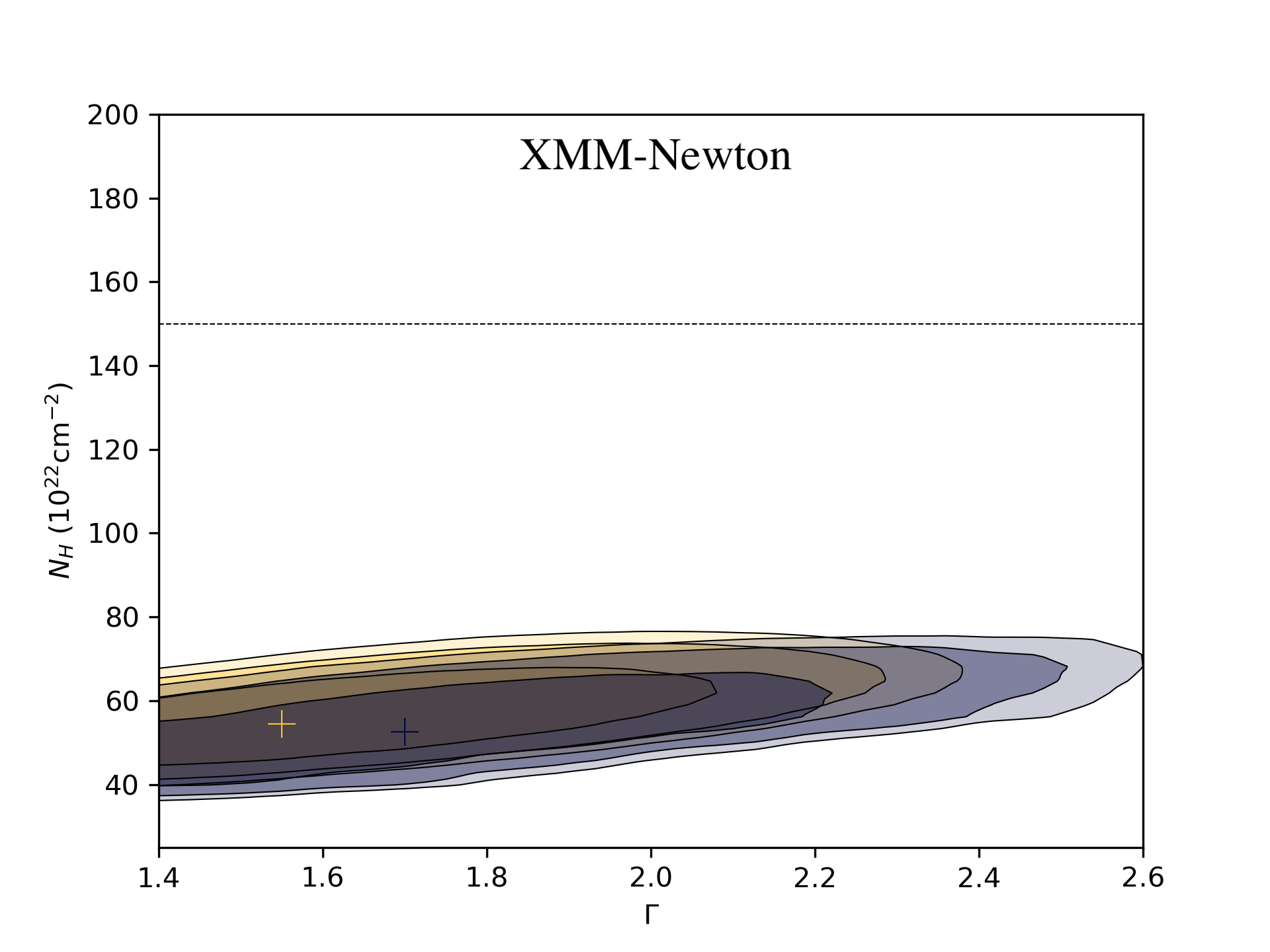}
    \caption{CGCG\,1822.3+2053. Contour plots for line-of-sight column density and photon index obtained with the \texttt{steppar} command in \texttt{XSPEC}. \textit{Left Panel}: Results for the \cha\ observation. The \texttt{borus02} model results are in orange and the \texttt{UXCLUMPY} model results are in purple. The contours correspond to the 68\,\%, 90\,\%, and 95\,\% confidence levels. \textit{Right Panel}: Results for the \xmm\ observation. The \texttt{borus02} model results are in yellow and the \texttt{UXCLUMPY} model results are in blue. The contours correspond to the 68\,\%, 90\,\%, and 95\,\% confidence levels. The dashed horizontal line indicates the Compton-thick threshold. }
    \label{fig:cgc_contours}
\end{figure*}

CGCG\,1822.3+2053 was observed in July of 2022 with \cha\ for 10\,ks and in September of 2021 with \xmm\ for 10\,ks. The \cha\ data were fitted in the energy range 1-7\,keV while the \xmm\ data were fitted in the energy range 1-9\,keV. The spectrum shows a moderately obscured nature. The spectral fits with \texttt{borus02} and \texttt{UXCLUMPY} are shown in Figure \ref{fig:cgc_spectra}. The $N_{\rm H,l.o.s}$ posteriors are shown in Figure \ref{fig:cgc_results} and the contours are shown in Figure \ref{fig:cgc_contours}. Table \ref{tab:cgc_results} shows the best-fit parameters and their errors for both models and both methods along with the intrinsic luminosity and the observed flux in cgs units. The corner plots for both models are shown in Figure \ref{fig:cgc_corner}.

The measured $N_{\rm H,l.o.s}$ for the \cha\ data is consistent with being Compton-thick at 90\,\% confidence. The torus parameters are unconstrained, however, the obscured nature of the source is confidently confirmed.  Based on the current \cha\ posteriors, there is a 18.3\,\% and 10.4\,\% probability of the source being Compton-thick according to the \texttt{borus02} and \texttt{UXCLUMPY} models, respectively. Both model posteriors confidently assert a Compton-thin scenario for the \xmm\ data with a 0.0\,\% probability of being Compton-thick. We classify this source as an obscured, Compton-thin AGN. Future \nustar\ data from Cycle 10 will help determine whether the source is Compton-thin or Compton-thick as well as provide tighter constraints on the torus parameters.  

\begin{deluxetable*}{|c|cccc|cccc|}

\tablecaption{CGCG\,1822.3+2053 best-fit results.\label{tab:cgc_results}}

\tablehead{Model & MYTorus & borus02* & borus02 & UXCLUMPY & MYTorus & borus02* & borus02 & UXCLUMPY \\ 
Algorithm & \multicolumn{4}{c|}{Levenberg-Marquardt} & \multicolumn{4}{c|}{Nested Sampling }
}

\startdata
red $\chi^2$  & 1.13 & 1.22 & 1.13 & 1.14 
& 1.13 & 1.26 & 1.16 & 1.17 \\
$\chi^2$/d.o.f. & 122/108 & 132/108 & 119/105  & 120/105 
& 122/108 & 136/108 & 122/105 & 123/105 \\
\hline
$\Gamma$ & 1.60$^{+0.43}_{-u}$ & 1.40$^{+0.37}_{-u}$ & 1.55$^{+0.62}_{-u}$ & 1.54$^{+0.77}_{-u}$ 
& 1.73$^{+0.40}_{-0.29}$ & 1.64$^{+0.31}_{-0.19}$ & 1.81$^{+0.49}_{-0.35}$ &  2.09$^{+0.34}_{-0.58}$ \\
$N_{H,\rm tor} (10^{22}\text{\,cm}^{-2})$ & 100* & 100* & 3200$^{+u}_{-u}$ & $-$ 
& 100* & 100* & 37$^{+2300}_{-31}$ & $-$ \\
$CF_{\rm tor}$ & $-$ & 0.67* & 0.20$^{+u}_{-u}$ & $-$  
& $-$ & 0.67* & 0.15$^{+0.74}_{-0.04}$ & $-$ \\
$\theta_{\rm inc}(^{\circ})$ & 45* & 45* & 78$^{+u}_{-u}$ & 90$^{+u}_{-u}$ 
& 45* & 45* & 79$^{+7}_{-42}$ & 64$^{+21}_{-59}$  \\
$\sigma_{\rm tor}$ & $-$ & $-$ & $-$ &  18$^{+63}_{-u}$ 
& $-$ & $-$ &  $-$ & 14$^{+64}_{-5}$ \\
CTKcover & $-$ & $-$ & $-$ &  0$^{+u}_{-u}$ 
& $-$ & $-$ &  $-$ & 0.39$^{+0.17}_{-0.36}$ \\
norm$(10^{-3})$ & 0.9$^{+1.8}_{-0.5}$ & 0.4$^{+0.6}_{-0.1}$ & 1.61$^{+6.78}_{-1.01}$ & 1.63$^{+37.3}_{-1.05}$ 
& 1.7$^{+2.7}_{-1.0}$ & 1.0$^{+0.6}_{-0.5}$ & 2.14$^{+7.90}_{-1.22}$ & 5.82$^{+13.0}_{-4.41}$  \\
$f_{\rm s}$$(10^{-3})$ & 4.26$^{+5.30}_{-3.08}$ & 5.97$^{+4.76}_{-4.95}$ & 3.26$^{+3.49}_{-2.25}$ & $-$ 
& 3.01$^{+2.93}_{-2.88}$ & 2.12$^{+2.75}_{-2.10}$ & 2.33$^{+2.72}_{-2.25}$ & $-$\\
$f_{\rm s, uxclumpy}$$(10^{-3})$ & $-$ & $-$ & $-$ & 7.67$^{+u}_{-6.48}$ 
& $-$ & $-$ & $-$ & 7.72$^{+21.1}_{-7.61}$ \\
$C_{\rm xmm}$ & 0.94$^{+0.67}_{-0.37}$ & 1.08$^{+0.60}_{-0.26}$ & 0.78$^{+0.58}_{-0.36}$ & 0.83$^{+0.59}_{-0.50}$ 
& 0.75$^{+0.62}_{-0.28}$ & 0.85$^{+0.62}_{-0.24}$ & 0.69$^{+0.67}_{-0.26}$ & 0.68$^{+0.63}_{-0.33}$ \\
$C_{\rm BAT}$ & 0.74$^{+0.92}_{-0.39}$ & 0.69$^{+0.80}_{-0.23}$ &  0.50$^{+0.91}_{-0.31}$ & 0.47$^{+1.01}_{-0.27}$ 
& 0.82$^{+1.02}_{-0.41}$ & 1.05$^{+1.07}_{-0.50}$ & 0.77$^{+2.13}_{-0.48}$ & 0.69$^{+0.96}_{-0.39}$ \\
\hline 
$N_{\rm H,cha} (10^{22}\text{\,cm}^{-2})$ & 76$^{+58}_{-27}$ & 63$^{+39}_{-21}$ & 87$^{+35}_{-27}$ & 77$^{+48}_{-22}$ 
& 87$^{+220}_{-25}$ & 77$^{+360}_{-22}$ & 87$^{+300}_{-25}$ & 96$^{+100}_{-30}$  \\
$N_{\rm H,xmm} (10^{22}\text{\,cm}^{-2})$ & 46$^{+11}_{-10}$ & 39$^{+10}_{-6}$ & 54$^{+16}_{-11}$ & 50$^{+21}_{-12}$ 
& 48$^{+12}_{-8}$ & 43$^{+9}_{-7}$ & 53$^{+16}_{-8}$ & 55$^{+16}_{-11}$  \\
\hline 
$\log(L_{\text{2-10\,keV, cha}})$ & 42.4$^{+0.2}_{-0.3}$ & 42.2$^{+0.2}_{-0.3}$ & 42.7$^{+0.2}_{-0.2}$ & 42.8$^{+0.5}_{-0.4}$ 
& 42.4$^{+0.2}_{-0.2}$ & 42.2$^{+0.2}_{-0.3}$ & 42.9$^{+0.2}_{-0.3}$ & 43.0$^{+0.3}_{-0.4}$  \\
$\log(L_{\text{2-10\,keV, xmm}})$ & 42.45$^{+0.05}_{-0.05}$ & 42.2$^{+0.05}_{-0.06}$ & 42.70$^{+0.04}_{-0.05}$ & 42.7$^{+0.5}_{-0.4}$ 
& 42.40$^{+0.05}_{-0.05}$ & 42.3$^{+0.05}_{-0.06}$ & 42.83$^{+0.05}_{-0.05}$ & 42.9$^{+0.3}_{-0.4}$  \\
$\log(L_{\text{15-150\,keV}})$ & 43.0$^{+0.1}_{-0.2}$ & 42.9$^{+0.2}_{-0.2}$ & 43.2$^{+0.1}_{-0.2}$ & 43.0$^{+0.5}_{-0.7}$ 
& 43.0$^{+0.1}_{-0.2}$ & 42.7$^{+0.2}_{-0.2}$ & 43.0$^{+0.1}_{-0.1}$ & 43.0$^{+0.4}_{-0.5}$  \\
\hline 
$\log(F_{\text{2-10\,keV, cha}})$ & -12.28$^{+0.09}_{-0.09}$ & -12.29$^{+0.09}_{-0.09}$ & -12.24$^{+0.09}_{-0.09}$ & -12.21$^{+0.09}_{-0.09}$ 
& -12.28$^{+0.09}_{-0.09}$ & -12.30$^{+0.09}_{-0.09}$ & -12.30$^{+0.09}_{-0.09}$ & -12.25$^{+0.09}_{-0.09}$  \\
$\log(F_{\text{2-10\,keV, xmm}})$ & -12.07$^{+0.04}_{-0.04}$ & -12.08$^{+0.04}_{-0.04}$ & -12.04$^{+0.04}_{-0.04}$ & -12.04$^{+0.04}_{-0.04}$ 
& -12.07$^{+0.04}_{-0.04}$ & -12.09$^{+0.04}_{-0.04}$ & -12.06$^{+0.04}_{-0.04}$ & -12.05$^{+0.04}_{-0.04}$ \\
$\log(F_{\text{15-150\,keV}})$ & -11.01$^{+0.09}_{-0.12}$ & -11.00$^{+0.09}_{-0.12}$ & -11.02$^{+0.09}_{-0.12}$ & -11.04$^{+0.09}_{-0.12}$ 
& -11.01$^{+0.10}_{-0.12}$ & -11.06$^{+0.10}_{-0.13}$ & -11.08$^{+0.10}_{-0.13}$ & -11.11$^{+0.10}_{-0.13}$  \\
\hline 
$P_{\rm ch}(\text{CT})$ & $-$ & $-$ & $-$ & $-$ 
& 16.9\,\% & 33.8\,\% & 18.3\,\% & 10.4\,\%  \\
$P_{\rm xmm}(\text{CT})$ & $-$ & $-$ & $-$ & $-$ 
& 0.0\,\% & 0.0\,\% & 0.0\,\% & 0.0\,\%  \\
\enddata
\tablecomments{A $u$ in the upper or lower limit indicates the parameter is unconstrained in that particular direction. A * indicates the parameter was frozen at that particular value during the fit. The luminosities shown are the intrinsic luminosities while the fluxes are observed.}
\end{deluxetable*}

\subsection{\texttt{MCG\,+2-57-2}} \label{sec:MCG+2-57-2}

MCG\,+2-57-2 was classified as a Seyfert 1.9 by \cite{smith2020bat}. It was observed with \cha\ in November of 2021 for 10\,ks. The spectrum is piled up which we account for using the algorithm developed by \cite{Davis2001} and implemented in \texttt{XSPEC} as \texttt{pileup}. The \cha\ data were fitted in the energy range 1-7\,keV. Both models and fitting methods agree that the spectrum is unobscured. The best-fit parameters and errors for both models and methods are shown in Table \ref{tab:mcg_results}. The $N_{\rm H,l.o.s}$ posteriors from the nested sampling and the contours from the LM fit along with the best-fit spectra are shown in Figure \ref{fig:mcg_results}. The corner plots for both models are shown in Figure \ref{fig:mcg_corner}. There is a 0.0\,\% probability of the data being Compton-thick for both models according to the $N_{\rm H,l.o.s}$ posteriors. We classify this source as an unobscured AGN.


\section{\texttt{Discussion}} \label{sec:discussion}

We have used two different models (\texttt{borus02} and \texttt{UXCLUMPY}) and two different methods (least squares and Bayesian regression) to estimate the line-of-sight $N_H$ in six local AGN. In this section, we compare the results of the two models and methods and show that the $N_{\rm H,l.o.s}$ estimates are reliable. Furthermore, we check the sources with multiple soft X-ray observations for variability in absorption. We also discuss the discrepancy between the most probable parameter values obtained from the posterior distributions and the parameter values resulting in the maximum likelihood spectral fit. 

\subsection{Comparison of $N_{\rm H,l.o.s}$ determinations}

The \texttt{borus02} model only includes the reprocessed emission and does not include the primary powerlaw that gets absorbed by the torus. We obtain the line-of-sight $N_H$ by adding an absorbed powerlaw to the \texttt{borus02} model and untying this $N_H$ from the average column density $N_{H\rm,avg}$ to account for the fact the torus is likely inhomogeneous. In the \texttt{UXCLUMPY} model, the line-of-sight $N_H$ can be calculated within the model since the total absorption depends on the number of clouds in the line of sight. 

\begin{figure}[h!]
    \centering
    \includegraphics[width=0.5\textwidth]{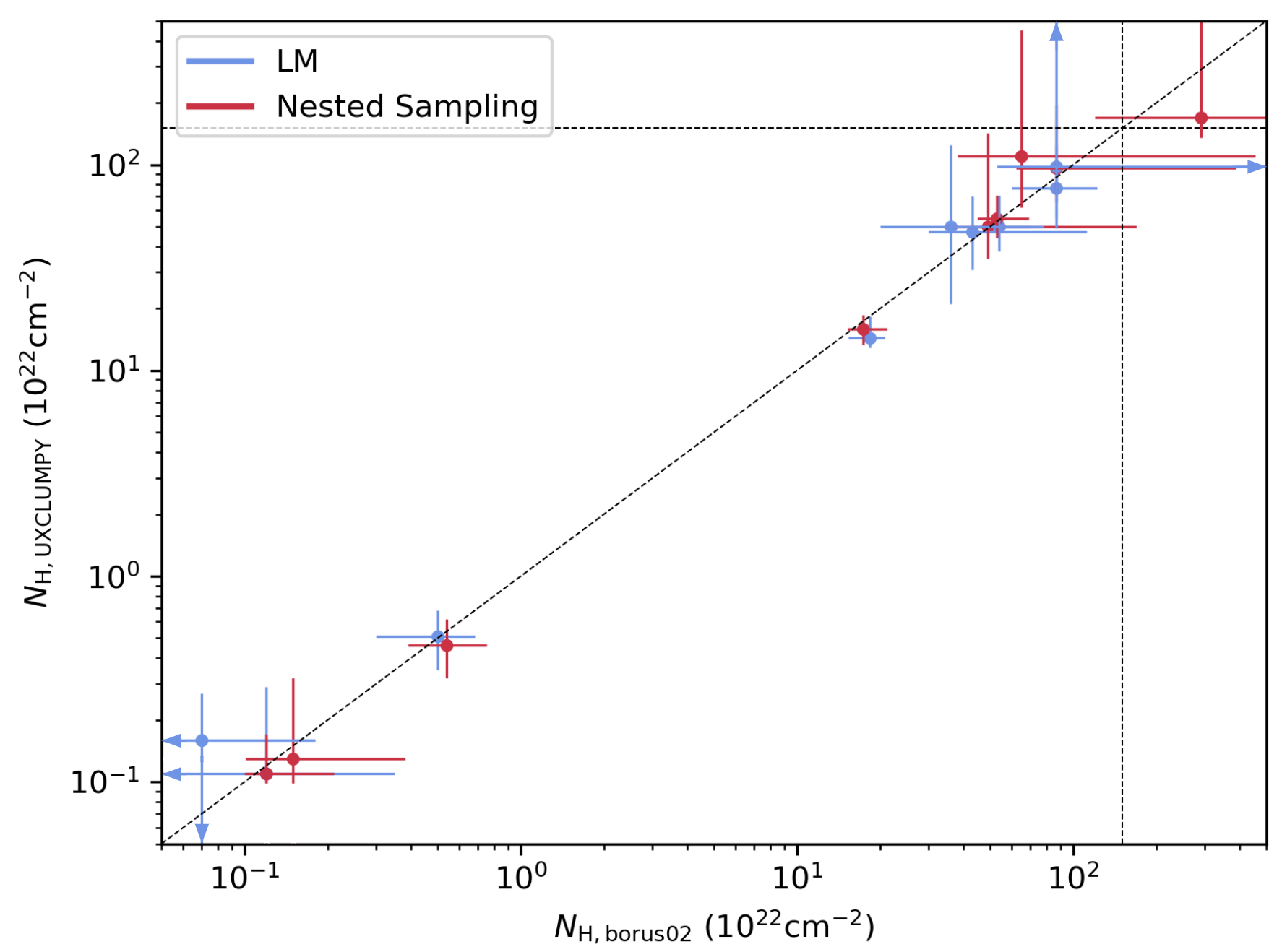}\hfill
    \includegraphics[width=0.5\textwidth]{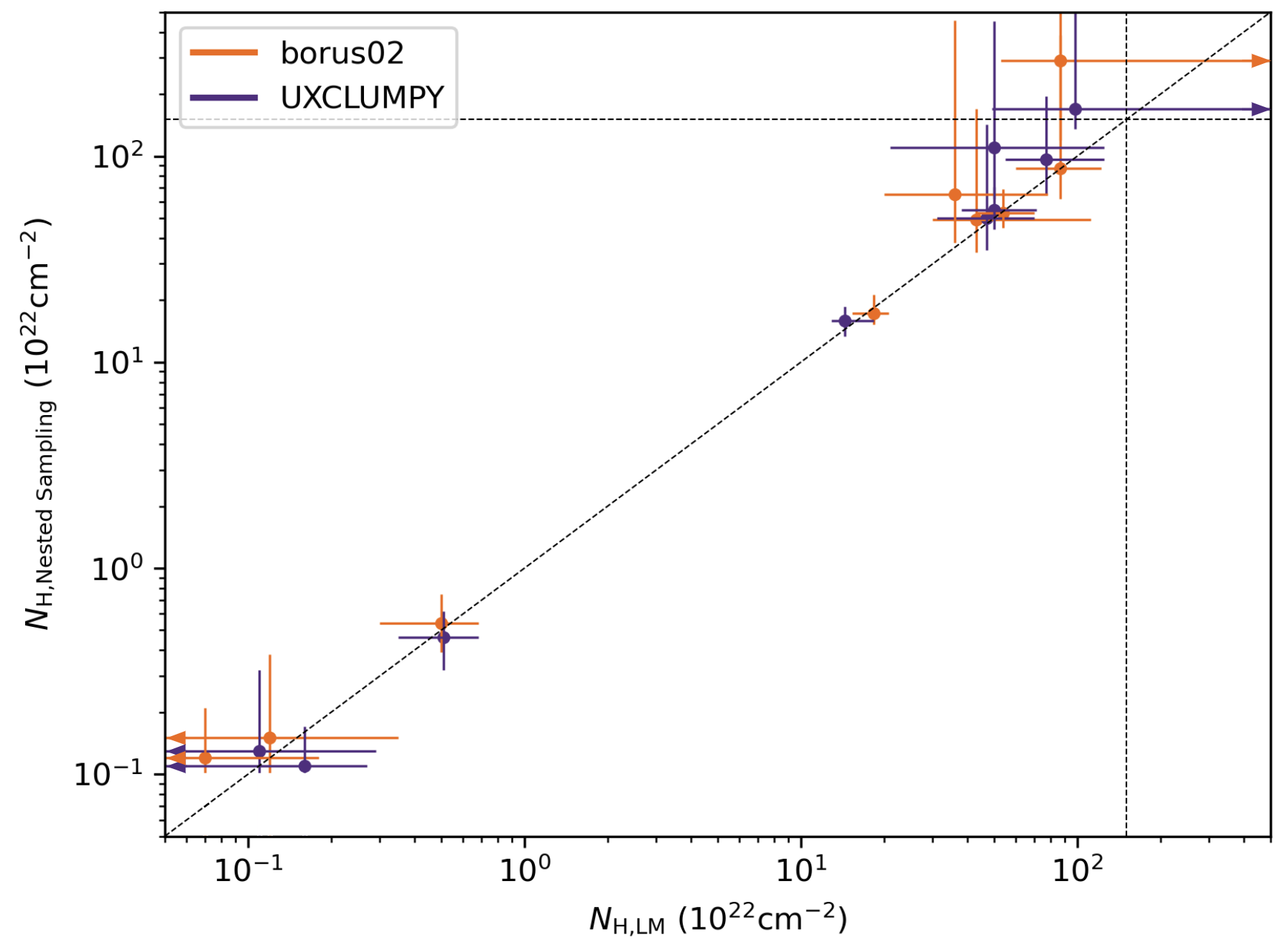}\hfill
    \caption{\textit{Top panel}: Comparison of the $N_{\rm H,l.o.s}$ measured with the \texttt{UXCLUMPY} model against the measured $N_{\rm H,l.o.s}$ with the \texttt{borus02} model. The values obtained with least squares fitting are shown in blue while the most probable values obtained through Bayesian fitting are shown in red. \textit{Bottom panel}: Comparison of the $N_{\rm H,l.o.s}$ measured with least squares fitting against the most probable $N_{\rm H,l.o.s}$ from Bayesian fitting. The \texttt{borus02} values are shown in orange and the \texttt{UXCLUMPY} values are shown in purple. The dashed lines in both panels indicate the Compton-thick threshold.}
    \label{fig:comparisons}
\end{figure}

Figure \ref{fig:comparisons} shows the measured $N_{\rm H,l.o.s}$ values between the two different models and methods. The top panel shows excellent agreement in $N_{\rm H,l.o.s}$ between the \texttt{borus02} and \texttt{UXCLUMPY} models, despite the different assumptions. This is in agreement with the results of, for example, \cite{pizzetti_multi-epoch_2022} and \cite{NTA23} which show similar consistency in the measurements of $N_{\rm H,l.o.s}$ from different models even when using large data samples and multiple observations. 

The bottom panel shows a relatively good agreement between methods, however, there are a few instances where the lower limit of the nested sampling algorithm is higher than the best-fit value from the LM algorithm. This is likely due to the way we define the `best' value and credible interval with the Bayesian technique. In general, the highest probability value in the posterior distribution is not the same as the value that gives the maximum likelihood (see Section \ref{subsec:posterior_values}). For example, the highest orange point in the bottom panel of Figure \ref{fig:comparisons} shows the \texttt{borus02} measurement for the \xmm\ observation of NGC\,5759. The posterior provides a 90\,\% credible interval that is inconsistent with the $N_{\rm H,l.o.s}$ value that provides the maximum likelihood. The maximum likelihood value of $N_{\rm H,xmm}=94\times10^{22}$\,cm$^{-2}$ is in much better agreement with the value obtained through least squares fitting. Furthermore, the LM fit value is a lower limit, so it is still consistent with the nested sampling determined value. 

There also seems to be a hint that the Bayesian regression favors higher $N_{\rm H,l.o.s}$, however, these cases are all at high obscuration where it is easier to constrain the lower limit on the obscuration leaving more mass in the upper end of the posterior. Furthermore, there are not enough data points here to indicate that this is even a real difference in the methods, so we conclude that the methods are consistent.

\subsection{$N_{\rm H,l.o.s}$ variability}

We conduct an obscuration variability analysis on the three sources with two soft X-ray observations. We obtain a `probability of variation,' $P_{\text{var}}$, from the $N_{\rm H,l.o.s}$ posterior distributions for each model. We define this probability as 
$$
P_{\text{var}} = \left(1 - \frac{\vert N_{\text{H,cha}}\cap N_{\text{H,xmm}}\vert}{\vert N_{\text{H,cha}}\vert + \vert N_{\text{H,xmm}}\vert}\right)\times100
$$
where $N_{\text{H,cha}}$ is the set of points in the $N_{\rm H,l.o.s}$ posterior returned by \texttt{BXA} for the \cha\ observation and $N_{\text{H,cha}}$ is for the \xmm\ observation. In words, this is the percentage of posterior mass uncommon to the two distributions. 

Using this definition, we find that 2MASX\,J17253053--4510279 has the highest probability of being variable with $P_{\text{var}}=89\,\%$ for both \texttt{borus02} and \texttt{UXCLUMPY}. For CGCG\,1822.3+2053 we obtain $P_{\text{var}}=28\,\%, 30\,\%$ for \texttt{borus02} and \texttt{UXCLUMPY}, respectively. For NGC\,5759 we obtain $P_{\text{var}}=4\,\%, 3\,\%$. All three of these sources will be observed simultaneously with \nustar\ and \xmm\ giving us a third observation to test for variability.

\subsection{Bayesian posterior distributions}\label{subsec:posterior_values}

The final result of the Bayesian analysis is the posterior distribution for each parameter. This result contains all of the information that we can obtain about a given parameter of interest with the available data and prior knowledge of that parameter. However, it is not immediately clear how to interpret this distribution. For example, can a single value (e.g., the mean, median, or mode) with associated uncertainty describe a given posterior distribution well? For well behaved parameter spaces with mono-modal, symmetric (e.g., Gaussian) posterior distributions, this question does not matter since the mean, median, and mode will be equivalent. However, as can be seen in the 1-D histograms on the diagonal of the corner plots in Figures \ref{fig:2mfgc_corner}-\ref{fig:2masx_corner}, many of the posterior distributions do not have gaussian shapes. We decided to report the mode and associated inter-quartile range of the posterior distribution for all parameters, but caution the reader against over-interpreting these values without visually inspecting the 1-D histogram shapes first.

\begin{figure}[h!]
    \centering
    \includegraphics[width=0.47\textwidth]{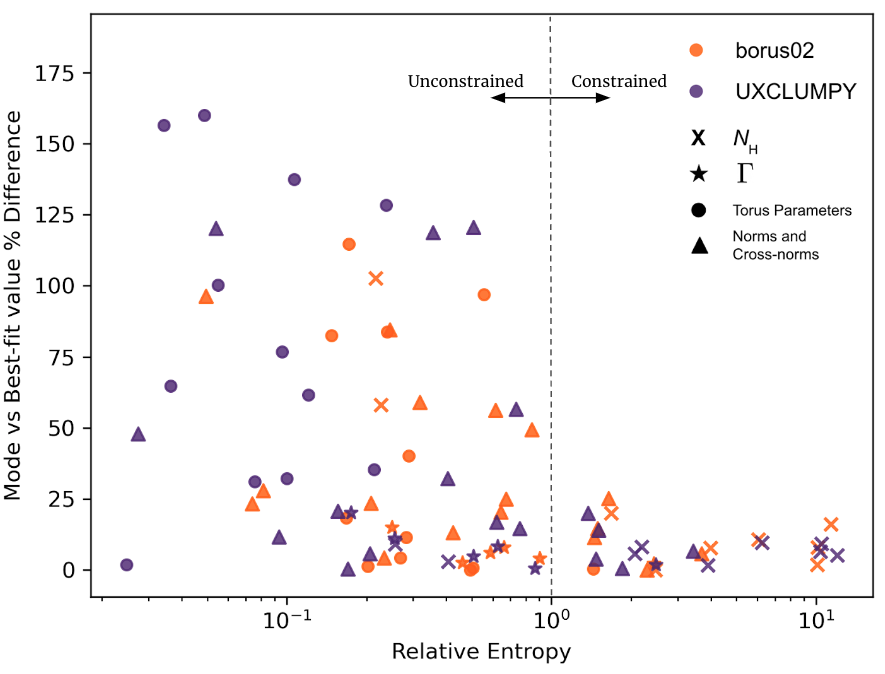}\hfill
    \caption{The percent difference between the mode of the posterior distribution and the best-fit value for each parameter plotted against the relative entropy of the posterior and prior distributions in bits. The relative entropy quantifies how tightly constrained the measurement of a parameter is relative to its prior. The orange points are parameters in the \texttt{borus02} model and the purple points are parameters in the \texttt{UXCLUMPY} model. The `x' markers indicate the $N_{\rm H,l.o.s}$, the star markers indicate the photon index, the triangle markers indicate normalization and cross-normalization parameters, and the circle markers indicate the torus geometry parameters.}
    \label{fig:diff_width}
\end{figure}

One related issue we would like to highlight in particular arises from using the mode of a given posterior parameter distribution as a “best-fit” when the parameter is in fact unconstrained. The immunity of BXA to local minima means that if a posterior is found to be unconstrained, i.e. not significantly different from its corresponding prior, a single parameter value should not be favored as a best-fit relative to the remaining unconstrained portion of the parameter space. There are cases where using the mode of all (including unconstrained) posteriors results in a bad fit for this reason, and underlines the importance of mapping out inter-parameter dependencies fully rather than relying on individual parameter vectors to describe a fit.

To illustrate this, we calculated the percent difference between the mode of the posterior distribution and the best-fit value for each parameter. In Figure \ref{fig:diff_width}, we plotted this difference against the relative entropy (also called the Kullback-Leibler divergence or the information gain) \citep{kullback1951information} between the posterior and the prior. We use the equation 
\begin{equation*}
\text{Relative Entropy} = \int_{-\infty}^{\infty} P(x)\log_{2}\left(\frac{P(x)}{\pi(x)}\right)dx
\end{equation*}
where $P(x)$ is the posterior distribution and $\pi(x)$ is the prior distribution. The logarithm is base 2 which provides the relative entropy in units of bits. We obtain $P(x)$ by applying a gaussian kernel density estimator with the bandwidth selected using Scott's Rule\footnote{Scott's Rule can lead to oversmoothing for multimodal distributions, however, none of the parameters shown in Figure \ref{fig:diff_width} demonstrate obvious multimodal behavior so this should not be an issue.} \citep{scott2015multivariate}. We implement this with the $\texttt{gaussian\_kde}$ function in SciPy \citep{2020SciPy-NMeth} and the code and data used for the values in Figure \ref{fig:diff_width} are provided in \cite{cox_2024_14225570}. We define the uniform prior distribution as $\pi(x)=(b-a)^{-1}$ and the log-uniform prior distribution as $\pi(x)=[x\log(b/a)]^{-1}$ where $a$ and $b$ are the lower and upper bounds of the prior range, respectively. Essentially, the relative entropy quantifies the difference between two distributions, which in this context means the degree to which the data were able to constrain a parameter relative to the initially assumed prior. High relative entropy values indicate a well constrained parameter while low relative entropy values mean the data are consistent with a large percentage of the prior range. At a glance, Figure \ref{fig:diff_width} shows that when the parameter is well constrained (relative entropy $\gtrsim$ 1\,bit\footnote{For reference, a Gaussian distribution whose width shrinks by a factor of 3 between the prior and posterior would have a relative entropy of 1\,bit \citep{buchner_infogain}.}), there is good agreement between the posterior mode and the best-fit value. However, as the parameter becomes more unconstrained, the scatter increases greatly. An important conclusion that we can draw from this is that one should be wary of any singular number obtained from the posterior distribution, especially when the relative entropy is less than a few bits. In the extreme cases (relative entropy\,$\ll$\,1\,bit), the most probable value is simply a random draw from the prior and contains no useful information. The advantage of having the posterior distribution and relative entropy is that it shows clearly when the fit is insensitive to the parameter, whereas LM can struggle to do this.

Furthermore, these results show that in most cases, the data are able to greatly constrain the $N_{\rm H,los}$ despite not being able to constrain any of the torus geometry parameters, which need high quality \nustar\ data to properly measure.


\section{Summary} \label{sec:conclusions}

In this paper, we have provided the first analysis of soft X-ray data for five hard X-ray sources selected from the Palermo BAT 150 month catalog and one from the 157 month catalog. These sources were selected as likely obscured candidates due to the lack of a soft X-ray ROSAT counterpart. We obtained snapshot images with \cha\ and estimated the line-of-sight column density, $N_{\rm H,l.o.s}$, to find Compton-thick candidates. These candidates will be followed up with simultaneous \nustar\ and \xmm\ data in an effort to catalog all of the Compton-thick AGN in the local universe and to better understand the nature of the obscuring medium colloquially known as the torus. We find that two of our sources (2MASX\,J17253053--4510279 and MCG\,+2-57-2) are likely unobscured. The rest of our sources are obscured with two of them (CGCG\,1822.3+2053 and NGC\,5759) showing a nonzero probability of being Compton-thick. We classify these two sources as Compton thick candidates with NGC\,5759 being the strongest between the two. 

We used two physically motivated models to fit the hard and soft X-ray data for each source and we compare the $N_{\rm H,l.o.s}$ measurements obtained with both. Since the data quality did not allow strong constraints in the torus parameters of these two models, we fitted with simpler models as well to confirm the $N_{\rm H,los}$ measurements. We also use least squares and Bayesian fitting methods for each source and model to compare the results. We find that the $N_{\rm H,l.o.s}$ measurements are in good agreement between all models and both fitting methods. 

We also consider the interpretation and reliability of values obtained from the posterior distribution obtained by nested sampling. We find that for unconstrained parameters with a low relative entropy between the posterior distribution and prior distribution, the most probable value according to the posterior distribution can be very different from the value that provides the best fit. Therefore, in most cases, the posterior distribution should be considered in conjunction with the best-fit value, and individual numbers derived from the posterior should be treated with caution. However, in general, the posterior distribution is still consistent with the best-fit values within a 90\,\% credible interval.

Three sources have two soft X-ray observations and we look for variability in the $N_{\rm H,l.o.s}$ between the two observations. We find little evidence of $N_{\rm H,l.o.s}$ variability in CGCG\,1822.3+2053 and NGC\,5759, however, 2MASX\,J17253053--4510279 does show a strong hint of variability despite being only marginally obscured.

Three sources have been approved for simultaneous \nustar\ and \xmm\ observations (2MASX\,J17253053--4510279, CGCG\,1822.3+2053, and NGC\,5759) in \nustar\ Cycle 10 (ID: 10209, P.I. Cox). These data, in conjunction with \bat, will provide continuous spectral coverage from 1-150\,keV, with a much clearer view at energies $\sim20$\,keV where the reflection is more important. These observations will help confirm the results shown here, provide a third observation for each of the sources to study variability, and better constrain the torus parameters. These results will be presented in a future publication (Cox, et al. in preparation).

\section{Acknowledgments}

I.C. and N.T.A. acknowledge support under contracts GO2-23072X, G04-25065X, and 80NSSC23K1611. The scientific results reported in this article are based on observations made by the X-ray observatories \cha\, \xmm\, and \bat, and has made use of the NASA/IPAC Extragalactic Database (NED), which is operated by the Jet Propulsion Laboratory, California Institute of Technology under contract with NASA. We acknowledge the use of the software package HEASoft. 

\bibliography{references}{}
\bibliographystyle{aasjournal}

\appendix

\section{Best fits}

In this section, we present the results for all sources in the sample. The following tables and plots are the same as Table \ref{tab:cgc_results} and Figures \ref{fig:cgc_spectra}$-$\ref{fig:cgc_contours}. The luminosities shown are intrinsic and in units of erg\,/s while the fluxes are observed and in units of erg\,/cm$^{2}$\,/s. In all plots, the color orange corresponds to the \texttt{borus02} fit to \cha\ data. The color purple corresponds to the \texttt{UXCLUMPY} fit to \cha\ data. The color yellow corresponds to the \texttt{borus02} fit to \xmm\ data. The color blue corresponds to the \texttt{UXCLUMPY} fit to \xmm\ data.

\begin{deluxetable*}{|c|cccc|cccc|}[h!]

\tablecaption{2MFGC\,9836 best-fit results. \label{tab:2mfgc_results}}

\tablehead{Model & MYTorus & borus02* & borus02 & UXCLUMPY & MYTorus & borus02* & borus02 & UXCLUMPY \\ 
Algorithm & \multicolumn{4}{c|}{Levenberg-Marquardt} & \multicolumn{4}{c|}{Nested Sampling}
}

\startdata
red $\chi^2$  & 0.95 & 1.01 & 0.95 & 0.96 
& 0.99 & 1.01 & 0.95 & 0.96 \\
$\chi^2$/d.o.f. & 134/141 & 142/141 & 131/138  & 132/138 
& 139/141 & 142/141 & 131/138 & 132/138 \\
\hline
$\Gamma$ & 1.92$^{+0.35}_{-0.31}$ & 1.99$^{+0.20}_{-0.31}$ & 1.84$^{+0.42}_{-0.19}$  & 1.77$^{+0.42}_{-0.22}$ 
& 1.91$^{+0.35}_{-0.28}$ & 1.96$^{+0.22}_{-0.29}$ & 1.79$^{+0.43}_{-0.27}$ &  1.89$^{+0.43}_{-0.30}$ \\
$N_{H,\rm tor} (10^{22}\text{\,cm}^{-2})$ & $N_{H,ch}$* & 100* & 3.79$^{+u}_{-u}$ & $-$ 
& $N_{H,ch}$* & 100* & 5.27$^{+1870}_{-3.87}$ & $-$ \\
$CF_{\rm tor}$ & $-$ & 0.67* & 0.56$^{+u}_{-u}$ & $-$  
& $-$ & 0.67* & 0.16$^{+0.78}_{-0.04}$ & $-$ \\
$\theta_{\rm inc}(^{\circ})$ & 90* & 45* & 81.4$^{+u}_{-42.6}$ & 55.6$^{+u}_{-u}$ 
& 90* & 45* & 73$^{+13}_{-42}$ & 31.6$^{+54.2}_{-25.9}$  \\
$\sigma_{\rm tor}$ & $-$ & $-$ & $-$ &  6.99$^{+54.62}_{-}$ 
& $-$ & $-$ &  $-$ & 13.8$^{+66}_{-5}$ \\
CTKcover & $-$ & $-$ & $-$ &  0.25$^{+u}_{-u}$ 
& $-$ & $-$ &  $-$ & 0.44$^{+0.13}_{-0.42}$ \\
norm$(10^{-3})$ & 3.30$^{+3.10}_{-1.46}$ & 3.42$^{+1.31}_{-1.55}$ & 3.19$^{+3.30}_{-1.52}$ & 2.77$^{+2.30}_{-1.09}$ 
& 3.46$^{+2.84}_{-1.55}$ & 2.75$^{+2.16}_{-0.92}$ &  2.98$^{+3.25}_{-1.30}$ & 4.25$^{+5.41}_{-2.05}$  \\
$f_{\rm s}$$(10^{-3})$ & 1.67$^{+2.18}_{-1.30}$ & 0.01$^{+1.18}_{-u}$ & 0.01$^{+2.78}_{-u}$ & $-$ 
& 1.48$^{+1.63}_{-1.45}$ & 0.04$^{+0.69}_{-0.03}$ & 0.03$^{+2.1}_{-0.02}$ & $-$\\
$f_{\rm s, uxclumpy}$$(10^{-3})$ & $-$ & $-$ & $-$ & 1.92$^{+u}_{-u}$ 
& $-$ & $-$ & $-$ & 2.5$^{+9.8}_{-2.5}$ \\
$C_{\rm BAT}$ & 0.79$^{+0.58}_{-0.33}$ & 0.80$^{+0.30}_{-0.33}$ & 0.75$^{+0.42}_{-0.33}$ & 0.58$^{+0.33}_{-0.22}$ 
& 0.86$^{+0.50}_{-0.41}$ & 0.72$^{+0.36}_{-0.27}$ & 0.65$^{+0.62}_{-0.25}$ & 0.61$^{+0.58}_{-0.26}$ \\
\hline 
$N_{H,ch} (10^{22}\text{\,cm}^{-2})$ & 17.2$^{+2.5}_{-2.3}$ & 17.4$^{+1.6}_{-2.7}$ & 18.3$^{+2.5}_{-3.0}$ & 14.4$^{+3.7}_{-1.5}$ 
& 16.9$^{+2.7}_{-2.0}$ & 16.8$^{+2.8}_{-2.0}$ & 17.3$^{+3.9}_{-2.1}$ & 15.9$^{+2.7}_{-2.5}$  \\
\hline 
$\log(L_{\text{2-10\,keV, cha}})$ & 43.20$^{+0.03}_{-0.03}$ & 43.17$^{+0.03}_{-0.03}$ & 43.23$^{+0.03}_{-0.03}$ & 43.25$^{+0.05}_{-0.05}$ 
& 43.21$^{+0.03}_{-0.03}$ & 43.17$^{+0.03}_{-0.03}$ & 43.25$^{+0.03}_{-0.03}$ & 43.3$^{+0.1}_{-0.1}$  \\
$\log(L_{\text{15-150\,keV}})$ & 43.44$^{+0.07}_{-0.08}$ & 43.26$^{+0.09}_{-0.11}$ & 43.45$^{+0.06}_{-0.07}$ & 43.3$^{+0.3}_{-0.4}$ 
& 43.45$^{+0.07}_{-0.08}$ & 43.27$^{+0.09}_{-0.11}$ & 43.40$^{+0.06}_{-0.08}$ & 43.4$^{+0.3}_{-0.4}$  \\
\hline 
$\log(F_{\text{2-10\,keV, cha}})$ & -11.45$^{+0.03}_{-0.03}$ & -11.45$^{+0.03}_{-0.03}$ &  -11.44$^{+0.03}_{-0.03}$ & -11.43$^{+0.03}_{-0.03}$ 
& -11.45$^{+0.03}_{-0.03}$ & -11.45$^{+0.03}_{-0.03}$ & -11.44$^{+0.03}_{-0.03}$ & -11.43$^{+0.03}_{-0.03}$  \\
$\log(F_{\text{15-150\,keV}})$ & -10.93$^{+0.06}_{-0.07}$ & -10.98$^{+0.06}_{-0.07}$ &  -10.96$^{+0.06}_{-0.07}$ & -10.95$^{+0.06}_{-0.07}$ 
& -10.93$^{+0.06}_{-0.07}$ & -10.98$^{+0.06}_{-0.07}$ & -10.97$^{+0.06}_{-0.07}$ & -10.94$^{+0.06}_{-0.07}$  \\
\hline 
$P_{\rm ch}(\text{CT})$ & $-$ & $-$ & $-$ & $-$ 
& 0.0\,\% & 0.0\,\% & 0.0\,\% & 0.0\,\%  \\
\enddata
\end{deluxetable*}


\begin{deluxetable*}{|c|cccc|cccc|}

\tablecaption{NGC\,5759 best-fit results. \label{tab:ngc_results}}

\tablehead{Model & MYTorus & borus02* & borus02 & UXCLUMPY & MYTorus & borus02* & borus02 & UXCLUMPY \\ 
Algorithm & \multicolumn{4}{c|}{Levenberg-Marquardt} & \multicolumn{4}{c|}{Nested Sampling}
}

\startdata
red $\chi^2$  & 1.20 & 1.13 & 1.13 & 1.18 
& 1.11 & 1.14 & 1.13 & 1.21 \\
$\chi^2$/d.o.f. & 85/71 & 80/71 & 77/68  & 81/68 
& 79/71 & 81/71 & 77/68 & 82/68 \\
\hline
$\Gamma$ & 1.93$^{+0.46}_{-0.47}$ & 1.70$^{+0.41}_{-u}$ & 1.73$^{+0.38}_{-0.28}$ & 1.72$^{+0.45}_{-u}$ 
& 1.85$^{+0.41}_{-0.36}$ & 1.67$^{+0.36}_{-0.22}$ & 1.76$^{+0.47}_{-0.28}$ &  1.71$^{+0.54}_{-0.26}$ \\
$N_{H,\rm tor} (10^{22}\text{\,cm}^{-2})$ & 100* & 100* & 65$^{+100}_{-u}$ & $-$ 
& 100* & 100* & 41$^{+1800}_{-17}$ & $-$ \\
$CF_{\rm tor}$ & $-$ & 0.67* & 1.00$^{+u}_{-0.54}$ & $-$  
& $-$ & 0.67* & 0.90$^{+0.07}_{-0.68}$ & $-$ \\
$\theta_{\rm inc}(^{\circ})$ & 45* & 45* & 33$^{+u}_{-u}$ & 0$^{+u}_{-u}$ 
& 45* & 45* & 83$^{+3}_{-49}$ & 79$^{+7}_{-74}$  \\
$\sigma_{\rm tor}$ & $-$ & $-$ & $-$ &  7$^{+46}_{-u}$ 
& $-$ & $-$ &  $-$ & 22$^{+51}_{-11}$ \\
CTKcover & $-$ & $-$ & $-$ &  0$^{+u}_{-u}$ 
& $-$ &  $-$ & $-$ & 0.06$^{+0.49}_{-0.04}$ \\
norm$(10^{-3})$ & 0.38$^{+1.01}_{-u}$ & 0.14$^{+0.27}_{-u}$ & 0.17$^{+0.35}_{-0.09}$& 13$^{+16}_{-10}$ 
& 0.78$^{+1.49}_{-0.55}$ & 0.27$^{+0.38}_{-0.17}$ & 0.57$^{+3.66}_{-0.42}$ & 1.65$^{+4.79}_{-1.16}$  \\
$f_{\rm s}$$(10^{-3})$ & 19$^{+31}_{-12}$ & 36$^{+49}_{-21}$ &  32$^{+22}_{-18}$ & $-$ 
& 9$^{+14}_{-6}$ & 24$^{+19}_{-14}$ & 14$^{+21}_{-12}$ & $-$\\
$f_{\rm s, uxclumpy}$$(10^{-3})$ & $-$ & $-$ & $-$ & 0.7$^{+u}_{-0.5}$ 
& $-$ & $-$ & $-$ & 30$^{+43}_{-26}$ \\
$C_{\rm xmm}$ & 0.92$^{+0.98}_{-0.41}$ & 1.01$^{+1.06}_{-0.45}$ & 1.10$^{+0.83}_{-0.43}$ & 1.01$^{+u}_{-0.49}$ 
& 0.66$^{+0.66}_{-0.21}$ & 0.66$^{+0.89}_{-0.21}$ & 0.63$^{+0.88}_{-0.22}$ & 0.62$^{+0.64}_{-0.17}$ \\
$C_{\rm BAT}$ & 5.73$^{+9.86}_{-3.76}$ & 5.76$^{+8.41}_{-3.41}$ & 5.23$^{+7.19}_{-3.86}$ & 3.39$^{+3.51}_{-2.22}$ 
& 2.71$^{+4.85}_{-1.33}$ & 5.18$^{+3.45}_{-2.98}$ & 5.38$^{+3.78}_{-4.06}$ & 1.79$^{+2.92}_{-1.08}$ \\
\hline 
$N_{\rm H,cha} (10^{22}\text{\,cm}^{-2})$ & 50$^{+63}_{-24}$ & 36$^{+44}_{-20}$ & 36$^{+42}_{-16}$ & 50$^{+75}_{-29}$ 
& 81$^{+350}_{-30}$ & 51$^{+390}_{-22}$ & 65$^{+390}_{-27}$ & 110$^{+340}_{-48}$  \\
$N_{\rm H,xmm} (10^{22}\text{\,cm}^{-2})$ & 85$^{+u}_{-33}$ & 80$^{+u}_{-32}$ & 87$^{+u}_{-34}$ & 98$^{+u}_{-49}$ 
& 140$^{+330}_{-50}$ & 120$^{+360}_{-36}$ & 290$^{+190}_{-190}$ & 170$^{+310}_{-57}$  \\
\hline 
$\log(L_{\text{2-10\,keV, cha}})$ & 41.8$^{+0.2}_{-0.3}$ & 42.0$^{+0.2}_{-0.3}$ & 42.0$^{+0.2}_{-0.2}$ & 43.9$^{+0.3}_{-0.4}$ 
& 42.3$^{+0.2}_{-0.4}$ & 41.9$^{+0.2}_{-0.3}$ & 42.0$^{+0.2}_{-0.2}$ & 43.1$^{+0.3}_{-0.5}$  \\
$\log(L_{\text{2-10\,keV, xmm}})$ & 41.8$^{+0.2}_{-0.4}$ & 42.0$^{+0.3}_{-0.8}$ & 42.0$^{+0.3}_{-0.9}$ & 43.9$^{+0.3}_{-0.4}$ 
& 42.4$^{+0.3}_{-1.7}$ & 42.0$^{+0.3}_{-0.7}$ & 42.1$^{+0.3}_{-1.6}$ & 43.0$^{+0.3}_{-0.5}$  \\
$\log(L_{\text{15-150\,keV}})$ & 42.1$^{+0.1}_{-0.1}$ & 42.4$^{+0.1}_{-0.2}$ & 42.4$^{+0.1}_{-0.2}$ & 44.8$^{+0.7}_{-0.6}$ 
& 42.7$^{+0.1}_{-0.1}$ & 42.4$^{+0.1}_{-0.2}$ & 42.3$^{+0.1}_{-0.2}$ & 43.7$^{+0.4}_{-0.6}$  \\
\hline
$\log(F_{\text{2-10\,keV, cha}})$ & -12.7$^{+0.1}_{-0.1}$ & -12.8$^{+0.1}_{-0.1}$ & -12.7$^{+0.1}_{-0.1}$ & -12.7$^{+0.1}_{-0.1}$ 
& -12.7$^{+0.1}_{-0.1}$ & -12.8$^{+0.1}_{-0.1}$ & -12.7$^{+0.1}_{-0.1}$ & -12.7$^{+0.1}_{-0.1}$  \\
$\log(F_{\text{2-10\,keV, xmm}})$ & -13.0$^{+0.1}_{-0.1}$ & -13.0$^{+0.1}_{-0.1}$ & -13.0$^{+0.1}_{-0.1}$ & -13.0$^{+0.1}_{-0.1}$ 
& -13.0$^{+0.1}_{-0.1}$ & -13.0$^{+0.1}_{-0.1}$ & -13.0$^{+0.1}_{-0.1}$ & -13.0$^{+0.1}_{-0.1}$  \\
$\log(F_{\text{15-150\,keV}})$ & -11.0$^{+0.1}_{-0.1}$ & -11.0$^{+0.1}_{-0.1}$ & -11.0$^{+0.1}_{-0.1}$ & -11.0$^{+0.1}_{-0.1}$ 
& -11.0$^{+0.1}_{-0.1}$ & -11.0$^{+0.1}_{-0.1}$ & -11.0$^{+0.1}_{-0.1}$ & -11.0$^{+0.1}_{-0.1}$  \\
\hline 
$P_{\rm ch}(\text{CT})$ & $-$ & $-$ & $-$ & $-$ 
& 42.6\,\% & 38.7\,\% & 59.1\,\% & 60.6\,\%  \\
$P_{\rm xmm}(\text{CT})$ & $-$ & $-$ & $-$ & $-$ 
& 80.4\,\% & 75.7\,\% & 84.1\,\% & 86.3\,\%  \\
\enddata
\end{deluxetable*}


\begin{deluxetable*}{|c|cccc|cccc|}

\tablecaption{IC\,1141 best-fit results. \label{tab:ic_results}}

\tablehead{Model & MYTorus & borus02* & borus02 & UXCLUMPY & MYTorus & borus02* & borus02 & UXCLUMPY \\ 
Algorithm & \multicolumn{4}{c|}{Levenberg-Marquardt} & \multicolumn{4}{c|}{Nested Sampling}
}

\startdata
red $\chi^2$  & 1.17 & 1.06 & 1.20 & 1.27 
& 1.28 & 1.06 & 1.27 & 1.27 \\
$\chi^2$/d.o.f. & 21/18 &  19/18  & 18/15 & 19/15 
& 23/18 & 19/18 & 19/15 & 19/15 \\
\hline
$\Gamma$ &1.98$^{+u}_{-u}$ & 1.95$^{+0.41}_{-u}$ & 1.96$^{+u}_{-u}$ & 1.83$^{+0.71}_{-u}$ 
& 2.03$^{+0.46}_{-0.51}$ & 1.88$^{+0.44}_{-0.36}$ & 1.79$^{+0.68}_{-0.30}$ &  2.35$^{+0.19}_{-0.83}$ \\
$N_{H,\rm tor} (10^{22}\text{\,cm}^{-2})$ & $N_{H,ch}$* & 100* & 107$^{+u}_{-93}$ & $-$ 
& $N_{H,ch}$* & 100* & 28$^{+2170}_{-24}$ & $-$ \\
$CF_{\rm tor}$ & $-$ & 0.67* & 0.86$^{+u}_{-u}$ & $-$  
& $-$ & 0.67* & 0.26$^{+0.65}_{-0.13}$ & $-$ \\
$\theta_{\rm inc}(^{\circ})$ & 90* & 45* & 32$^{+u}_{-u}$ & 90$^{+u}_{-u}$ 
& 90* & 45* & 39$^{+44}_{-14}$ & 14$^{+62}_{-14}$  \\
$\sigma_{\rm tor}$ & $-$ & $-$ & $-$ &  84$^{+u}_{-u}$ 
& $-$ & $-$ &  $-$ & 70$^{+10}_{-60}$ \\
CTKcover & $-$ & $-$ & $-$ &  0.0$^{+u}_{-y}$ 
& $-$ & $-$ &  $-$ & 0.10$^{+0.39}_{-0.10}$ \\
norm$(10^{-3})$ & 1.20$^{+4.36}_{-u}$ & 0.92$^{+1.15}_{-0.68}$ & 0.94$^{+6.72}_{-0.73}$& 1.22$^{+4.84}_{-0.93}$ 
& 1.24$^{+2.89}_{-0.80}$ & 0.86$^{+1.03}_{-0.52}$ & 1.12$^{+4.86}_{-0.78}$ & 6.69$^{+8.82}_{-6.05}$  \\
$f_{\rm s}$$(10^{-3})$ & 3.68$^{+7.65}_{-2.74}$ & 0.22$^{+9.16}_{-u}$ & 1.58$^{+10}_{-u}$ & $-$ 
& 3.28$^{+5.53}_{-2.36}$ & 0.09$^{+3.80}_{-0.08}$ & 0.58$^{+3.8}_{-0.56}$ & $-$\\
$f_{\rm s, uxclumpy}$$(10^{-3})$ & $-$ & $-$ & $-$ & 2.4$^{+u}_{-u}$ 
& $-$ & $-$ & $-$ & 9.1$^{+30}_{-u}$ \\
$C_{\rm BAT}$ & 2.01$^{+5.06}_{-1.46}$ & 2.04$^{+3.05}_{-u}$ & 1.82$^{+2.93}_{-u}$ & 1.20$^{+u}_{-0.90}$ 
& 2.77$^{+2.36}_{-2.12}$ & 1.90$^{+2.68}_{-1.20}$ & 1.29$^{+3.39}_{-0.72}$ & 0.94$^{+1.70}_{-0.71}$ \\
\hline 
$N_{H,ch} (10^{22}\text{\,cm}^{-2})$ & 44$^{+15}_{-12}$ & 45$^{+16}_{-15}$ & 43$^{+69}_{-13}$ & 47$^{+23}_{-16}$ 
& 45$^{+14}_{-13}$ & 43$^{+22}_{-13}$ & 49$^{+120}_{-15}$ & 50$^{+92}_{-15}$  \\
\hline 
$\log(L_{\text{2-10\,keV, cha}})$ & 42.2$^{+0.1}_{-0.1}$ & 42.1$^{+0.1}_{-0.1}$ & 42.1$^{+0.1}_{-0.1}$ & 42.3$^{+0.6}_{-0.4}$ 
& 42.2$^{+0.1}_{-0.1}$ & 42.1$^{+0.1}_{-0.1}$ & 42.2$^{+0.1}_{-0.1}$ & 42.7$^{+0.4}_{-0.6}$  \\
$\log(L_{\text{15-150\,keV}})$ & 42.4$^{+0.1}_{-0.1}$ & 42.2$^{+0.2}_{-0.3}$ & 42.2$^{+0.1}_{-0.3}$ & 42.6$^{+0.6}_{-0.4}$ 
& 42.3$^{+0.1}_{-0.2}$ & 42.2$^{+0.2}_{-0.3}$ & 42.2$^{+0.2}_{-0.3}$ & 42.7$^{+0.5}_{-0.6}$  \\
\hline 
$\log(F_{\text{2-10\,keV, cha}})$ & -12.27$^{+0.08}_{-0.08}$ & -12.29$^{+0.08}_{-0.08}$ & -12.28$^{+0.08}_{-0.08}$ & -12.24$^{+0.08}_{-0.08}$ 
& -12.28$^{+0.08}_{-0.08}$ & -12.29$^{+0.08}_{-0.08}$ & -12.29$^{+0.08}_{-0.08}$ & -12.25$^{+0.08}_{-0.08}$  \\
$\log(F_{\text{15-150\,keV}})$ & -11.1$^{+0.1}_{-0.2}$ & -11.1$^{+0.1}_{-0.2}$ & -11.1$^{+0.1}_{-0.1}$ & -11.1$^{+0.1}_{-0.1}$ 
& -11.1$^{+0.1}_{-0.2}$ & -11.1$^{+0.1}_{-0.2}$ & -11.2$^{+0.1}_{-0.2}$ & -11.1$^{+0.1}_{-0.1}$  \\
\hline 
$P_{\rm ch}(\text{CT})$ & $-$ & $-$ & $-$ & $-$ 
& 0.0\,\% & 0.1\,\% & 5.5\,\% & 4.4\,\%  \\
\enddata
\end{deluxetable*}


\begin{deluxetable*}{|c|ccc|ccc|}

\tablecaption{2MASX\,J17253053--4510279 best-fit results.\label{tab:2masx_results}}

\tablehead{Model & Simple & borus02 & UXCLUMPY & Simple & borus02 & UXCLUMPY \\ 
Algorithm & \multicolumn{3}{c|}{Levenberg-Marquardt} & \multicolumn{3}{c|}{Nested Sampling}
}

\startdata
red $\chi^2$  & 0.80 & 0.93 & 0.92 
& 0.93 & 0.93 & 0.93 \\
$\chi^2$/d.o.f. & 252/314 & 291/313  & 289/313 
& 292/314 & 292/313 & 292/313 \\
\hline
$\Gamma$ & 1.76$^{+0.15}_{-0.09}$ & 1.79$^{+0.11}_{-0.12}$ & 1.70$^{+0.13}_{-0.06}$ 
& 1.78$^{+0.10}_{-0.08}$ & 1.85$^{+0.10}_{-0.08}$ &  1.83$^{+0.10}_{-0.08}$ \\
$N_{H,\rm tor} (10^{22}\text{\,cm}^{-2})$ & $-$ & 100$^*$ & $-$ 
& $-$ & 100$^*$ & $-$ \\
$CF_{\rm tor}$ & $-$ & 0.67$^*$ & $-$  
& $-$ & 0.67$^*$ & $-$ \\
$\theta_{\rm inc}(^{\circ})$ & 90* & 45$^*$ & 90$^*$ 
& 90* & 45$^*$ & 90$^*$  \\
$\sigma_{\rm tor}$ & $-$ & $-$ &  28$^*$ 
& $-$ & $-$ & 28$^*$ \\
CTKcover & $-$ & $-$ &  0$^*$ 
& $-$ & $-$ & 0$^*$ \\
norm$(10^{-3})$ & 0.26$^{+0.06}_{-0.05}$ & 0.32$^{+0.06}_{-0.05}$& 0.38$^{+0.08}_{-0.09}$ 
& 0.33$^{+0.05}_{-0.04}$ & 0.35$^{+0.06}_{-0.05}$ & 0.44$^{+0.07}_{-0.09}$  \\
$C_{\rm xmm}$ & 2.16$^{+0.27}_{-0.23}$ & 1.95$^{+0.22}_{-0.20}$ & 1.92$^{+0.31}_{-0.21}$ 
& 1.99$^{+0.23}_{-0.18}$ & 1.96$^{+0.24}_{-0.19}$ & 1.86$^{+0.41}_{-0.18}$ \\
$C_{\rm BAT}$ & 5.74$^{+3.02}_{-1.54}$ & 3.78$^{+1.45}_{-1.29}$ & 3.13$^{+1.62}_{-0.96}$ 
& 4.83$^{+1.63}_{-1.33}$ & 4.30$^{+1.42}_{-1.24}$ & 4.21$^{+1.49}_{-1.13}$ \\
\hline 
$N_{\rm H,cha} (10^{22}\text{\,cm}^{-2})$ & 0.38$^{+0.21}_{-0.18}$ & 0.50$^{+0.18}_{-0.20}$ & 0.51$^{+0.17}_{-0.16}$ 
& 0.53$^{+0.17}_{-0.17}$ & 0.54$^{+0.21}_{-0.15}$ & 0.46$^{+0.16}_{-0.14}$  \\
$N_{\rm H,xmm} (10^{22}\text{\,cm}^{-2})$ & 0.02$^{+0.13}_{-u}$ & 0.07$^{+0.11}_{-u}$ & 0.16$^{+0.11}_{-u}$ 
& 0.11$^{+0.09}_{-0.01}$ & 0.12$^{+0.12}_{-0.02}$ & 0.11$^{+0.08}_{-0.01}$  \\
\hline 
$\log(L_{\text{2-10\,keV, cha}})$ & 41.87$^{+0.03}_{-0.03}$ & 41.95$^{+0.03}_{-0.03}$ & 42.42$^{+0.05}_{-0.08}$ 
& 41.96$^{+0.03}_{-0.03}$ & 41.95$^{+0.03}_{-0.03}$ & 42.39$^{+0.05}_{-0.06}$  \\
$\log(L_{\text{2-10\,keV, xmm}})$ & 41.87$^{+0.02}_{-0.02}$ & 41.95$^{+0.02}_{-0.02}$ & 42.71$^{+0.05}_{-0.08}$ 
& 41.96$^{+0.02}_{-0.02}$ & 41.94$^{+0.02}_{-0.02}$ & 42.68$^{+0.05}_{-0.06}$  \\
$\log(L_{\text{15-150\,keV}})$ & 42.2$^{+0.1}_{-0.1}$ & 42.2$^{+0.1}_{-0.1}$ & 43.3$^{+0.2}_{-0.2}$ 
& 42.3$^{+0.1}_{-0.1}$ & 42.2$^{+0.1}_{-0.1}$ & 43.3$^{+0.1}_{-0.1}$  \\
\hline 
$\log(F_{\text{2-10\,keV, cha}})$ & -12.06$^{+0.03}_{-0.03}$ & -11.96$^{+0.03}_{-0.03}$ & -11.96$^{+0.03}_{-0.03}$ 
& -11.97$^{+0.03}_{-0.03}$ & -11.96$^{+0.03}_{-0.03}$ & -11.98$^{+0.03}_{-0.03}$  \\
$\log(F_{\text{2-10\,keV, xmm}})$ & -11.71$^{+0.02}_{-0.02}$ & -11.65$^{+0.02}_{-0.02}$ & -11.65$^{+0.02}_{-0.02}$ 
& -11.66$^{+0.02}_{-0.02}$ & -11.65$^{+0.02}_{-0.02}$ & -11.66$^{+0.02}_{-0.02}$  \\
$\log(F_{\text{15-150\,keV}})$ & -10.96$^{+0.08}_{-0.10}$ & -10.95$^{+0.08}_{-0.10}$ & -10.95$^{+0.08}_{-0.10}$ 
& -10.96$^{+0.08}_{-0.10}$ & -10.97$^{+0.08}_{-0.10}$ & -10.97$^{+0.08}_{-0.10}$  \\
\hline 
$P_{\rm ch}(\text{CT})$ & $-$ & $-$ & $-$ 
& 0.0\,\% & 0.0\,\% & 0.0\,\%  \\
$P_{\rm xmm}(\text{CT})$ & $-$ & $-$ & $-$ 
& 0.0\,\% & 0.0\,\% & 0.0\,\%  \\
\enddata
\end{deluxetable*}


\begin{deluxetable*}{|c|ccc|ccc|}

\tablecaption{MCG\,+2-57-2 best-fit results. \label{tab:mcg_results}}

\tablehead{Model & Simple & borus02 & UXCLUMPY & Simple & borus02 & UXCLUMPY \\ 
Algorithm & \multicolumn{3}{c|}{Levenberg-Marquardt} & \multicolumn{3}{c|}{Nested Sampling}
}

\startdata
red $\chi^2$ & 0.97 & 1.01 & 1.01 
& 1.01 & 1.01 & 1.01 \\
$\chi^2$/d.o.f. & 199/205 &  207/205  & 208/205 
& 208/205 & 207/205 & 208/205 \\ 
\hline
$\alpha$ & 1$^{+}_{-0.23}$ & 1$^{+}_{-0.5}$ & 1$^{+}_{-0.5}$ 
& 0.5$^{+0.4}_{-0.1}$ & 0.5$^{+0.4}_{-0.1}$ &  0.5$^{+0.4}_{-0.1}$ \\
$\Gamma$ & 1.90$^{+0.24}_{-0.23}$ & 2.00$^{+0.07}_{-0.27}$ & 1.99$^{+0.13}_{-0.24}$ 
& 1.99$^{+0.43}_{-0.24}$ & 1.93$^{+0.43}_{-0.16}$ &  1.97$^{+0.43}_{-0.20}$ \\
$N_{H,\rm tor} (10^{22}\text{\,cm}^{-2})$ & $-$ & 100$^*$ & $-$ 
& $-$ & 100$^*$ & $-$ \\
$CF_{\rm tor}$ & $-$ & 0.67$^*$ & $-$  
& $-$ & 0.67$^*$ & $-$ \\
$\theta_{\rm inc}(^{\circ})$ & 90* & 45$^*$ & 90$^*$ 
& 90* & 45$^*$ & 90$^*$  \\
$\sigma_{\rm tor}$ & $-$ & $-$ &  28$^*$ 
& $-$ &  $-$ & 28$^*$ \\
CTKcover & $-$ & $-$ &  0$^*$ 
& $-$ &  $-$ & 0$^*$ \\
norm$(10^{-3})$ & 0.72$^{+0.27}_{-0.18}$ & 0.99$^{+3.38}_{-0.17}$& 1.20$^{+3.90}_{-0.14}$ 
& 2.77$^{+1.45}_{-1.83}$ & 1.14$^{+2.93}_{-0.17}$ & 1.39$^{+3.75}_{-0.20}$  \\
$C_{\rm BAT}$ & 3.99$^{+3.27}_{-1.89}$ & 3.24$^{+0.92}_{-2.49}$ & 3.37$^{+2.30}_{-2.59}$ 
& 2.05$^{+3.13}_{-0.94}$ & 1.44$^{+2.26}_{-0.56}$ & 1.56$^{+2.43}_{-0.65}$ \\
\hline 
$N_{H,ch} (10^{22}\text{\,cm}^{-2})$ & 0.17$^{+0.26}_{-u}$ & 0.17$^{+0.20}_{-u}$ & 0.14$^{+0.17}_{-u}$ 
& 0.19$^{+0.23}_{-0.07}$ & 0.16$^{+0.26}_{-0.05}$ & 0.14$^{+0.22}_{-0.03}$  \\
\hline 
$\log(L_{\text{2-10\,keV, cha}})$ & 42.62$^{+0.03}_{-0.03}$ & 42.67$^{+0.03}_{-0.03}$ & 42.8$^{+0.4}_{-0.1}$ 
& 42.53$^{+0.02}_{-0.02}$ & 42.67$^{+0.03}_{-0.03}$ & 43.2$^{+0.1}_{-0.3}$  \\
$\log(L_{\text{15-150\,keV}})$ & 42.8$^{+0.1}_{-0.1}$ & 42.8$^{+0.1}_{-0.1}$ & 43.4$^{+0.6}_{-0.3}$ 
& 42.5$^{+0.1}_{-0.1}$ & 42.7$^{+0.1}_{-0.1}$ & 43.1$^{+0.5}_{-0.2}$  \\
\hline 
$\log(F_{\text{2-10\,keV, cha}})$ & -11.69$^{+0.03}_{-0.03}$ & -11.61$^{+0.03}_{-0.03}$ & -11.62$^{+0.03}_{-0.03}$ 
& -11.67$^{+0.03}_{-0.03}$ & -11.61$^{+0.03}_{-0.03}$ & -11.63$^{+0.03}_{-0.03}$  \\
$\log(F_{\text{15-150\,keV}})$ & -10.89$^{+0.07}_{-0.08}$ & -10.91$^{+0.07}_{-0.08}$ & -10.91$^{+0.07}_{-0.08}$ 
& -10.91$^{+0.07}_{-0.08}$ & -10.91$^{+0.07}_{-0.08}$ & -10.91$^{+0.07}_{-0.08}$  \\
\hline 
$P_{\rm ch}(\text{CT})$ & $-$ & $-$ & $-$ 
& 0.0\,\% & 0.0\,\% & 0.0\,\% \\
\enddata
\end{deluxetable*}



 \begin{figure*}[h!]
     \centering
     \includegraphics[scale=.45]{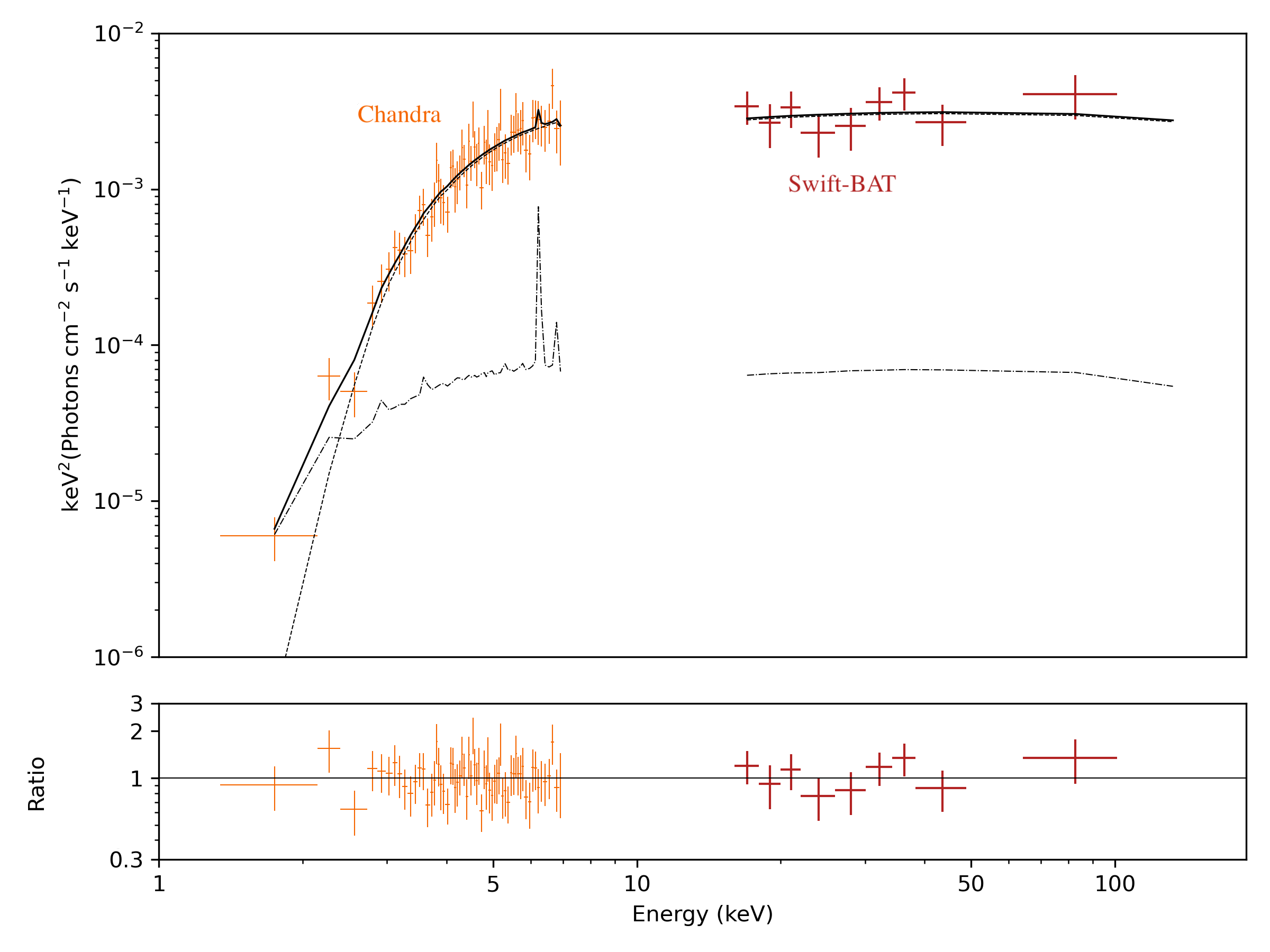}
     \includegraphics[scale=.45]{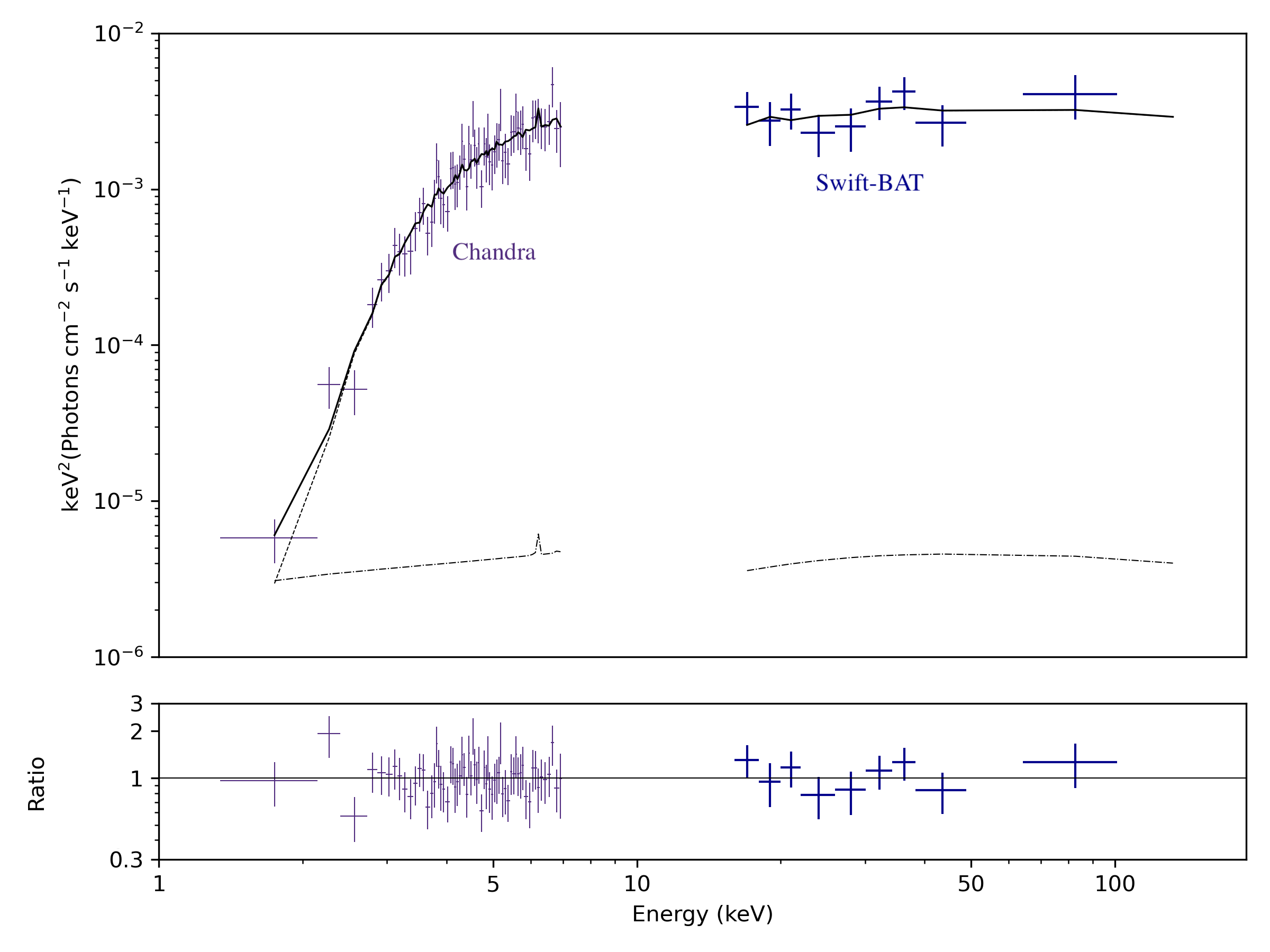} \\
     \includegraphics[scale=.58]{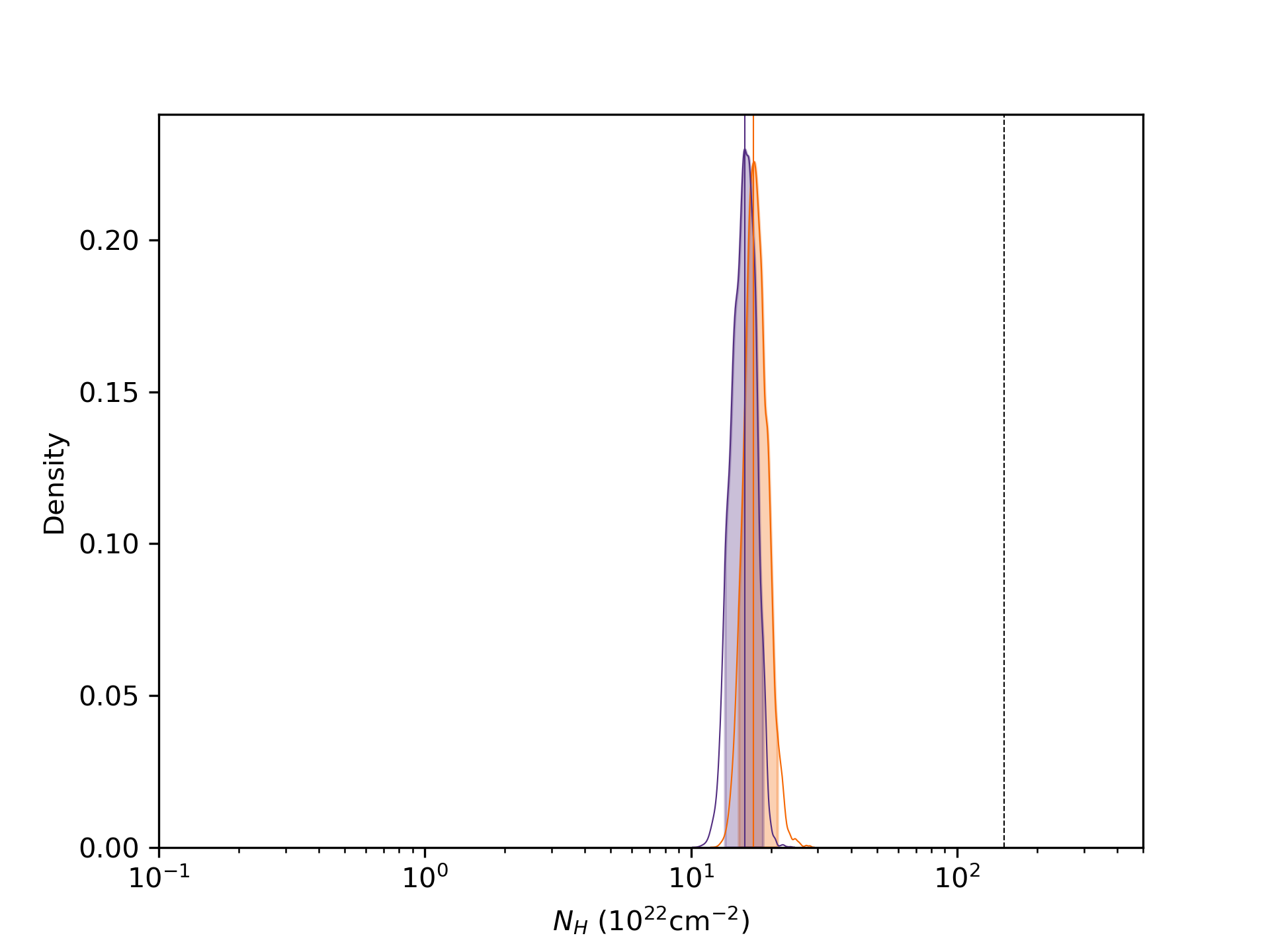}
     \includegraphics[scale=.58]{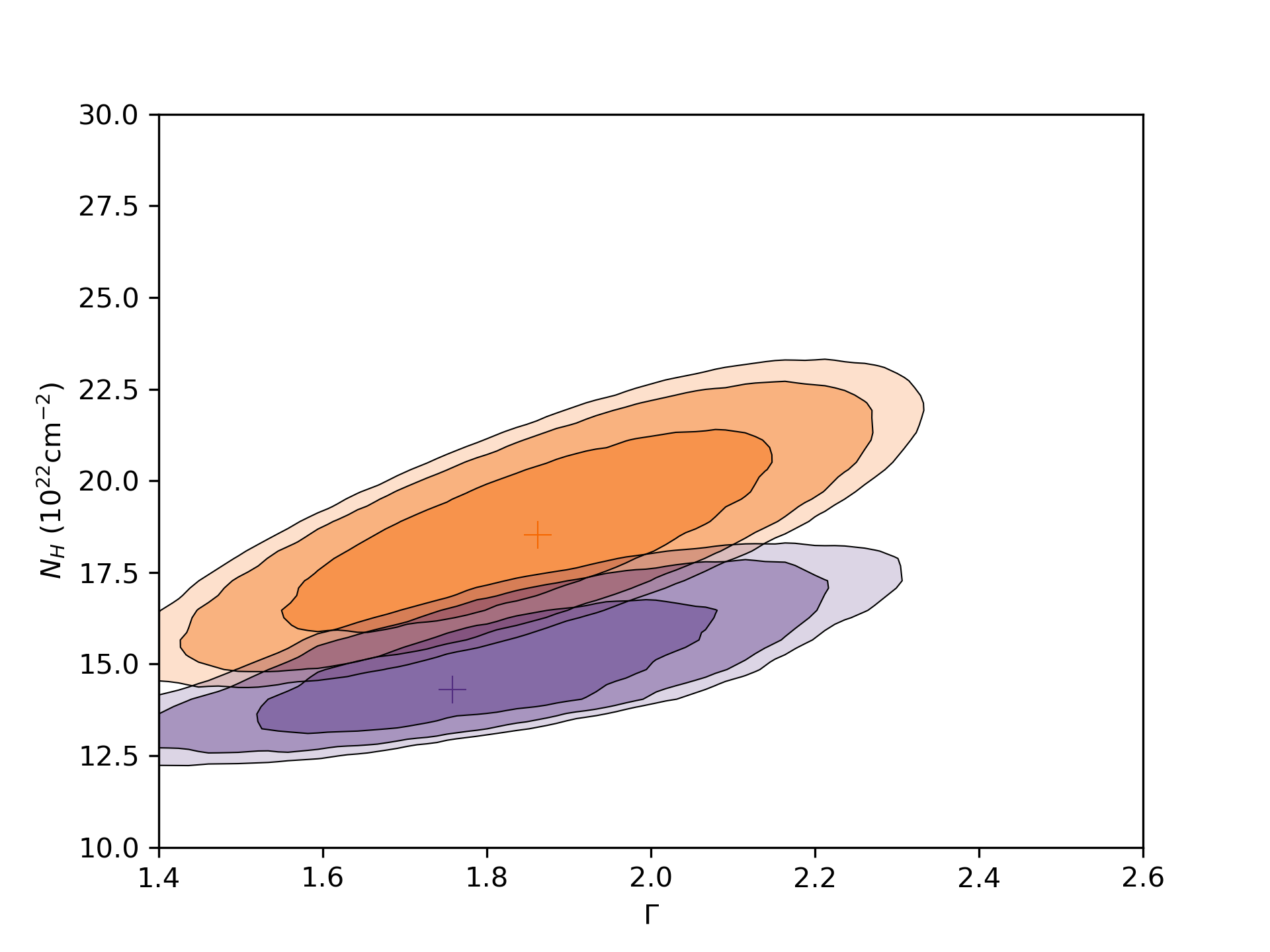}
     \caption{2MFGC\,9836. \textit{Top Left panel}: Spectral fit for the \texttt{borus02} model. \textit{Top Right panel}: Spectral fit for the \texttt{UXCLUMPY} model. \textit{Bottom Left panel}: $N_{\rm H,l.o.s}$ posteriors for 2MFGC\,9836. The \texttt{borus02} results are in orange and the \texttt{UXCLUMPY} results are in purple. \textit{Bottom Right panel}: Contour plots of line-of-sight column density and photon index for 2MFGC\,9836. The \texttt{borus02} results are in orange and the \texttt{UXCLUMPY} results are in purple.}
     \label{fig:2mfgc_results}
 \end{figure*}

  \begin{figure*}[h!]
     \centering
     \includegraphics[scale=.45]{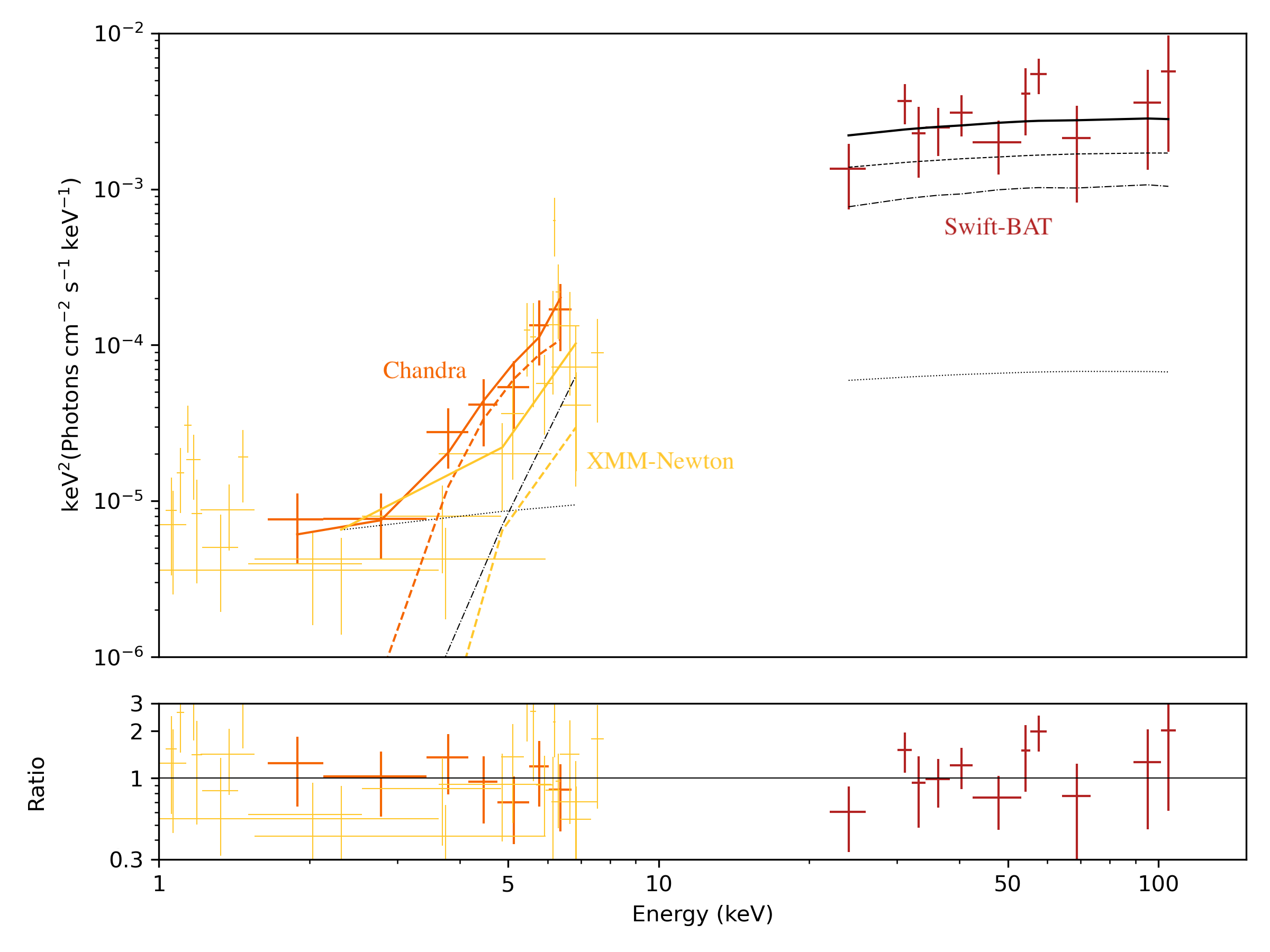}
     \includegraphics[scale=.45]{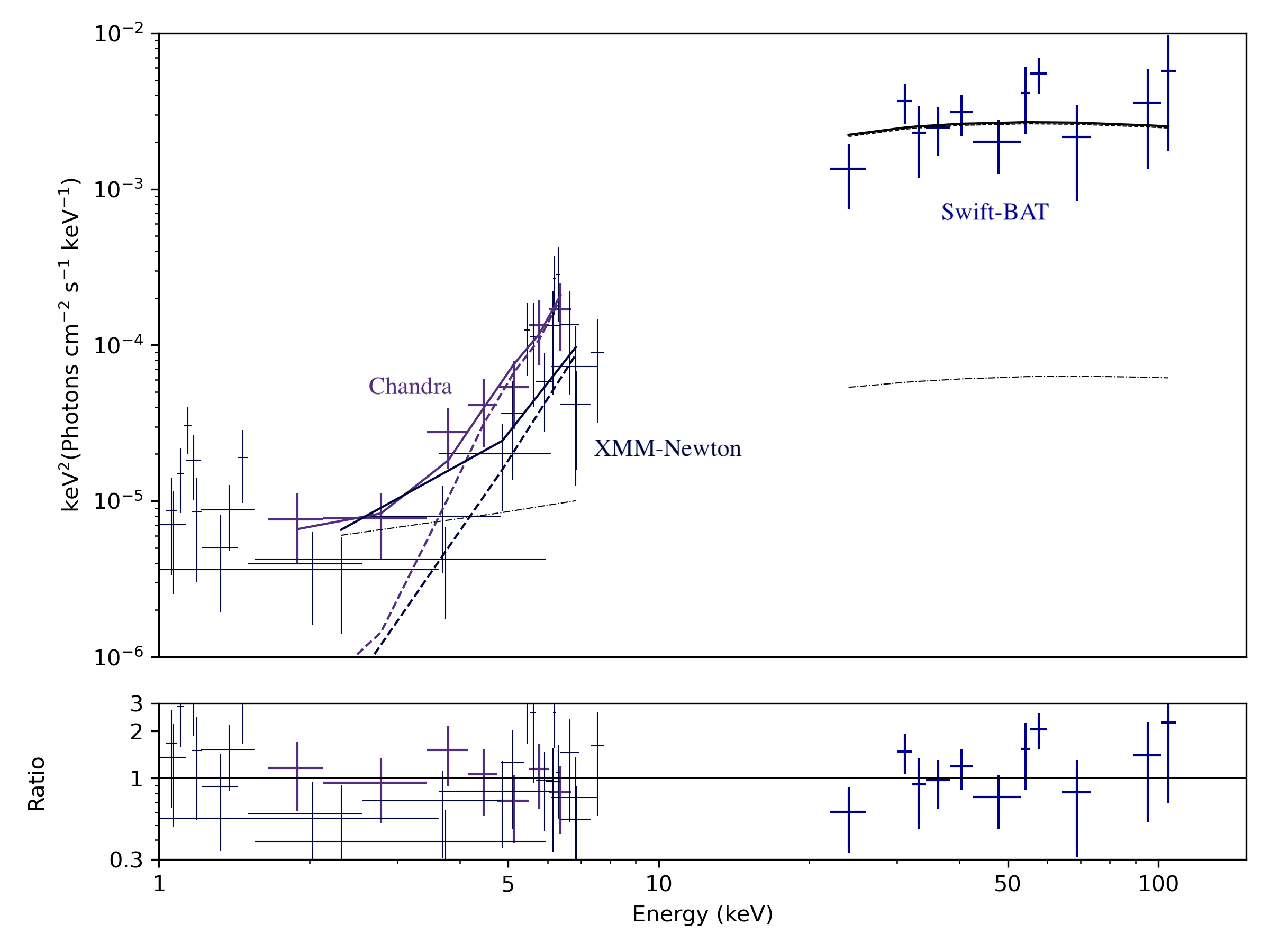} \\
     \includegraphics[scale=.58]{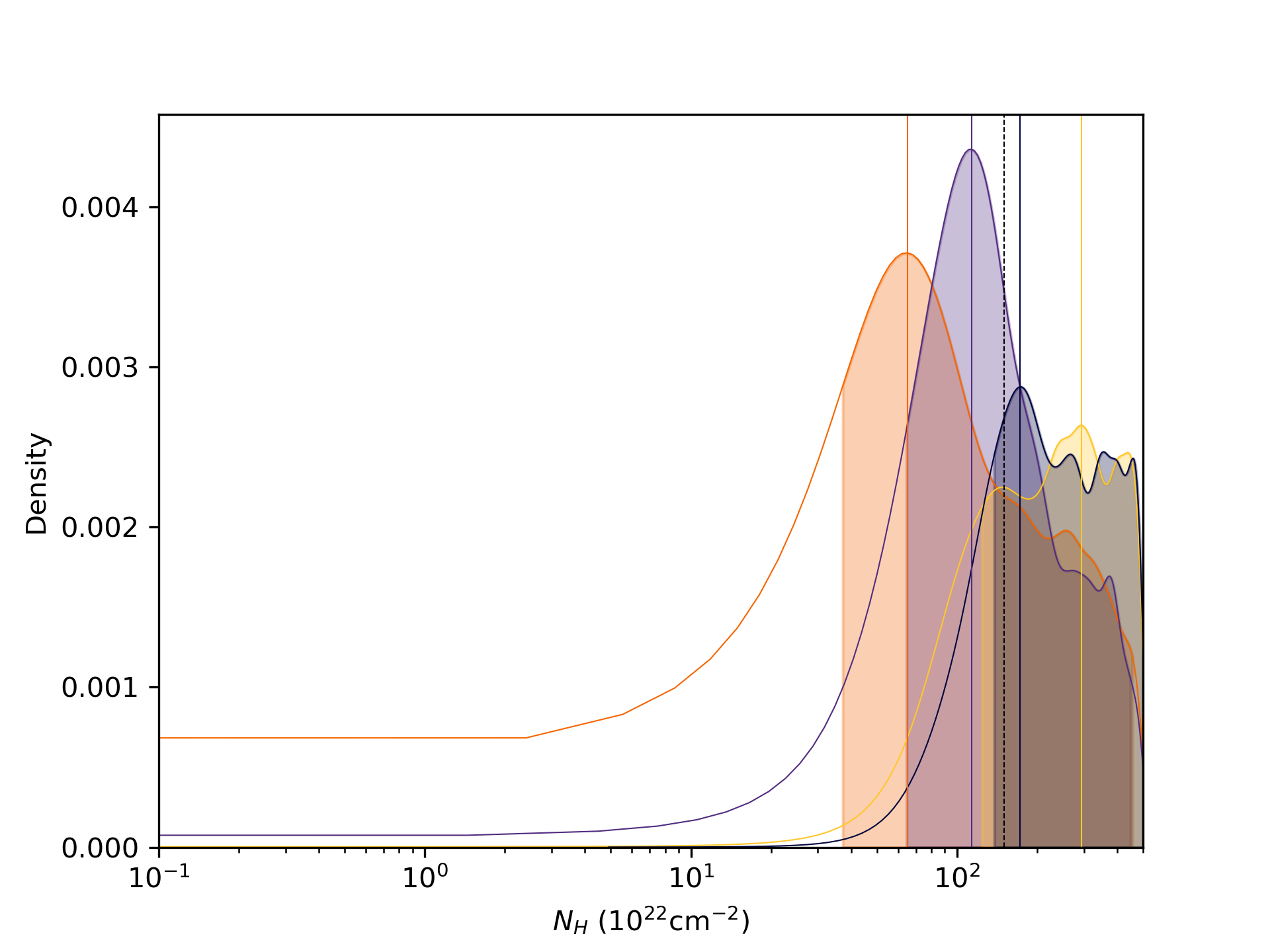}
     \includegraphics[scale=.58]{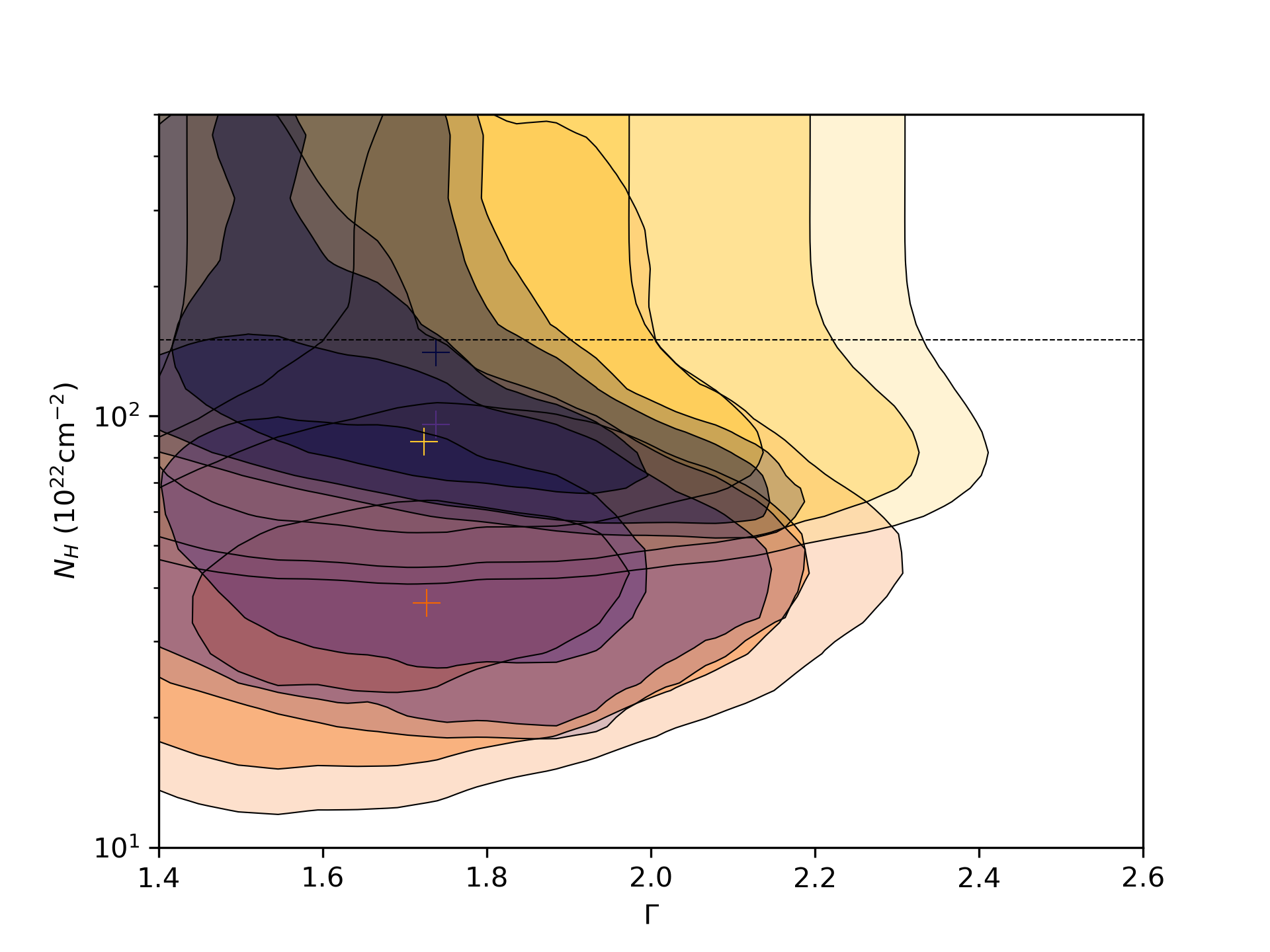}
     \caption{NGC\,5759. \textit{Top panels}: Same as Figure \ref{fig:cgc_spectra}. \textit{Bottom Left panel}: Same as Figure \ref{fig:cgc_results}. \textit{Bottom Right panel}: Same as Figure \ref{fig:cgc_contours}.}
     \label{fig:ngc_results}
 \end{figure*}

 \begin{figure*}[h!]
     \centering
     \includegraphics[scale=.45]{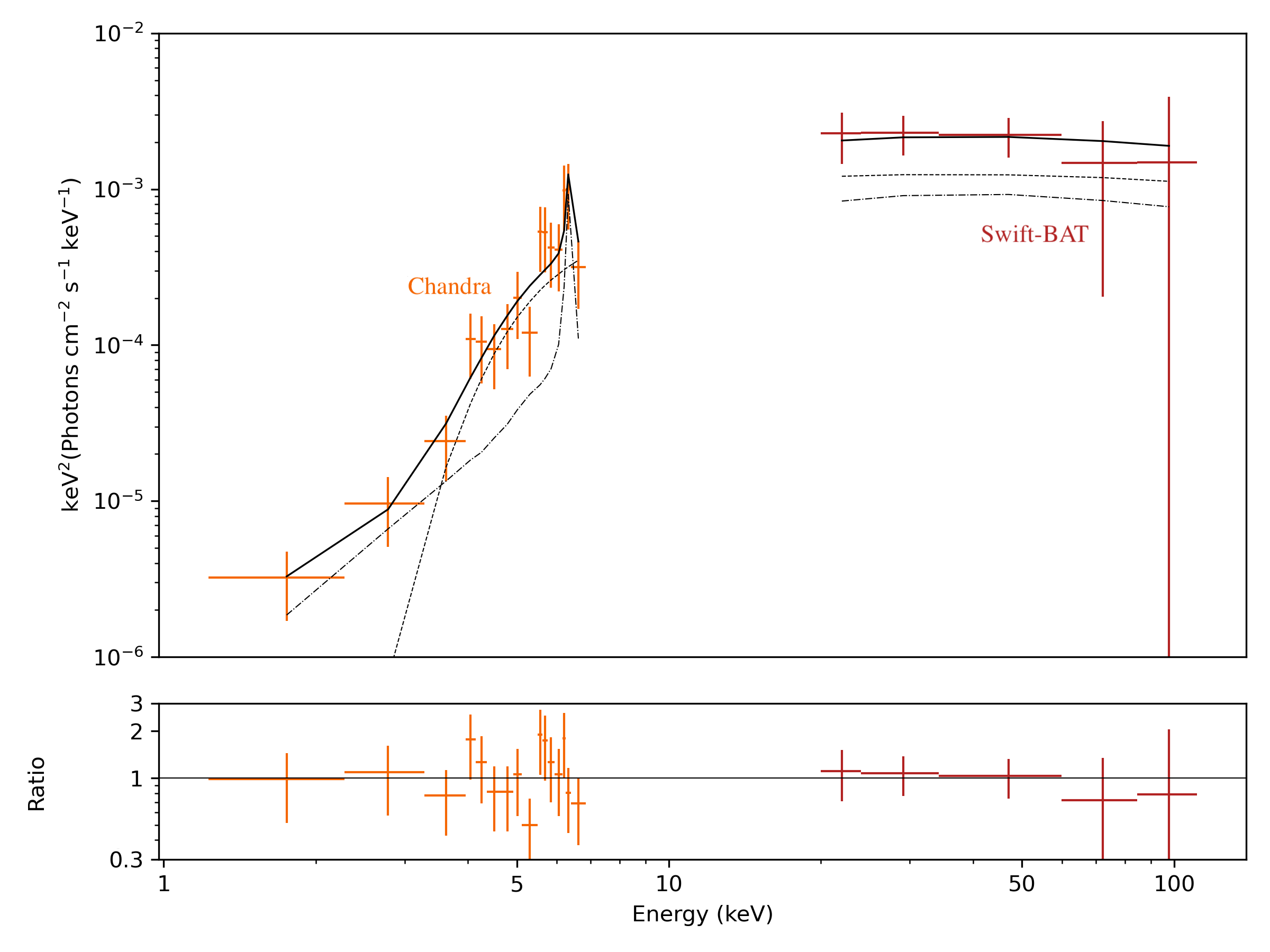}
     \includegraphics[scale=.45]{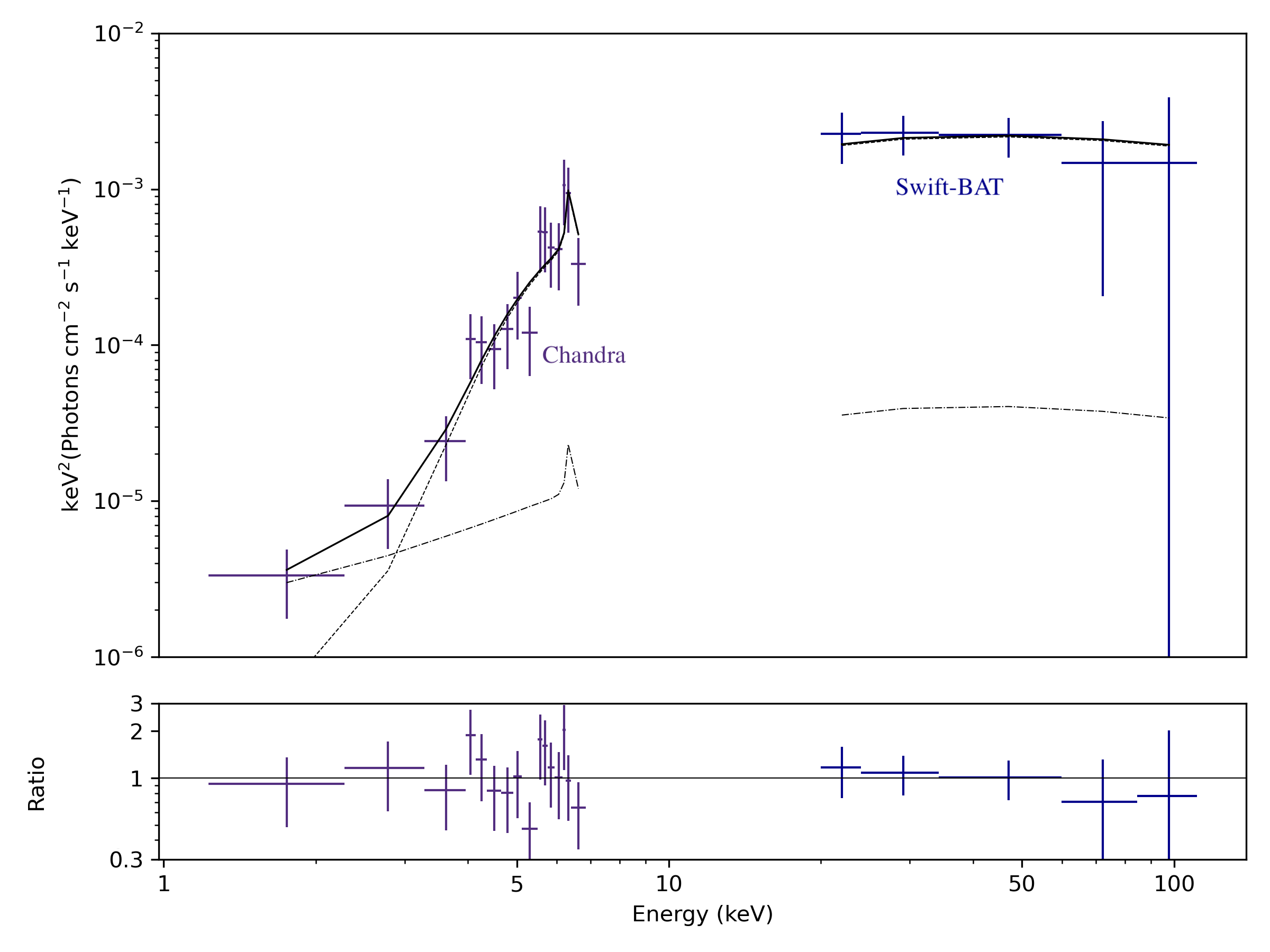} \\
     \includegraphics[scale=.58]{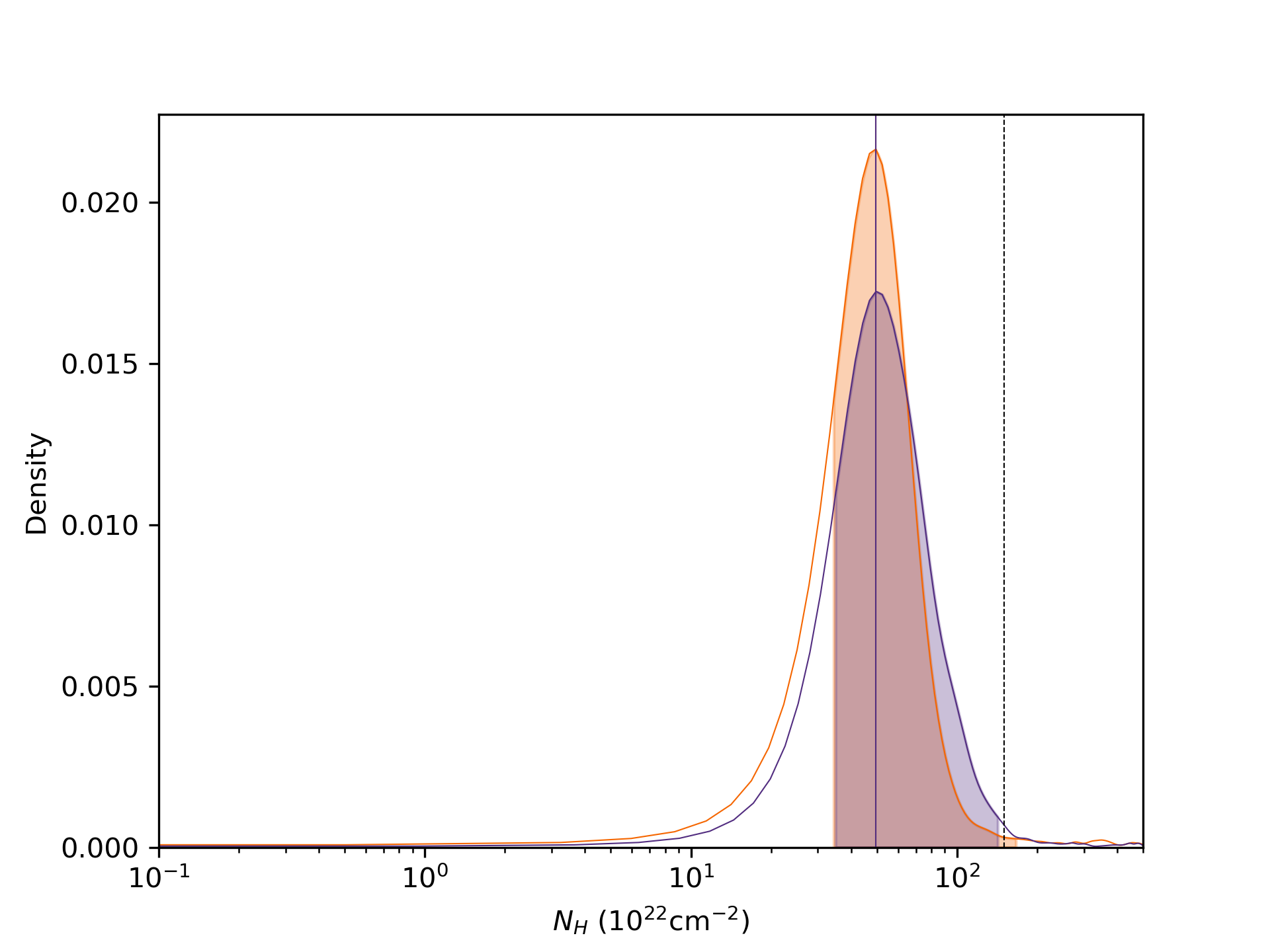}
     \includegraphics[scale=.58]{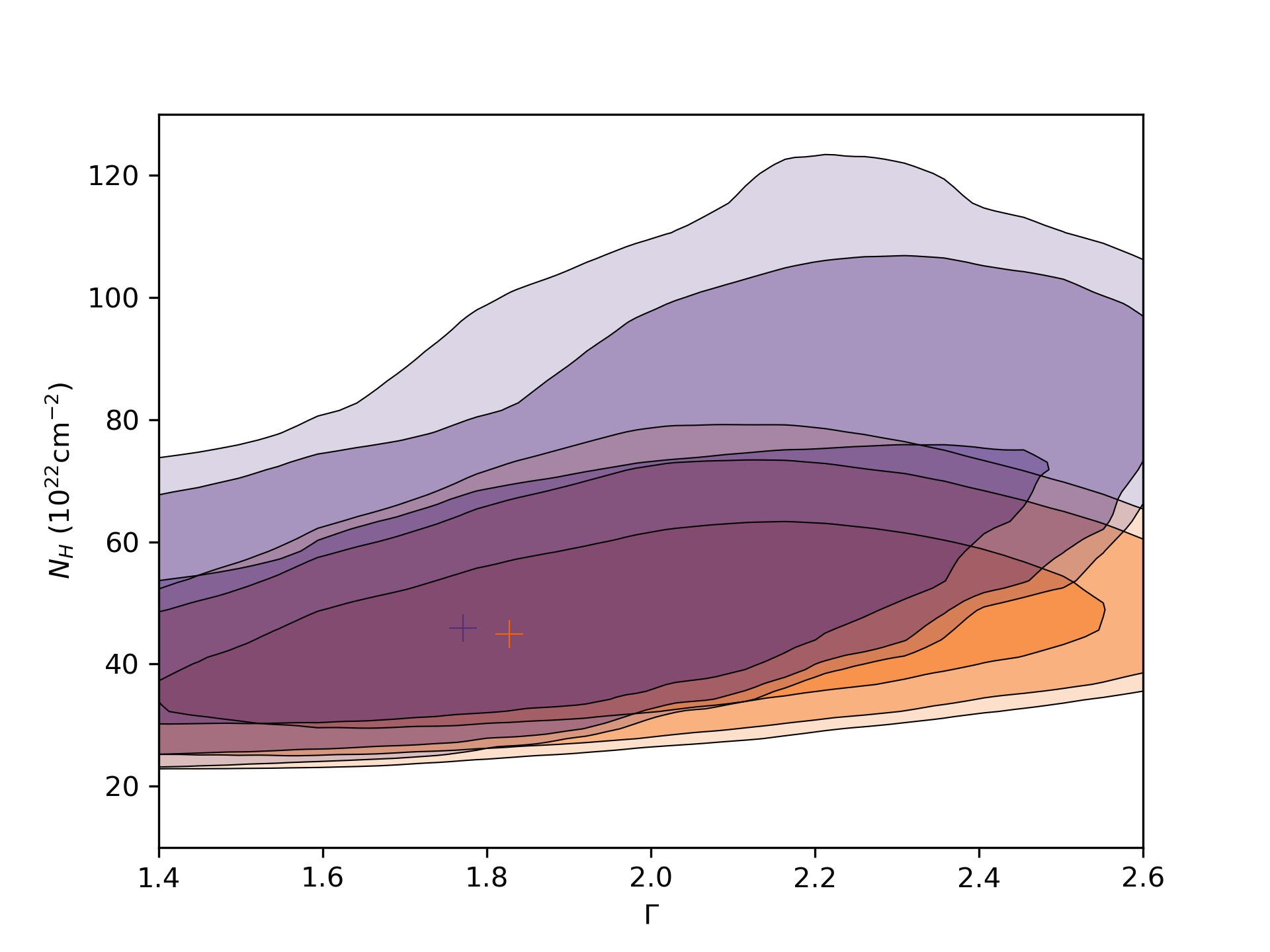}
     \caption{IC\,1141. \textit{Top Left panel}: Spectral fit for the \texttt{borus02} model. \textit{Top Right panel}: Spectral fit for the \texttt{UXCLUMPY} model. \textit{Bottom Left panel}: $N_{\rm H,l.o.s}$ posteriors for IC\,1141. The \texttt{borus02} results are in orange and the \texttt{UXCLUMPY} results are in purple. \textit{Bottom Right panel}: Contour plots of line-of-sight column density and photon index for IC\,1141. The \texttt{borus02} results are in orange and the \texttt{UXCLUMPY} results are in purple.}
     \label{fig:IC1141_results}
 \end{figure*}

 \begin{figure*}[h!]
     \centering
     \includegraphics[scale=.45]{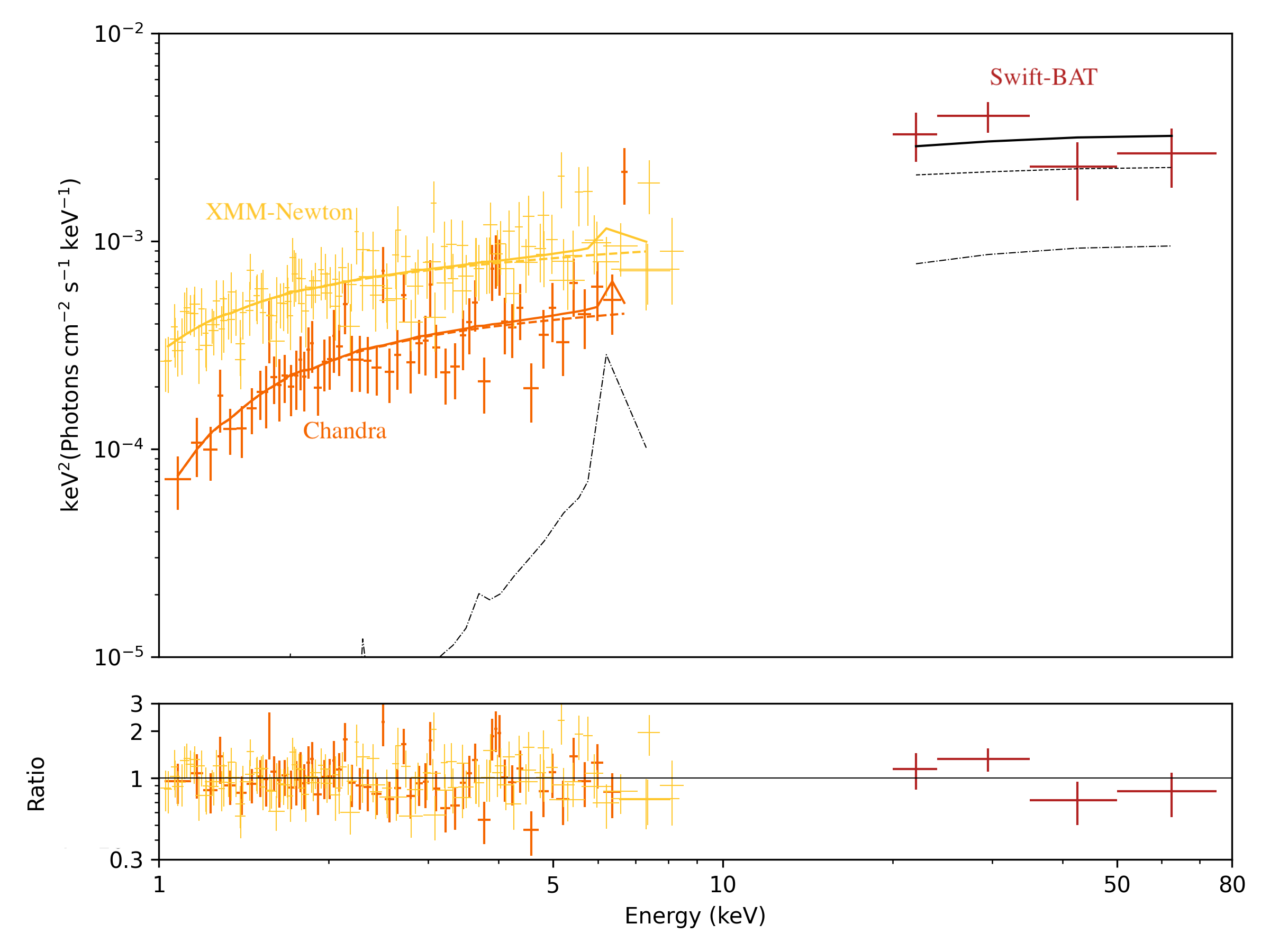}
     \includegraphics[scale=.45]{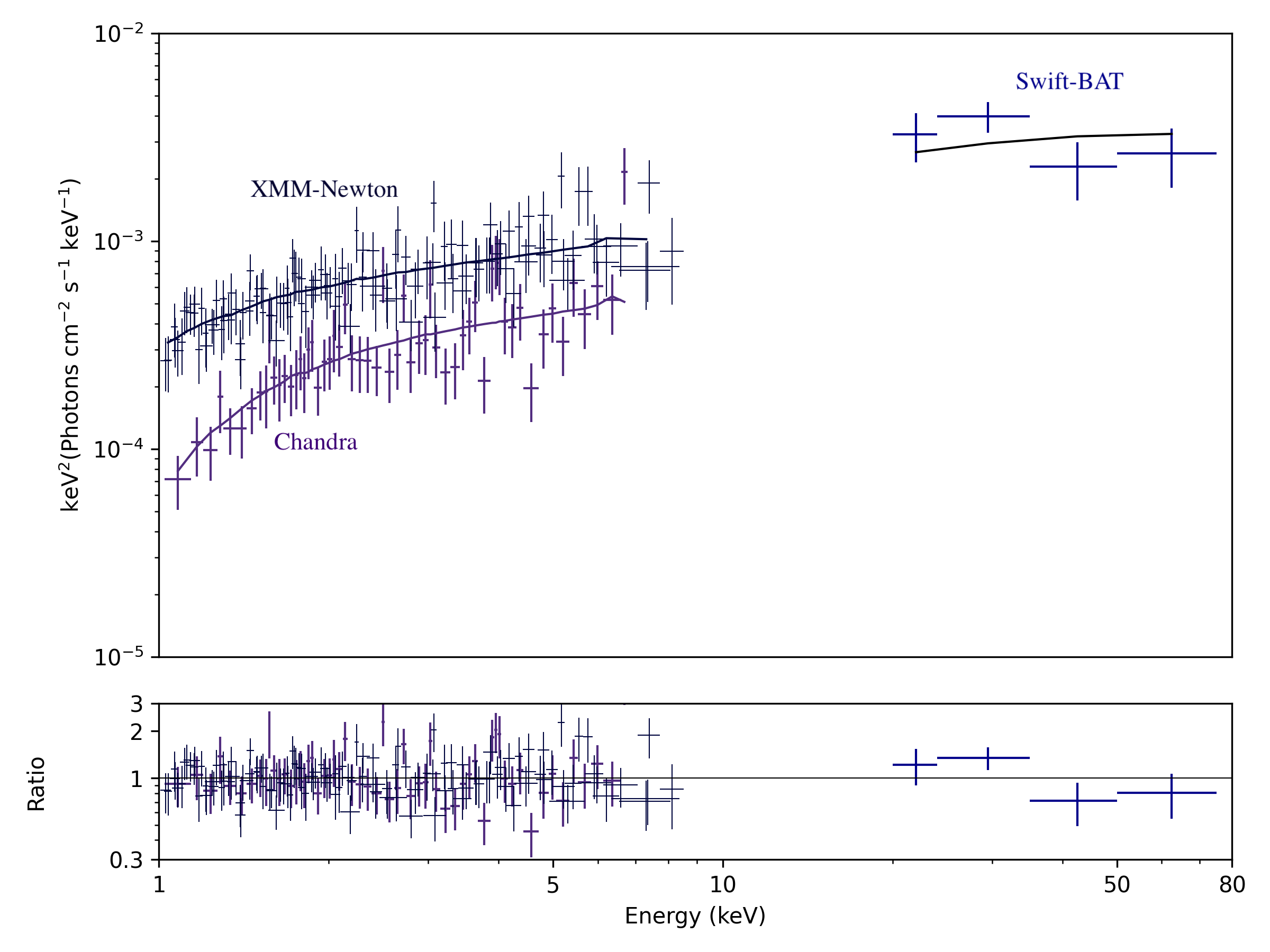} \\
     \includegraphics[scale=.58]{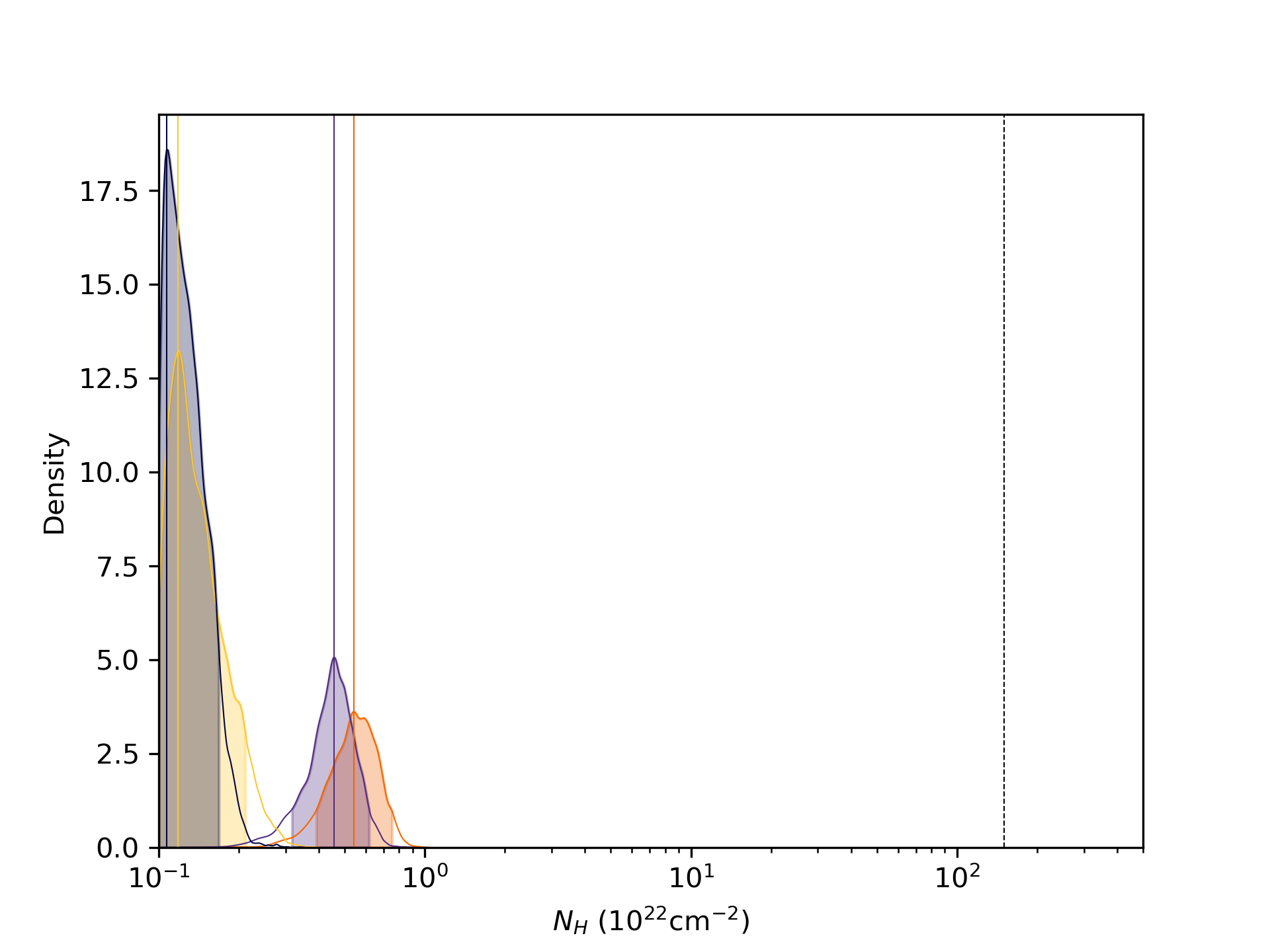}
     \includegraphics[scale=.58]{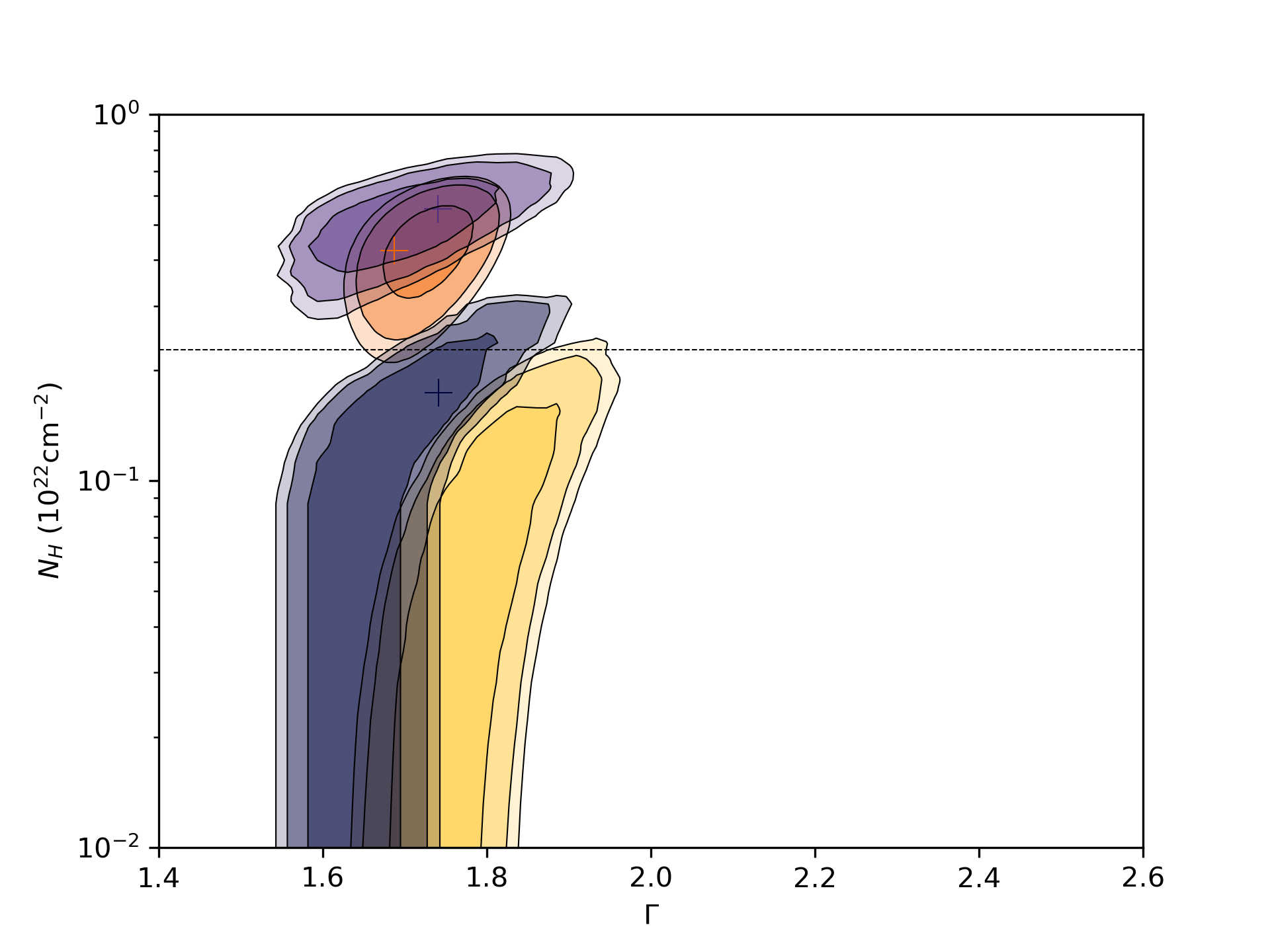}
     \caption{2MASX\,J17253053--4510279. \textit{Top panels}: Same as Figure \ref{fig:cgc_spectra}. \textit{Bottom Left panel}: Same as Figure \ref{fig:cgc_results}. \textit{Bottom Right panel}: Same as Figure \ref{fig:cgc_contours}.}
     \label{fig:2masx_results}
\end{figure*}

 \begin{figure*}[h!]
     \centering
     \includegraphics[scale=.45]{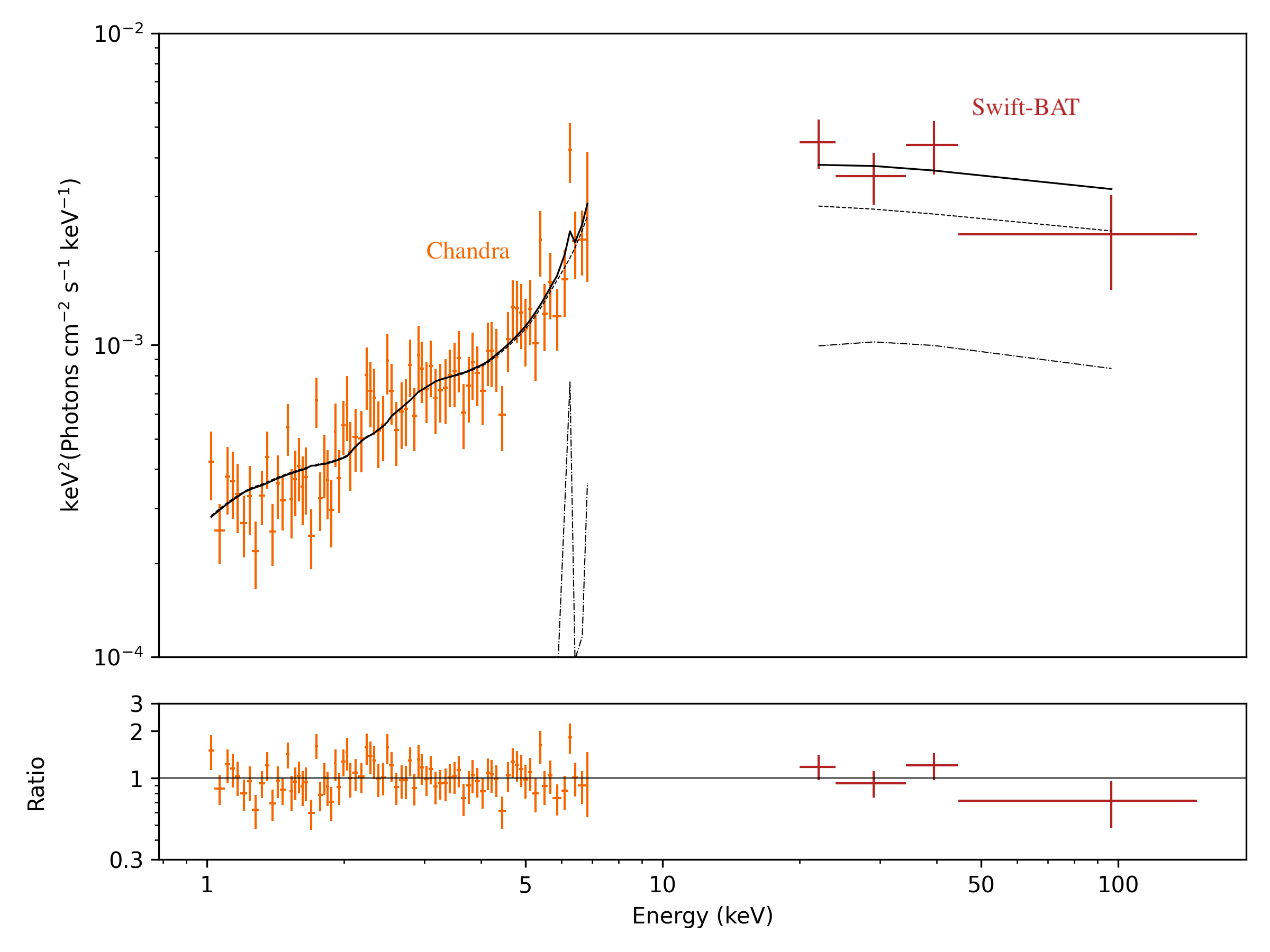}
     \includegraphics[scale=.45]{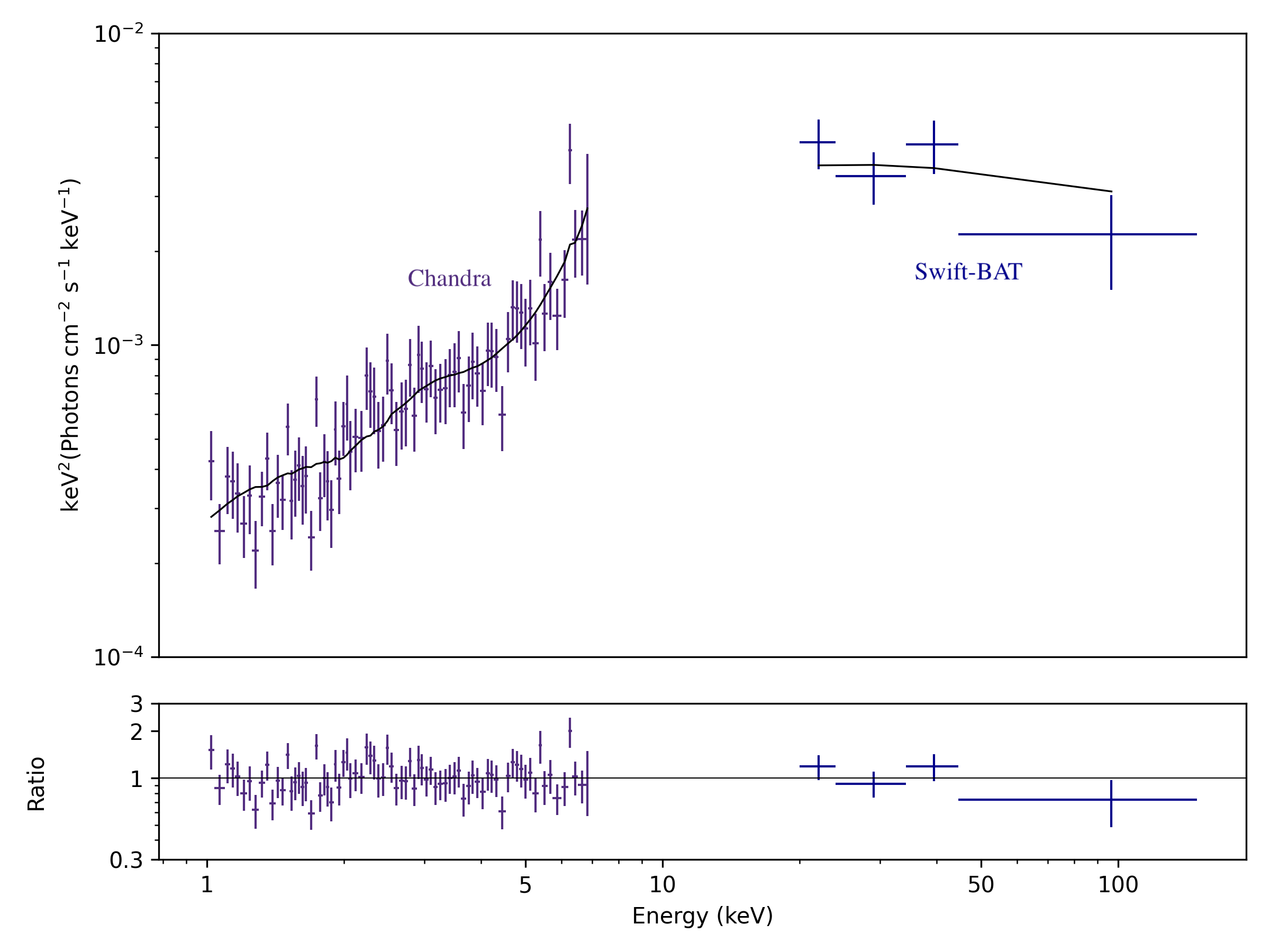} \\
     \includegraphics[scale=.58]{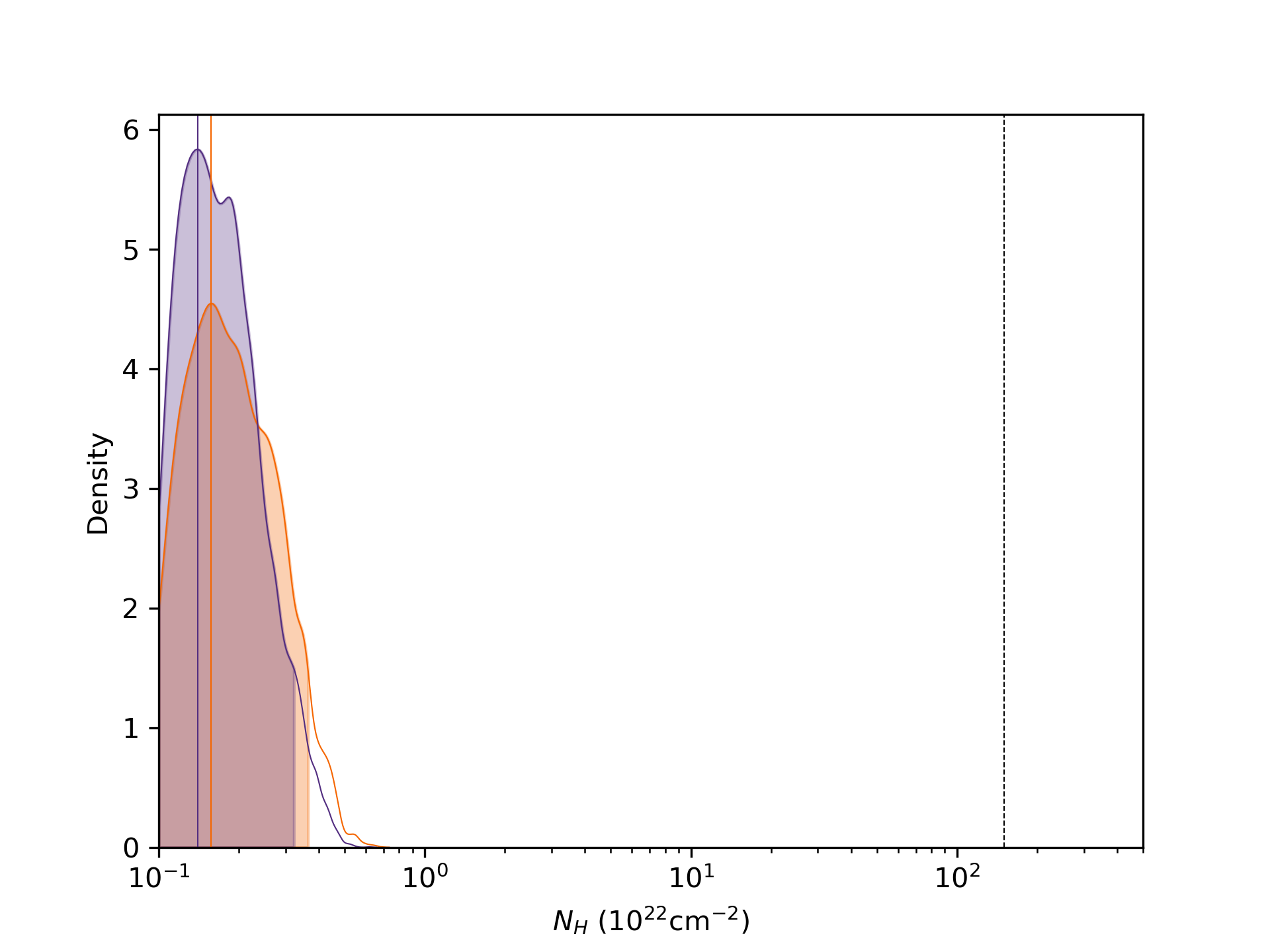}
     \includegraphics[scale=.58]{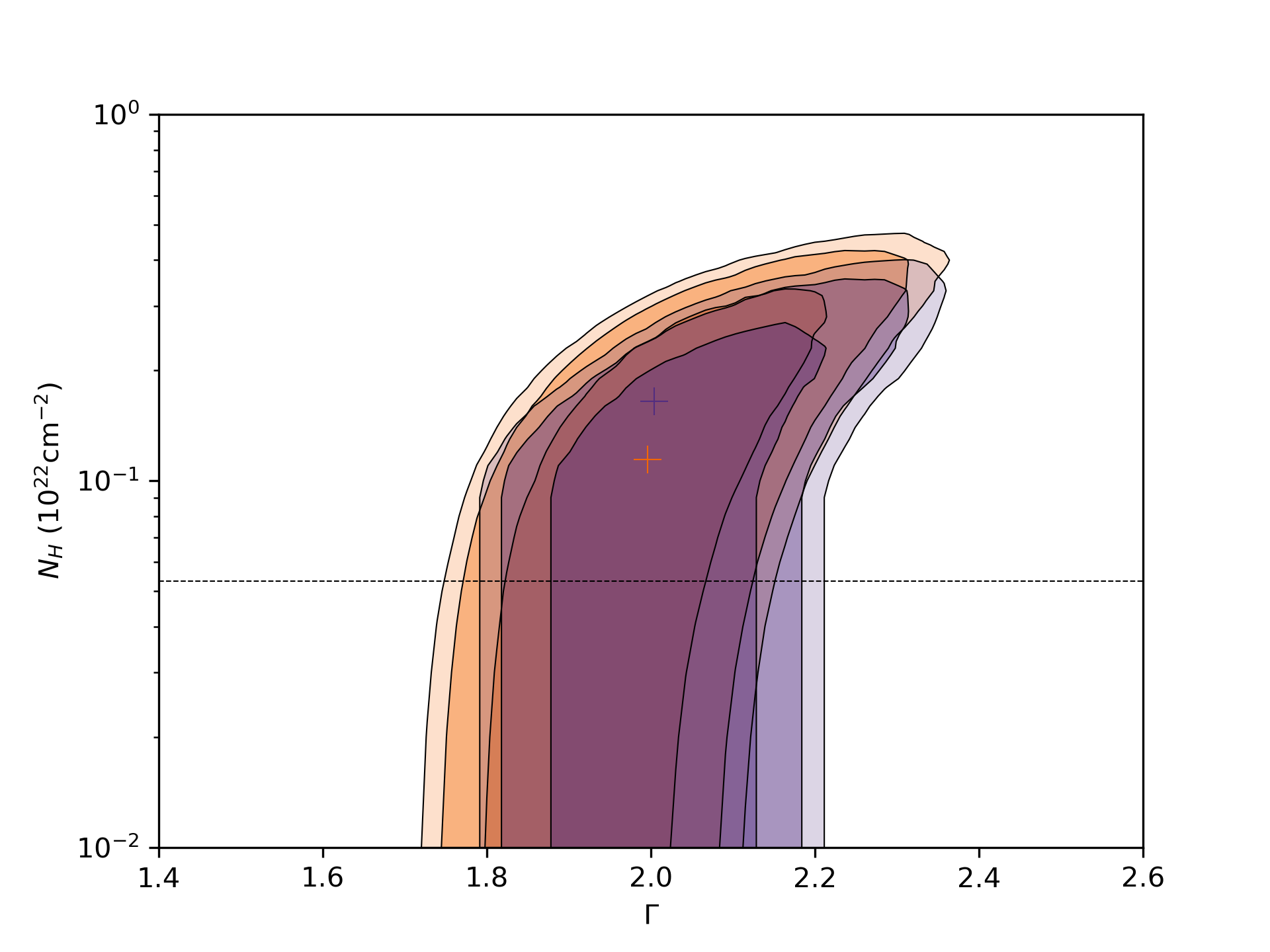}
     \caption{MCG\,+2-57-2. \textit{Top Left panel}: Spectral fit for the \texttt{borus02} model. \textit{Top Right panel}: Spectral fit for the \texttt{UXCLUMPY} model. \textit{Bottom Left panel}: $N_{\rm H,l.o.s}$ posteriors for MCG\,+2-57-2. The \texttt{borus02} results are in orange and the \texttt{UXCLUMPY} results are in purple. \textit{Bottom Right panel}: Contour plots of line-of-sight column density and photon index for MCG\,+2-57-2. The \texttt{borus02} results are in orange and the \texttt{UXCLUMPY} results are in purple. The hard spectral shape in the \cha\ data is due to pileup.}
     \label{fig:mcg_results}
\end{figure*}


\begin{figure*}[h!]
    \centering
    \includegraphics[scale=.25]{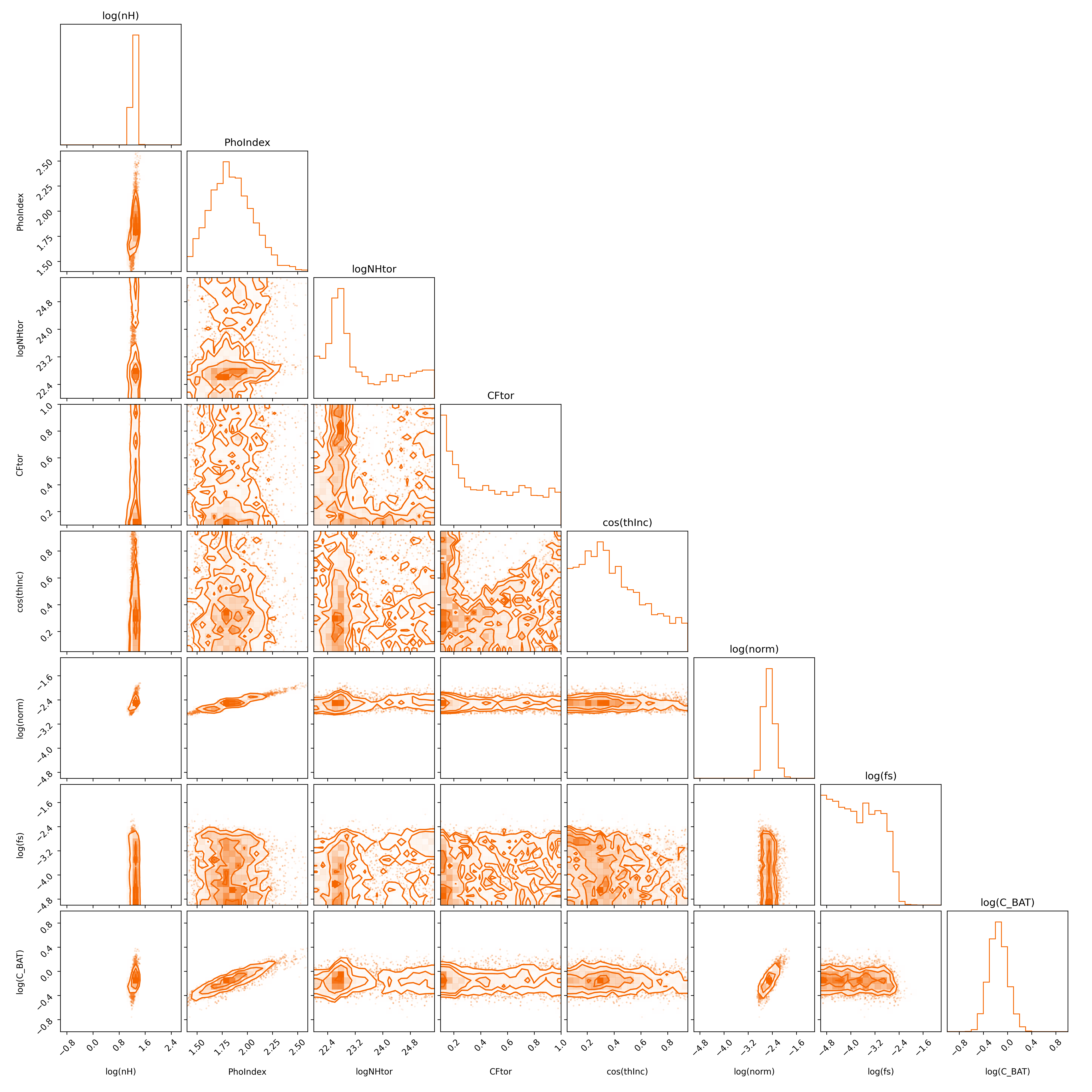}\hfill \\
    \includegraphics[scale=.25]{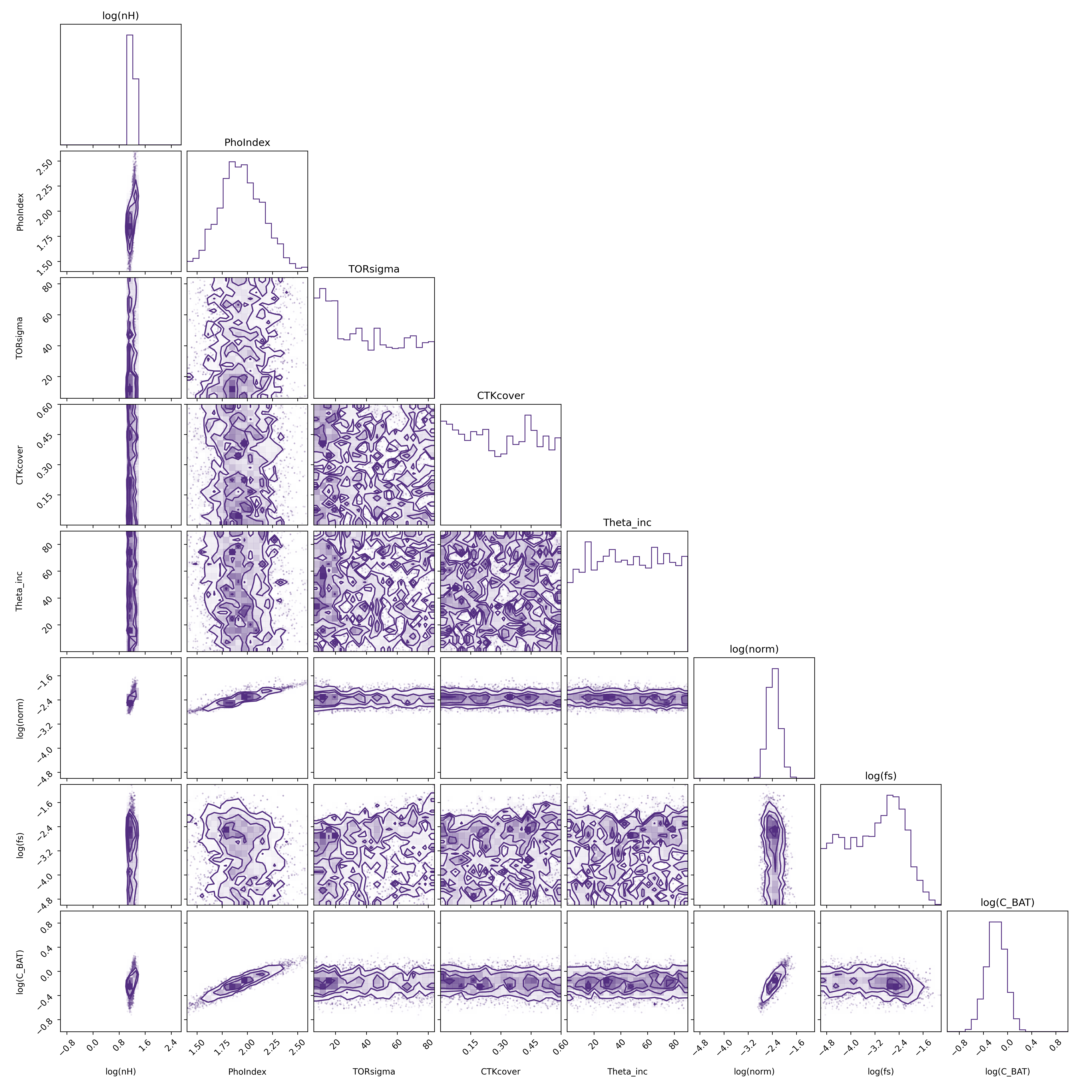}  
    \caption{2MFGC\,9836. \textit{Top panel}: Corner plot for the \texttt{borus02} model. \textit{Bottom Left panel}: Corner plot for the \texttt{UXCLUMPY} model.}
    \label{fig:2mfgc_corner}
\end{figure*}

\begin{figure*}[h!]
    \centering
    \includegraphics[scale=.20]{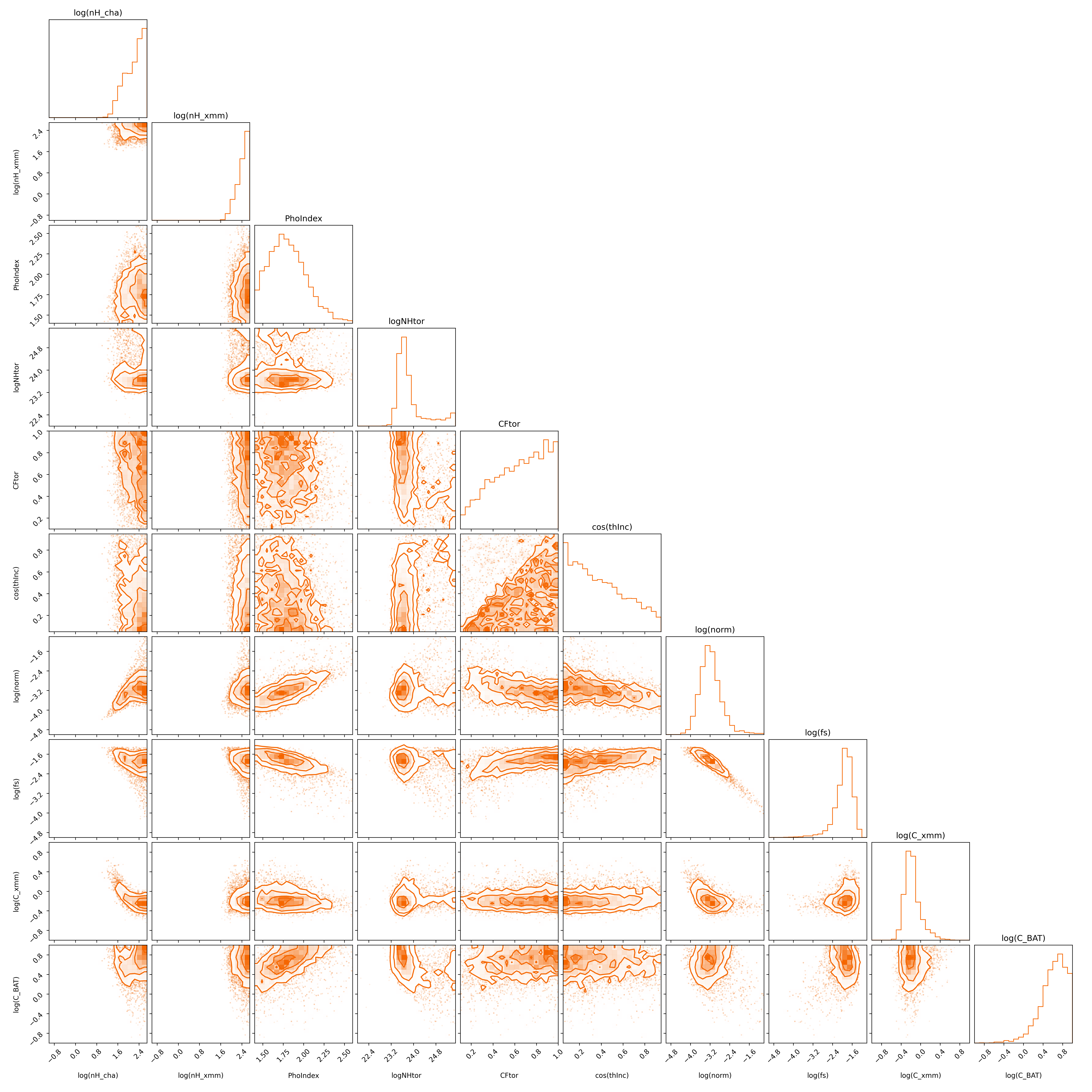}\hfill \\
    \includegraphics[scale=.20]{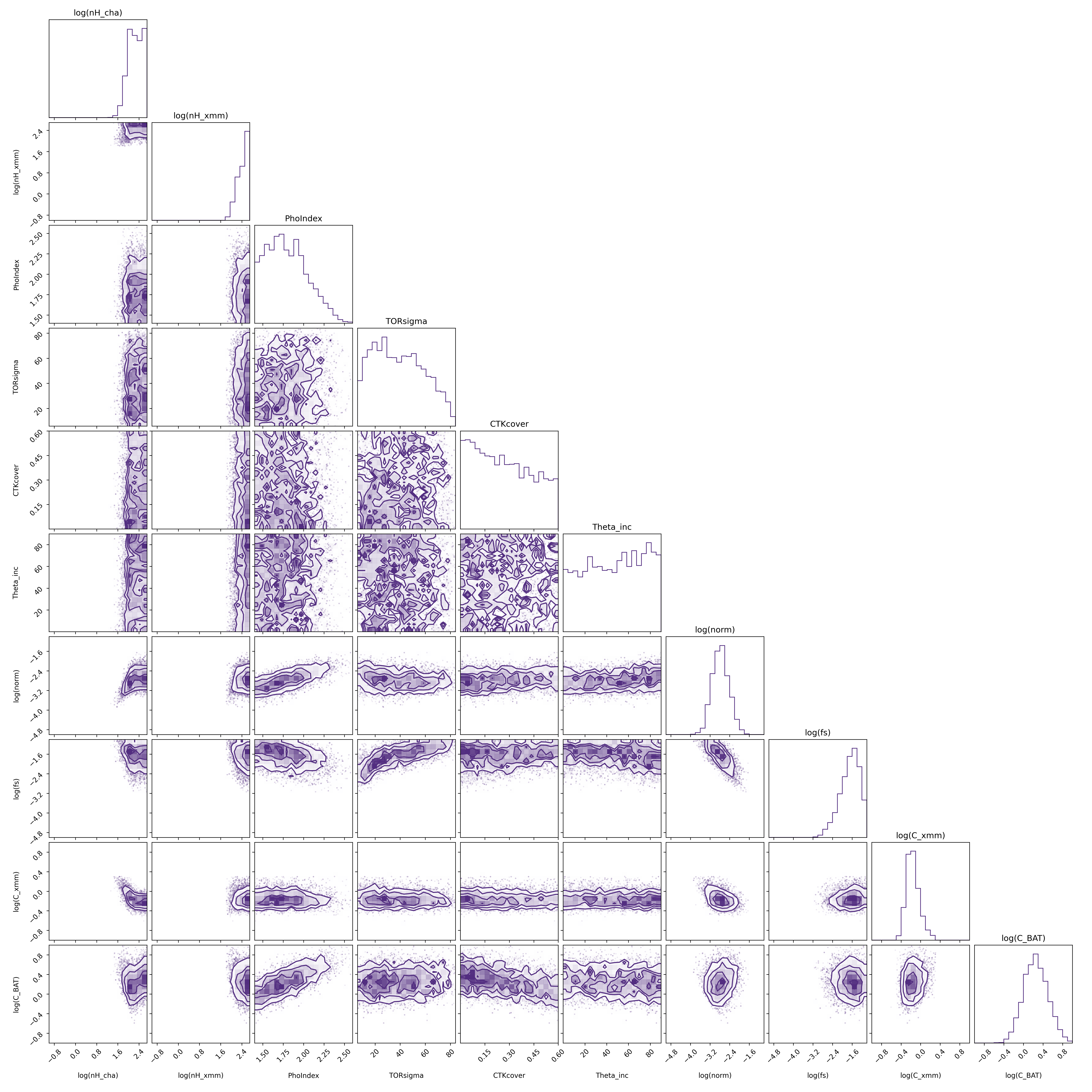}  
    \caption{NGC\,5759. \textit{Top panel}: Corner plot for the \texttt{borus02} model. \textit{Bottom Left panel}: Corner plot for the \texttt{UXCLUMPY} model.}
    \label{fig:ngc_corner}
\end{figure*}

\begin{figure*}[h!]
    \centering
    \includegraphics[scale=.25]{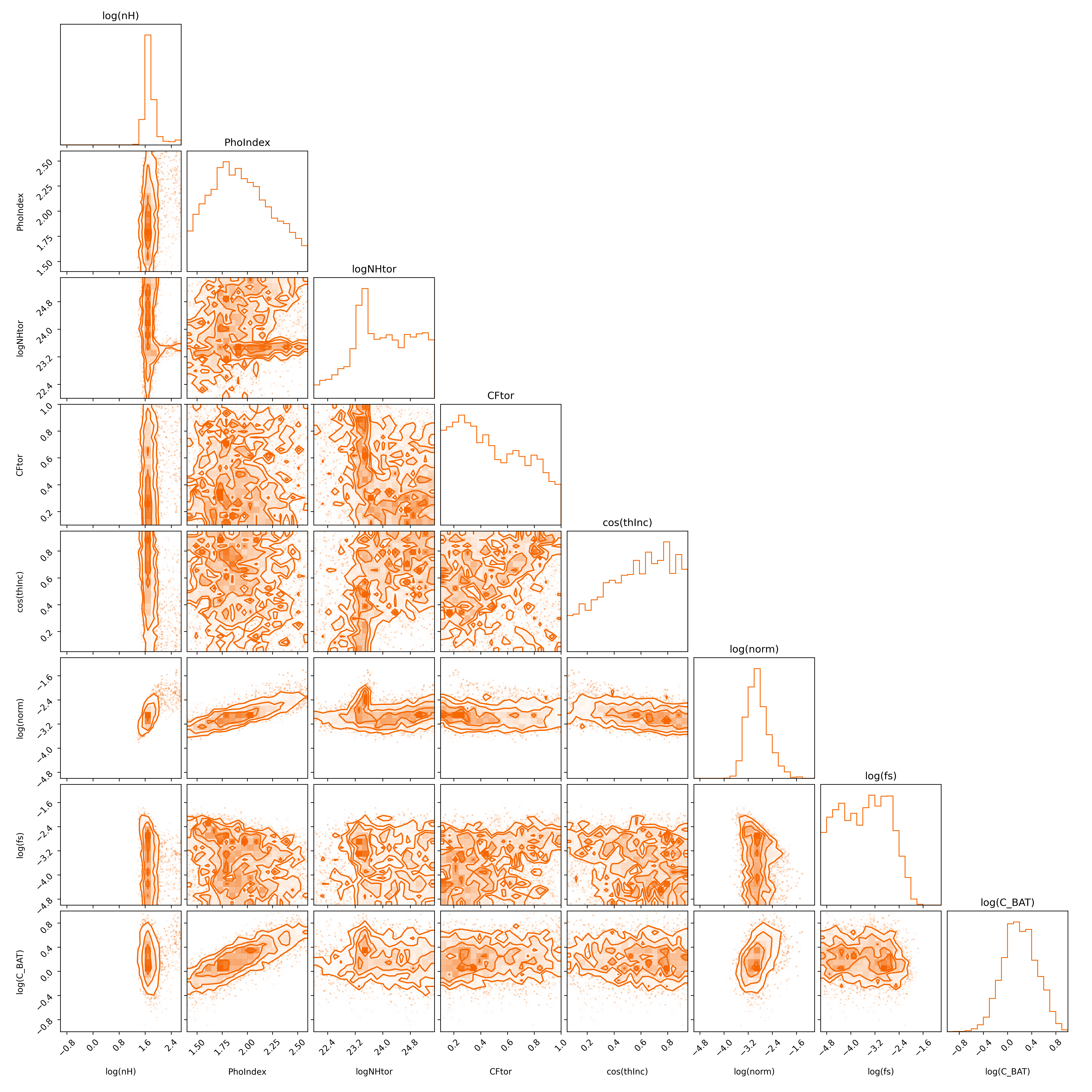}\hfill \\
    \includegraphics[scale=.25]{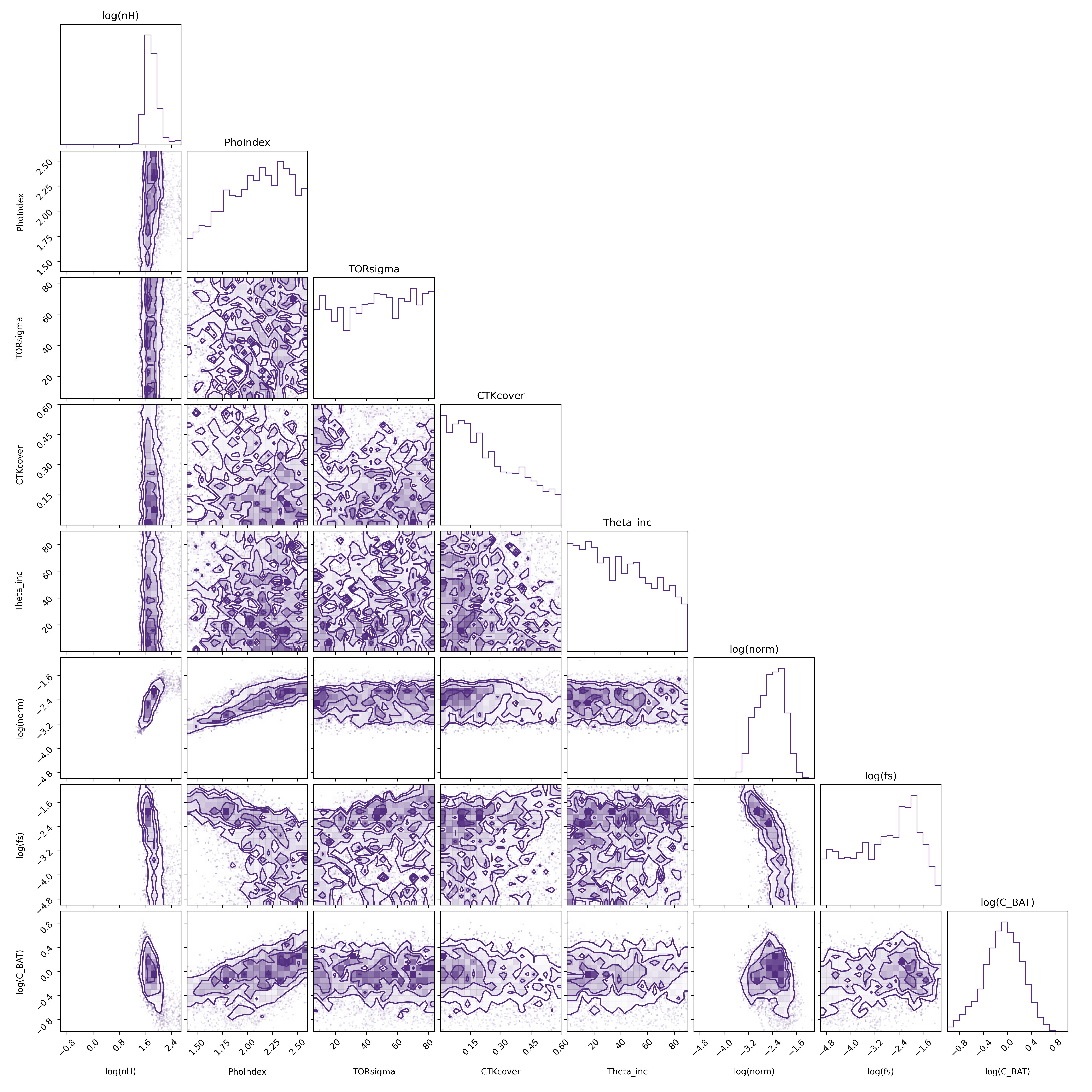}  
    \caption{IC\,1141. \textit{Top panel}: Corner plot for the \texttt{borus02} model. \textit{Bottom Left panel}: Corner plot for the \texttt{UXCLUMPY} model.}
    \label{fig:ic_corner}
\end{figure*}

\begin{figure*}[h!]
    \centering
    \includegraphics[scale=.3]{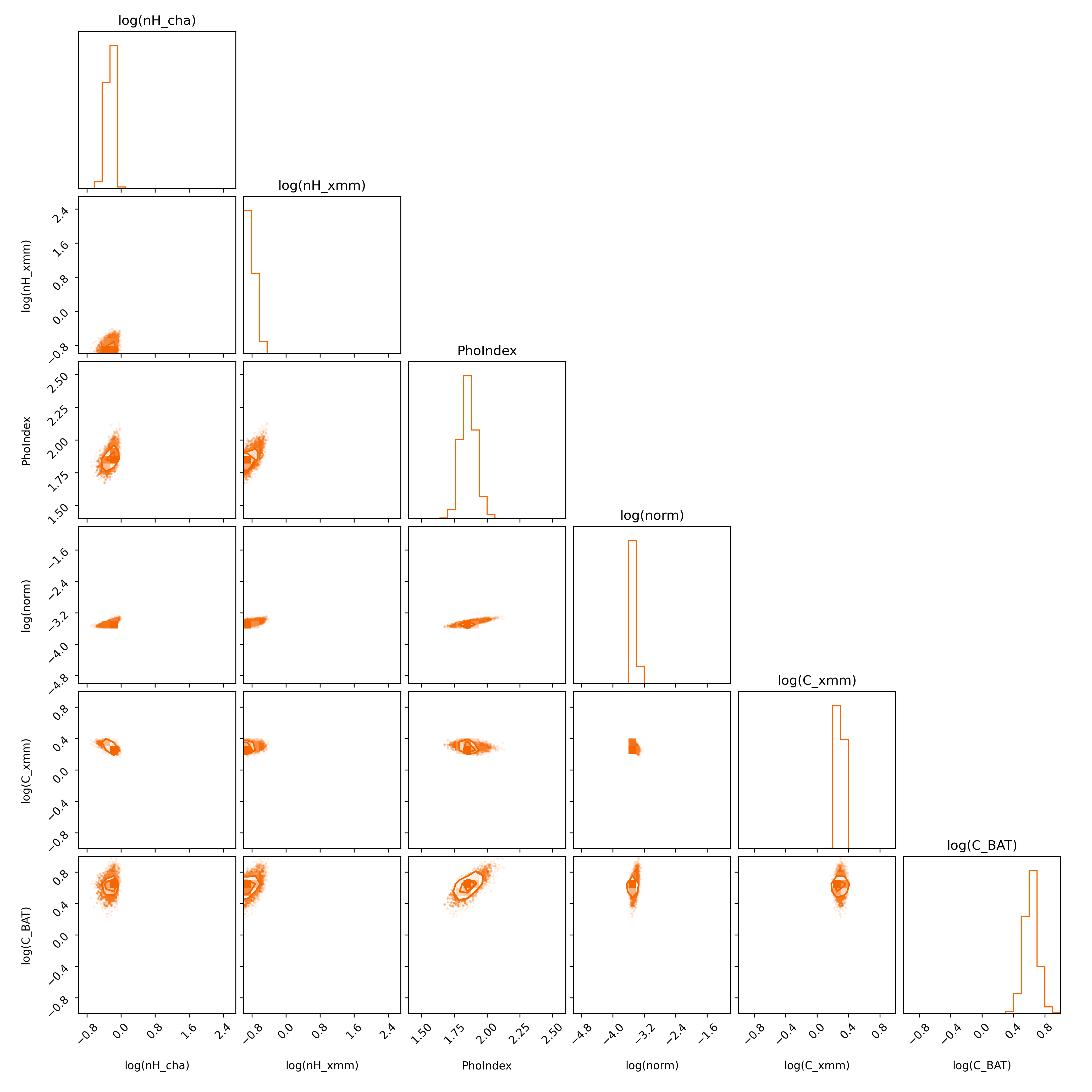}\hfill \\
    \includegraphics[scale=.3]{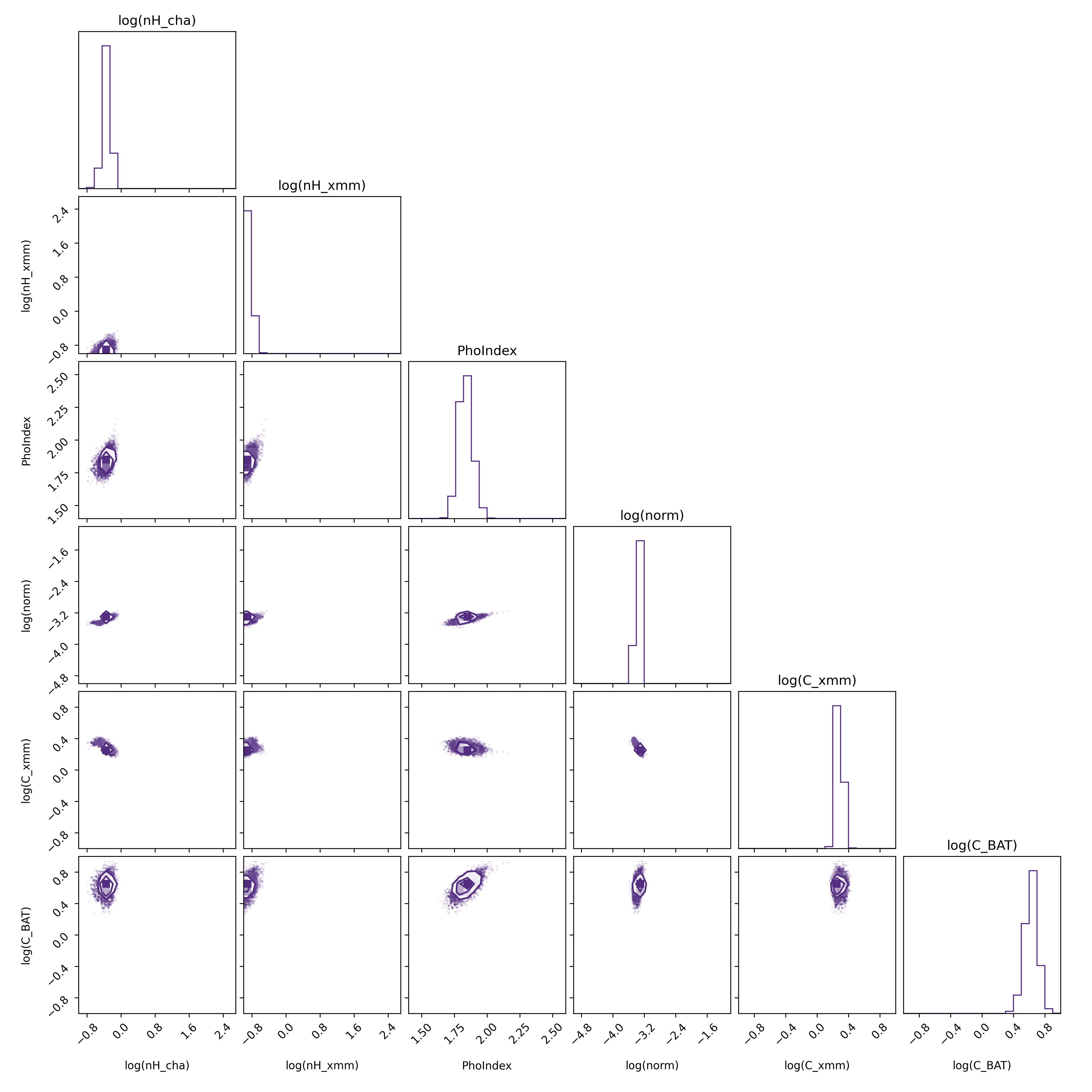}  
    \caption{2MASX\,J17253053--4510279. \textit{Top panel}: Corner plot for the \texttt{borus02} model. \textit{Bottom Left panel}: Corner plot for the \texttt{UXCLUMPY} model.}
    \label{fig:2masx_corner}
\end{figure*}

\begin{figure*}[h!]
    \centering
    \includegraphics[scale=.20]{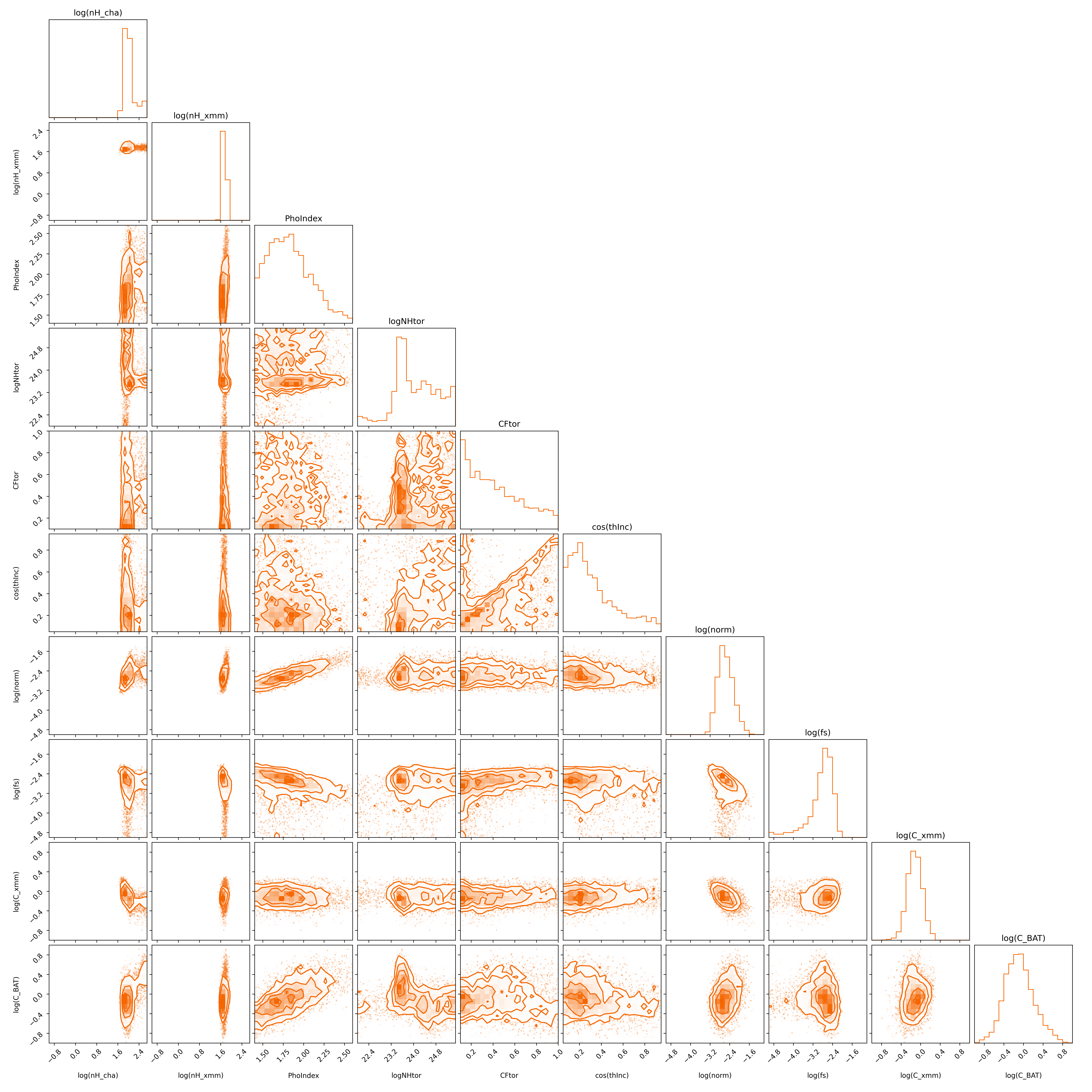}\hfill \\
    \includegraphics[scale=.20]{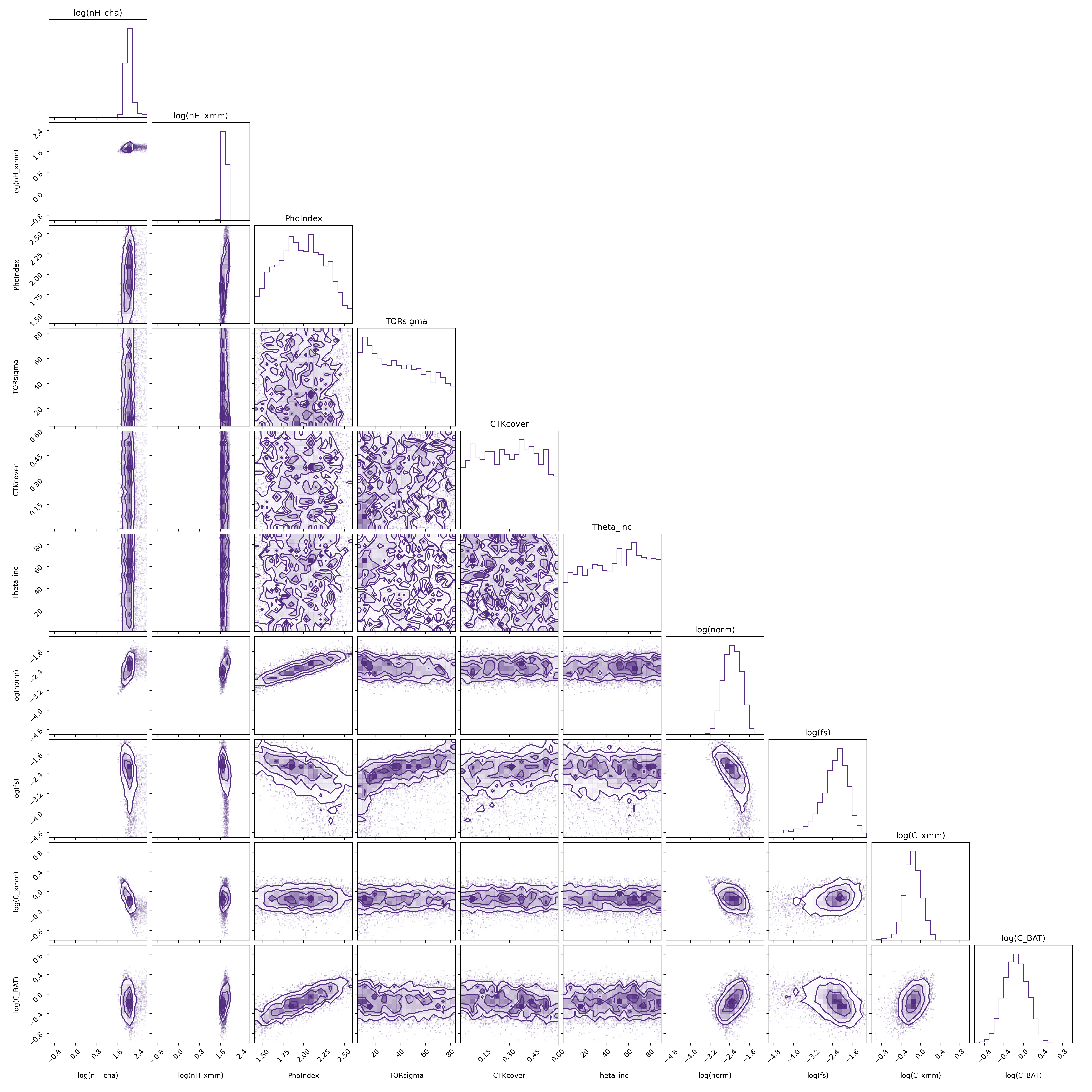}  
    \caption{CGCG\,1822.3+2053. \textit{Top panel}: Corner plot for the \texttt{borus02} model. \textit{Bottom Left panel}: Corner plot for the \texttt{UXCLUMPY} model.}
    \label{fig:cgc_corner}
\end{figure*}

\begin{figure*}[h!]
    \centering
    \includegraphics[scale=.33]{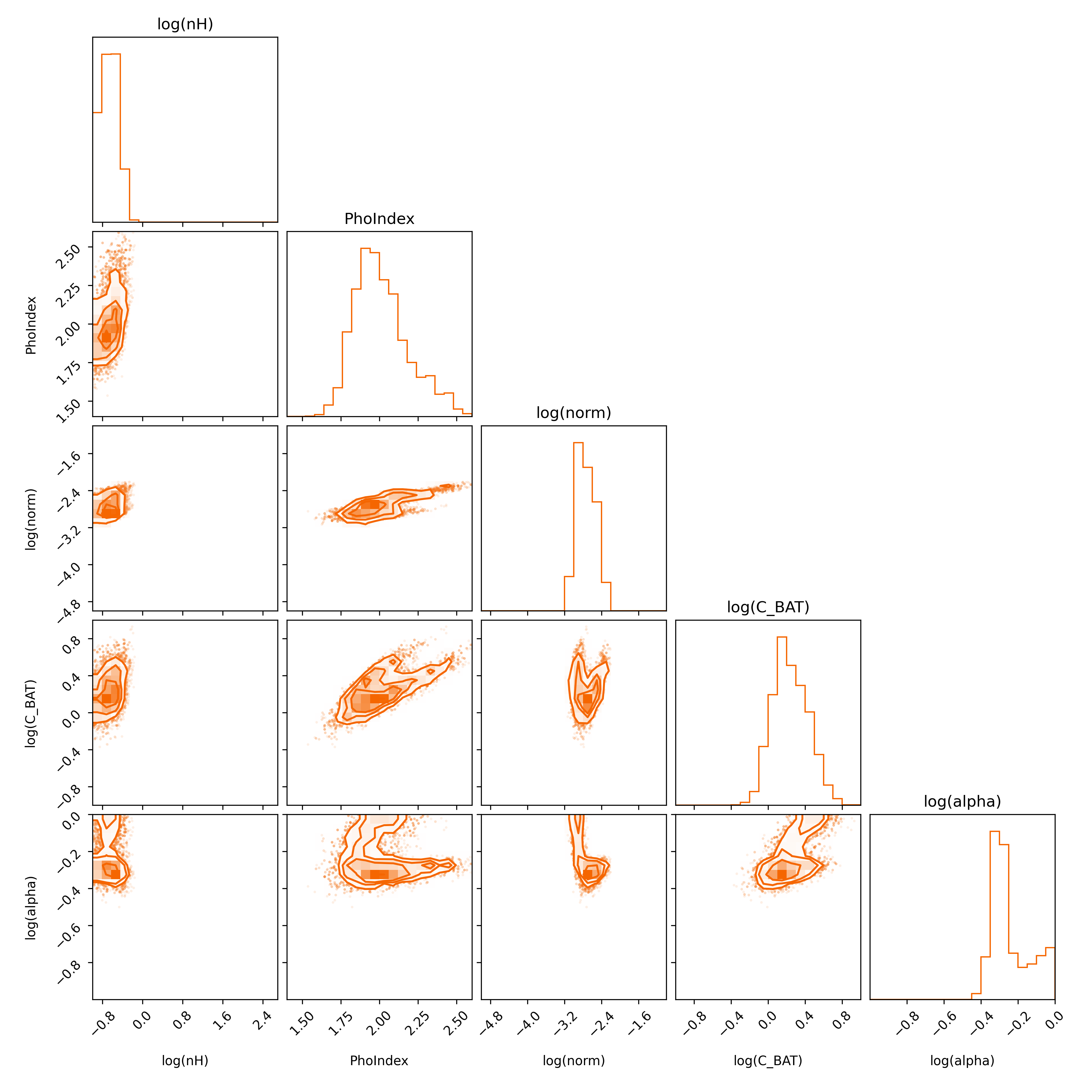}\hfill \\
    \includegraphics[scale=.33]{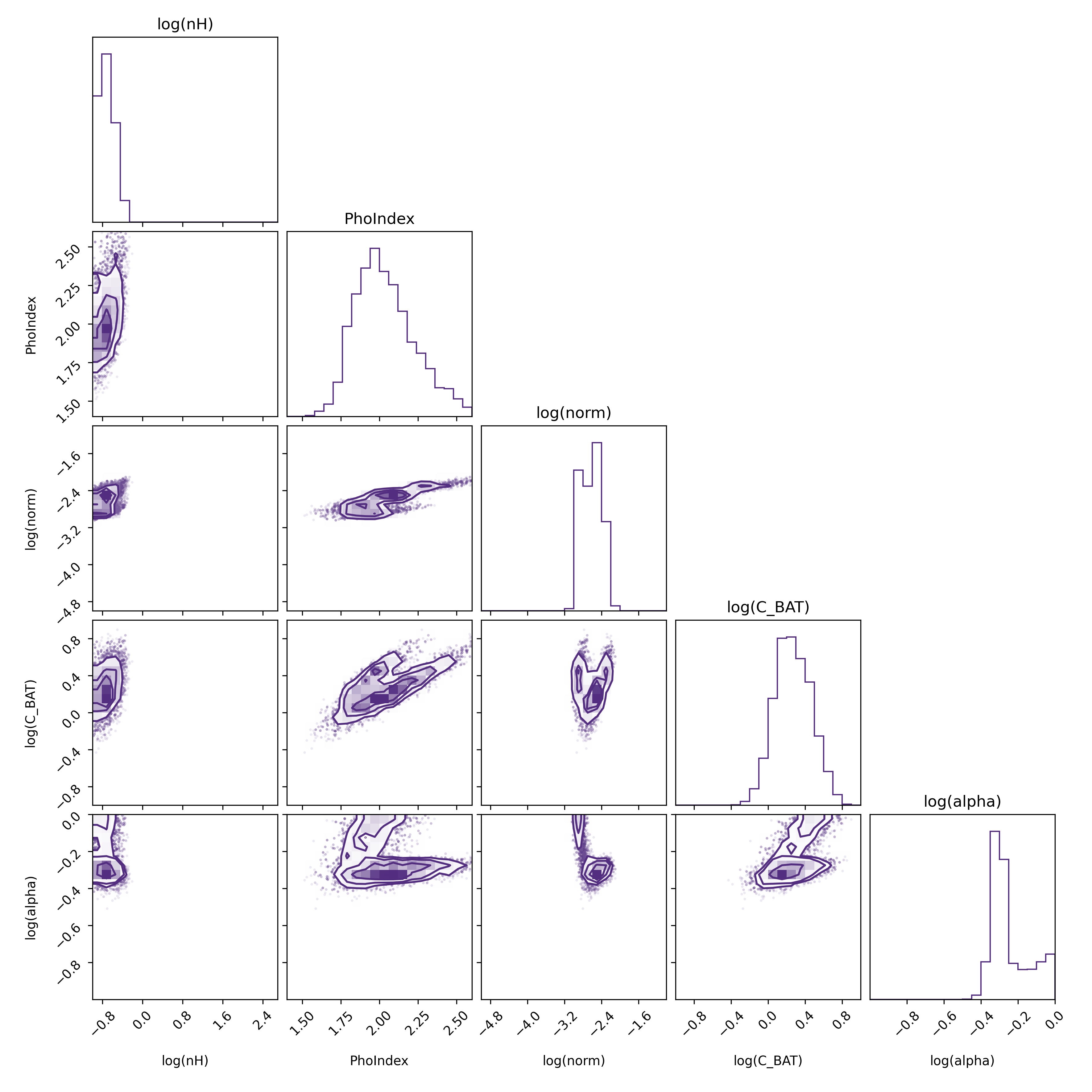}  
    \caption{MCG\,+2-57-2. \textit{Top panel}: Corner plot for the \texttt{borus02} model. \textit{Bottom Left panel}: Corner plot for the \texttt{UXCLUMPY} model.}
    \label{fig:mcg_corner}
\end{figure*}

\end{document}